\documentclass[a4paper,11pt]{article}
\pdfoutput=1
\usepackage[utf8]{inputenc}
\usepackage{jheppub}
\usepackage{empheq}
\usepackage{comment}
\usepackage{bm}
\usepackage[dvipsnames]{xcolor}
\usepackage{tikz}
\usepackage{titlesec}
\usepackage{accents}
\usepackage{simplewick}
\usepackage{simpler-wick}
\usepackage{multirow}

\titleformat*{\section}{\Large\bfseries}
\titleformat*{\subsection}{\large\bfseries}
\titleformat*{\subsubsection}{\bfseries}
\titlespacing*{\section}{0pt}{24pt}{14pt}
\titlespacing*{\subsection}{0pt}{18pt}{10pt}
\titlespacing*{\subsubsection}{0pt}{14pt}{6pt}

\newcommand{\I}{\mathrm{i}}
\newcommand{\D}{\mathrm{d}}

\newcommand{\C}{\mathbb{C}}

\newcommand{\Z}{\mathbb{Z}}

\renewcommand{\O}{\mathcal{O}}

\DeclareMathOperator{\Li}{Li}

\DeclareMathOperator{\K}{\mathcal{K}}

\DeclareMathOperator{\G}{\mathcal{G}}

\newcommand{\mf}{\chi}
\newcommand{\op}{\O^{[2]}}
\newcommand{\Op}{\O^{[3]}}
\newcommand{\src}{\phi^{[1]}}
\newcommand{\Src}{\phi^{[0]}}

\newcommand{\act}{\mathfrak{a}}

\newcommand{\<}{\langle}
\renewcommand{\>}{\rangle}
\newcommand{\lla}{\langle \! \langle}
\newcommand{\rra}{\rangle \! \rangle}

\newcommand{\bs}[1]{\boldsymbol{#1}}
\newcommand{\reg}[1]{\hat{#1}}
\newcommand{\dreg}{\reg{d}}
\newcommand{\Dreg}{\reg{\Delta}}
\newcommand{\areg}{\reg{\alpha}}
\newcommand{\breg}{\reg{\beta}}

\definecolor{darknavy}{RGB}{0,0,150} 
\definecolor{darkgreen}{rgb}{0,0.42,0.24}
\definecolor{darkred}{RGB}{177,35,35}  

\newcommand{\green}[1]{{\color{darkgreen} #1}}
\newcommand{\red}[1]{{\color{red} #1}}

\newcommand{\yellow}[1]{{\color{orange} #1}}

\newcommand{\nn}{\nonumber}

\newcommand{\ep}{\epsilon}
\newcommand{\z}{\zeta}

\newcommand{\ino}{i}
\newcommand{\ireg}{\reg{i}}
\newcommand{\iuv}{\ireg}
\newcommand{\idiv}{i^{\text{div}}}
\newcommand{\ifin}{i^{\text{fin}}}
\newcommand{\iren}{i^{\text{ren}}}

\newcommand{\s}[2]{\sigma_{(#1) #2}}
\newcommand{\m}[1]{l_{#1 -}}
\newcommand{\p}[1]{l_{#1 +}}

\newcommand{\Kreg}{\reg{\K}}
\newcommand{\Greg}{\reg{\G}}

\renewcommand{\[}{\begin{equation}}
\renewcommand{\]}{\end{equation}}

\newcommand{\indep}{\green{$0$ (indep)}}
\newcommand{\dep}{\yellow{$0$ (dep)}}
\newcommand{\one}{\red{1}}
\newcommand{\two}{\red{2}}

\newcommand{\nicebox}{\fboxsep=10pt\fbox}

\title{Handbook of derivative AdS  amplitudes} 

\author[a]{Adam Bzowski,}
\affiliation[a]{
Institute of Physics, 
University of Silesia,
75 Pu\l ku Piechoty 1, 
41-500 Chorz\'{o}w, 
Poland
}

\emailAdd{adam.bzowski@us.edu.pl}

\abstract{In the 2022 study, together with Paul McFadden and Kostas Skenderis, I analyzed tree-level 3- and 4-point Witten diagrams (amplitudes) of scalar operators in anti-de Sitter space in momentum space. This paper constitutes its extension to Witten diagrams with bulk interactions involving spacetime derivatives. In $d = 3$ boundary dimensions the Witten diagrams involving conformally coupled and massless scalars can be evaluated in closed form. Such cases are of interest in holographic cosmology and correspond to dual operators of conformal dimensions $\Delta = 2$ and $3$ respectively. I present explicit formulae for all such amplitudes and provide a Mathematica package serving as the repository of all the results. I discuss renormalization issues and show that, contrary to the expectation, even finite correlators may acquire non-trivial renormalization effects.
}

\begin{document}

\maketitle

\section{Introduction}

In \cite{Bzowski:2022rlz} we initiated a detailed study of regulated and renormalized, scalar, tree-level 3- and 4-point Witten diagrams in Euclidean anti-de Sitter (AdS) spacetime. The diagrams constitute building blocks of the correlation functions of the dual operators in the dual CFT. For this reason we refer to Witten diagrams as \emph{amplitudes}. 

\begin{figure}[thb]
\begin{tikzpicture}[scale=0.9]
\draw (0,0) circle [radius=3];
\draw [fill=black] (-2.121,-2.121) circle [radius=0.1];
\draw [fill=black] (-2.121, 2.121) circle [radius=0.1];
\draw [fill=black] ( 2.121,-2.121) circle [radius=0.1];
\draw [fill=black] ( 2.121, 2.121) circle [radius=0.1];
\draw [fill=black] ( 0, 0) circle [radius=0.1];
\draw (-2.121,-2.121) -- ( 2.121, 2.121);
\draw ( 2.121,-2.121) -- (-2.121, 2.121);
\node [left] at (-2.121, 2.2) {$\O_1(\bs{k}_1)$}; 
\node [left] at (-2.121,-2.2) {$\O_2(\bs{k}_2)$}; 	
\node [right] at ( 2.121,-2.2) {$\O_3(\bs{k}_3)$}; 
\node [right] at ( 2.121, 2.2) {$\O_4(\bs{k}_4)$}; 	
\node [above] at (-0.9, 1.06) {$\K_{[\Delta_1]}$};
\node [above] at (-1.3,-1.06) {$\K_{[\Delta_2]}$};
\node [above] at ( 1.2,-1.06) {$\K_{[\Delta_3]}$};
\node [above] at ( 0.8, 1.06) {$\K_{[\Delta_4]}$};
\node [right] at (0,0) {$\lambda_{1234}$};
\end{tikzpicture}
\qquad
\begin{tikzpicture}[scale=0.9]
\draw (0,0) circle [radius=3];
\draw [fill=black] (-2.121,-2.121) circle [radius=0.1];
\draw [fill=black] (-2.121, 2.121) circle [radius=0.1];
\draw [fill=black] ( 2.121,-2.121) circle [radius=0.1];
\draw [fill=black] ( 2.121, 2.121) circle [radius=0.1];
\draw [fill=black] (-1, 0) circle [radius=0.1];
\draw [fill=black] ( 1, 0) circle [radius=0.1];
\draw (-2.121,-2.121) -- (-1,0) -- (-2.121, 2.121);
\draw ( 2.121, 2.121) -- ( 1,0) -- ( 2.121,-2.121);
\draw (-1,0) -- (1,0);
\node [left] at (-2.121, 2.2) {$\O_1(\bs{k}_1)$}; 
\node [left] at (-2.121,-2.2) {$\O_2(\bs{k}_2)$}; 	
\node [right] at ( 2.121,-2.2) {$\O_3(\bs{k}_3)$}; 
\node [right] at ( 2.121, 2.2) {$\O_4(\bs{k}_4)$}; 	
\node [right] at (-1.5, 1.2) {$\K_{[\Delta_1]}$};
\node [right] at (-1.5, -1.2) {$\K_{[\Delta_2]}$};
\node [left] at ( 1.5, -1.2) {$\K_{[\Delta_3]}$};
\node [left] at ( 1.5, 1.2) {$\K_{[\Delta_4]}$};
\node [above] at (0,0) {$\G_{[\Delta_x]}$};
\node [right] at (1,0) {$\lambda_{34x}$};
\node [left] at (-1,0) {$\lambda_{12x}$};
\end{tikzpicture}
\centering
\caption{Witten diagram representing the contact and exchange 4-point amplitudes $\ino_{[\Delta_1 \Delta_2 \Delta_3 \Delta_4]}$ and $\ino_{[\Delta_1 \Delta_2; \Delta_3 \Delta_4 x \Delta_x]}$. External and internal lines correspond to bulk-to-boundary and bulk-to-bulk propagators in Euclidean AdS, while each interior point requires integration over its radial position. The bulk-to-boundary and bulk-to-bulk propagators are denoted by $\K_{[\Delta]}$ and $\G_{[\Delta]}$ respectively. The subscript $\Delta$ refers to the conformal dimension of the dual operator. Thus, the mass of the bulk scalar field $\Phi_{[\Delta]}$ is given by $m^2 = \Delta(\Delta - d)$, where the spacetime dimension of AdS equals $d+1$. All conventions are summarized in Appendix \ref{sec:AdS_conventions}.\label{fig:intro}}
\end{figure}
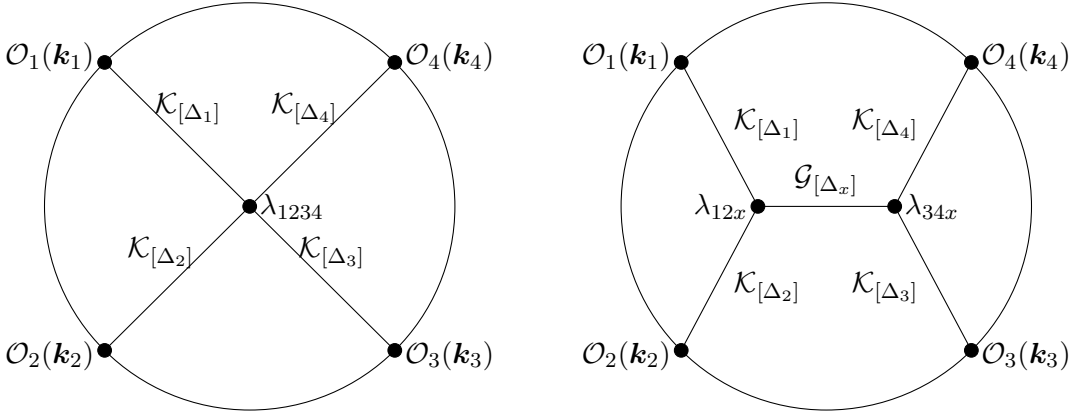

In \cite{Bzowski:2022rlz} we concentrated on 3- and 4-point amplitudes of the dual scalar operators of dimensions $\Delta=2,3$ in $d=3$ boundary dimensions (\textit{i.e.}, in 4-dimensional AdS spacetime). From the point of view of the bulk theory those cases correspond to conformally coupled and massless scalar fields respectively. We discussed and presented explicit formulae for all tree-level contact and exchange $4$-point amplitudes in momentum space, with exchanged scalar operators of dimension $\Delta=2,3$. Examples of such 4-point contact and exchange amplitudes are presented in Figure \ref{fig:intro}.

The amplitudes analyzed in \cite{Bzowski:2022rlz} are \emph{non-derivative amplitudes}, \textit{i.e.}, the bulk interactions of the fields do not involve any derivatives. For example, the Witten diagrams presented in Figure \ref{fig:intro} contain three interaction vertices between five bulk fields $\Phi_j$, with $j=1,2,3,4$ corresponding to the external operators and $j = x$ corresponding to the scalar field in the exchange channel. The interaction action leading to the diagrams in Figure \ref{fig:intro} reads
\begin{align} \label{intro_S4asym}
S_{\text{int}} & = \int \D^{d+1} x \sqrt{g} \left[ \lambda_{12x} \Phi_1 \Phi_2 \Phi_x + \lambda_{34x} \Phi_x \Phi_3 \Phi_4 - \lambda_{1234} \Phi_1 \Phi_2 \Phi_3 \Phi_4 \right].
\end{align}
Notice that there are no derivatives acting on the fields in the interaction action. For example, the 4-point contact amplitude in the left panel of Figure \ref{fig:intro} in momentum space is given by the integral of the product of four bulk-to-boundary propagators,
\begin{align} \label{intro_4ptC}
\ino_{[\Delta_1 \Delta_2 \Delta_3 \Delta_4]} = \int_0^\infty \D z \, z^{-d-1} \K_{[\Delta_1]}(z, k_1) \K_{[\Delta_2]}(z, k_2) \K_{[\Delta_3]}(z, k_3) \K_{[\Delta_4]}(z, k_4).
\end{align}
Here $\K_{[\Delta]}$ denotes the bulk-to-boundary propagator in Poincar\'{e} coordinates and $k_j = | \bs{k}_j |$ is the momentum length. See Appendix \ref{sec:conventions} for precise definitions.

In this paper I extend the analysis of \cite{Bzowski:2022rlz} to scalar fields with interactions involving derivatives. Since we are interested in 3- and 4-point functions only, we consider only cubic and quartic bulk couplings. We concentrate mostly on 2-derivative vertices; schematically, this means that we consider bulk AdS actions for scalar fields involving the following couplings,
\begin{align}
& \int \D^{d+1} x \sqrt{g} \, \Phi_1 \nabla_{\mu} \Phi_2 \nabla^{\mu} \Phi_3, && \int \D^{d+1} x \sqrt{g} \, \Phi_1 \Phi_2 \nabla_{\mu} \Phi_3 \nabla^{\mu} \Phi_4,
\end{align}
with the contractions by the Euclidean AdS metric $g_{\mu\nu}$. We will also consider the 4-derivative 4-point coupling
\begin{align}
\int \D^{d+1} x \sqrt{g} \, \nabla_{\mu} \Phi_1 \nabla^{\mu} \Phi_2 \nabla_{\nu} \Phi_3 \nabla^{\nu} \Phi_4.
\end{align}
Such interactions result in amplitudes with radial derivatives spread throughout expressions such as \eqref{intro_4ptC}. We will provide precise definitions in Section \ref{sec:amplitudes}.

This article is accompanied by the Mathematica package \verb|HandbooK|, which gathers all results from this paper as well as \cite{Bzowski:2022rlz} and the upcoming paper \cite{toappear} on cosmology. This includes all 2-, 3-, and 4-point amplitudes both in AdS as well as the de Sitter spacetime. Our hope is that the three papers together constitute the most comprehensive and useful set of tools available to a wide range of researchers. To this end, we provide full documentation including a set of accompanying Mathematica notebooks.

Apart from listing the results of our calculations, we discuss a number of new features and peculiarities discovered in the study of derivative amplitudes. In particular, we find the following:
\begin{enumerate}
\item All derivative 4-point exchange amplitudes reduce to combinations of non-derivative 4-point exchange amplitudes and derivative as well as non-derivative 4-point contact amplitudes. For this reason we concentrate on derivative contact amplitudes.
\item We find out that some amplitudes, despite being finite, remain scheme-dependent. This means that the memory of the regularization is retained in the finite amplitude, even though the regulator can be removed from the final expression. We discuss both the mathematical and physical reasons for this new behavior. This implies that some explicit expressions found in literature may be non-unique. In particular, additional caution is required when deriving precise numerical predictions, which can be scheme-dependent.
\item Consequently we show that even finite amplitudes may undergo non-trivial renormalization. This important discovery goes against the naive expectation that finite amplitudes are fixed and do not require renormalization. What is more, the local counterterms induced by renormalization may significantly affect higher-point functions. For example, in some amplitudes scale-dependence can be completely removed by a suitable choice of the renormalization scheme.
\end{enumerate}

The layout of this paper is as follows. In the remainder of the Introduction we present motivation and the review of the relevant literature. In Section \ref{sec:amplitudes} we present the objects of interest: Witten diagrams and their regularization. In Section \ref{sec:identities} we derive a number of important identities between various derivative and non-derivative amplitudes. In Section \ref{sec:action_to_amplitudes} we make a connection between amplitudes and boundary correlation functions they build. In Sections \ref{sec:3pt}, \ref{sec:4ptC} and \ref{sec:4ptX} we proceed to evaluate 3- and 4-point contact and exchange diagrams. Due to the sheer number of amplitudes, we present only a small fraction of the more interesting amplitudes. We carefully renormalize all the amplitudes and discover and discuss some of their peculiarities. In Section \ref{sec:symmetric} we apply the results to a more physical AdS theory with a Lagrangian containing a massless field interacting through derivative vertices only. Finally, Section \ref{sec:package} summarizes the Mathematica packages accompanying this paper, which contain a complete record of all our results. The paper contains two appendices. In Appendix \ref{sec:conventions} we summarize our conventions and definitions for QFT, AdS and momenta. Appendix \ref{sec:amp_defs} lists the definitions of all derivative amplitudes used in this paper and the Mathematica package.

\subsection{Motivation and review of the literature}

Recently, a renewed interest in amplitudes in momentum space in both anti-de Sitter (AdS) and de Sitter (dS) spacetimes can be observed. The main motivation stems from holographic cosmology, \cite{McFadden:2009fg,McFadden:2010na,Maldacena:2011nz,Bzowski:2011ab,Bzowski:2012ih,Mata:2012bx,McFadden:2013ria,Anninos:2014lwa}, where cosmological predictions at late times can be derived from the dual QFT. More recently, the rejuvenation of the subject came from the works on cosmological bootstrap, \cite{Arkani-Hamed:2015bza,Arkani-Hamed:2018kmz,Baumann:2019oyu,Baumann:2020dch,Sleight:2019hfp,Wang:2022eop,Arkani-Hamed:2023kig}, where late-time correlators in four-dimensional de Sitter are constructed by solving three-dimensional conformal Ward identities.

The approach \textit{via} cosmological bootstrap has already a deep impact on our understanding of both dS and AdS theories, including the general structure of correlators, their IR and UV properties, properties of the wavefunction of the universe or analyticity and unitarity of de Sitter theories, \cite{Baumann:2021fxj,Sleight:2019mgd,Sleight:2020obc,Sleight:2021plv,Stefanyszyn:2023qov,DiPietro:2021sjt,Meltzer:2021zin,Iacobacci:2022yjo,Goodhew:2022ayb,Salcedo:2022aal,Agui-Salcedo:2023wlq,Melville:2023kgd,Cespedes:2023aal,Armstrong:2022vgl}. It is not, however, free of some drawbacks. First, cosmological bootstrap is unable to identify the underlying bulk theory. With the amplitudes obtained by the solutions to conformal Ward identities, it is difficult to establish which bulk Lagrangian produces a given correlator. Second, with the amplitudes related to each other by an intricate web of identities, it is often difficult to obtain explicit analytic expressions. For this reason exact expressions for both dS and AdS amplitudes in momentum space are scarce. Third, the combinations of the amplitudes obtained by conformal bootstrap are usually less singular than the ones obtained by the direct calculations. In particular, conformal bootstrap may not deal with renormalization issues correctly.

This paper, together with \cite{Bzowski:2022rlz,toappear}, addresses the issues stated above by explicitly calculating 3- and 4-point amplitudes in scalar theories. The calculated amplitudes correspond to specific interaction vertices determined by the bulk actions. Motivated by holographic cosmology, we concentrate on correlation functions of the dual scalar operators with dimensions $\Delta=2,3$ in three spacetime dimensions. These operators are dual to conformally coupled and massless bulk scalars, with the latter modelling the inflaton. Our explicit calculations provide simple, exact, analytic expressions.

While we concentrate on AdS amplitudes, there exists the closed and direct relation between Witten diagrams in anti-de Sitter and de Sitter universes. The relation, worked out in \cite{McFadden:2009fg,McFadden:2010na,Sleight:2020obc,Sleight:2021plv,toappear}, provides a one-to-one correspondence between AdS and dS amplitudes. Whether in the setting of de Sitter cosmology or Euclidean AdS, exact expressions for amplitudes in momentum space remain scarce. Most approaches to the evaluation of the amplitudes are based on reduction schemes, weight-shifting or spin-changing operators, \cite{Raju:2010by,Raju:2012zr,Raju:2012zs,Arkani-Hamed:2015bza,Arkani-Hamed:2017fdk,Arkani-Hamed:2018kmz,Albayrak:2019asr,Baumann:2019oyu, Baumann:2020dch,Sleight:2019mgd,Sleight:2019hfp,Baumann:2021fxj,Armstrong:2022mfr,Chen:2023xlt,Lee:2022fgr}. While evaluating amplitudes in principle, those methods usually produce long, complicated expressions, unwieldy for practical purposes. Thus, very few explicit expressions for larger classes of 4-point amplitudes exist in the literature. Furthermore, those results usually include only finite amplitudes, often related to higher-spin cases or living in higher spacetime dimensions. Some examples can be found in \cite{Groote:2018rpb, Arkani-Hamed:2018kmz, Isono:2018rrb, Albayrak:2018tam, Albayrak:2019asr, Maglio:2019grh, Isono:2019wex, Coriano:2019nkw, Albayrak:2020isk, Bonifacio:2022vwa,Albayrak:2023kfk,Albayrak:2023jzl,Chowdhury:2023khl}. Most recently, a larger study of correlators, mostly in higher dimensions, and overlapping with this paper, was undertaken in \cite{Albayrak:2023kfk}.

Finally, momentum space approach is well-suited for the analysis of renormalization. Indeed, many of the amplitudes discussed here and in the previous paper \cite{Bzowski:2022rlz} exhibit singularities and require careful renormalization. The extraction of divergences is best performed in momentum space  \cite{Bzowski:2015pba,Bzowski:2015yxv,Bzowski:2017poo,Bzowski:2018fql}, and as such we work throughout in momentum space.

\section{Amplitudes} \label{sec:amplitudes}

In \cite{Bzowski:2022rlz} we calculated 3- and 4-point \emph{amplitudes}, \textit{i.e.}, the 3- and 4-point \emph{Witten diagrams}, exemplified by Figure \ref{fig:intro} in the Introduction. External and internal lines correspond to bulk-to-boundary and bulk-to-bulk propagators in Euclidean AdS, while each interior point requires integration over its radial position. The bulk-to-boundary and bulk-to-bulk propagators are denoted by $\K_{[\Delta]}$ and $\G_{[\Delta]}$ respectively. The subscript $\Delta$ refers to the conformal dimension of the dual operator. Thus, the mass of the bulk scalar field $\Phi_{[\Delta]}$ is given by
\begin{align}
m^2 = \Delta(\Delta - d),
\end{align}
where the spacetime dimension of AdS equals $d+1$. All conventions are summarized in Appendix \ref{sec:AdS_conventions}.

In this paper I extend the analysis to scalar fields with interactions involving derivatives. Since we are interested in 3- and 4-point functions only, we consider only cubic and quartic bulk couplings. This results in spacetime derivatives $\nabla_\mu$ acting upon propagators in Witten diagrams in position space. The spacetime indices must then be contracted to form a scalar. Finally, we Fourier transform the expressions to momentum space.

In the remainder of this section we define the most useful amplitudes for the remainder of the paper. The full list of all derivative amplitudes is presented in Appendix \ref{sec:amp_defs}.

\subsection{Definitions}

At this point, we will not make reference to any specific bulk action: rather, we will simply focus on the individual amplitudes defined by the expressions below, postponing consideration of their relation to the bulk action.

For example, the 3-point exchange amplitude with derivative vertices, in position space, is defined as
\begin{align} \label{i3D_position}
\ino_{[\Delta_1 \contraction[0.5ex]{}{\Delta}{{}_2}{\Delta} \Delta_2 \Delta_3]}(\bs{x}_1, \bs{x}_2, \bs{x}_3) & = \int \D^{d+1} x \, \sqrt{g} \, \K_{[\Delta_1]}(\bs{x}_1; x) \frac{\partial}{\partial x^\mu} \K_{[\Delta_2]}(\bs{x}_2; x) \frac{\partial}{\partial x_\mu} \K_{[\Delta_3]}(\bs{x}_3; x),
\end{align}
where $x = (z, \bs{x})$ with $z$ representing the radial variable and $\bs{x}$ the boundary coordinates. The `Wick contractions' in $[\Delta_1 \contraction[0.5ex]{}{\Delta}{{}_2}{\Delta} \Delta_2 \Delta_3]$ determine on which propagators the derivatives act and how are they contracted\footnote{In principle, all derivatives are covariant with respect to the AdS metric. However, in this paper we will almost exclusively consider single derivatives acting on scalar quantities, thus neither the order nor covariance actually matter. The only exception is the AdS Laplacian, $\Box = \nabla_\mu \nabla^\mu$.}. When Fourier transformed with respect to $\bs{x}_j$ this becomes
\begin{align} \label{i3D_example}
& \ino_{[\Delta_1 \contraction[0.5ex]{}{\Delta}{{}_2}{\Delta} \Delta_2 \Delta_3]}(k_1, k_2, k_3) = \int_0^{\infty} \D z \, z^{-d-1} \K_{[\Delta_1]}(z, k_1) \times\nn\\
& \qquad \times \left[ - \bs{k}_{2} \cdot \bs{k}_3 \K_{[\Delta_2]}(z, k_2) \K_{[\Delta_3]}(z, k_3) + \partial_z \K_{[\Delta_2]}(z, k_2) \partial_z \K_{[\Delta_3]}(z, k_3) \right],
\end{align}
where $k_j = | \bs{k}_j |$ is the length of the momentum. Thus, in momentum space, we can replace the derivative $\partial_\mu$ by the momentum-space operator
\begin{empheq}[box=\nicebox]{align} \label{Dm}
\mathcal{D}^{m}(z, \bs{k}) = z \left( \I \bs{k}^m + \hat{\bs{z}} \partial_z \right),
\end{empheq}
where $m = 1, \ldots, d$ represents the boundary coordinates, which are contracted with the Euclidean metric, $\delta_{mn}$. Furthermore, $\hat{\bs{z}}$ is a unit vector, $\hat{\bs{z}} \cdot \hat{\bs{z}} = 1$, orthogonal to the boundary directions, $\hat{\bs{z}} \cdot \bs{k} = 0$. In this way we can rewrite \eqref{i3D_example} as
\begin{empheq}[box=\nicebox]{align} \label{3pt2d}
\ino_{[\Delta_1 \contraction[0.5ex]{}{\Delta}{{}_2}{\Delta} \Delta_2 \Delta_3]} & = \int_0^{\infty} \D z \, z^{-d-1} \K_{[\Delta_1]}(z, k_1) [ \mathcal{D}^{m} \K_{[\Delta_2]}](z, \bs{k}_2) [ \mathcal{D}_{m} \K_{[\Delta_3]}](z, \bs{k}_3).
\end{empheq}
We dropped the arguments of the operators $\mathcal{D}^m$ as they correspond to the arguments of the bulk-to-boundary propagator they act on, \textit{e.g.}, $[ \mathcal{D}^{m} \K_{[\Delta_2]}](z, \bs{k}_2) = \mathcal{D}^{m}(z, \bs{k}_2) \K_{[\Delta_2]}(z, k_2)$ and so on.

We define the contact 4-point amplitudes analogously,
\begin{empheq}[box=\nicebox]{align} \label{4ptC2d}
& \ino_{[\Delta_1 \Delta_2 \contraction[0.5ex]{}{\Delta}{{}_3}{\Delta} \Delta_3 \Delta_4]}(k_1, k_2, k_3, k_4, s) \nn\\
& \qquad = \int_0^\infty \D z \, z^{-d-1} \K_{[\Delta_1]}(z, k_1) \K_{[\Delta_2]}(z, k_2) [\mathcal{D}^{m} \K_{[\Delta_3]}](z, \bs{k}_3) [\mathcal{D}_{m} \K_{[\Delta_4]}](z, \bs{k}_4)
\end{empheq}
Note that unlike the non-derivative contact amplitude $\ino_{[\Delta_1 \Delta_2 \Delta_3 \Delta_4]}$, the derivative amplitude does depend on the Mandelstam variable $s = | \bs{k}_1 + \bs{k}_2|$. We define the 4-derivative amplitude $\ino_{[\contraction[0.5ex]{}{\Delta}{{}_1}{\Delta} \Delta_1 \Delta_2 \contraction[0.5ex]{}{\Delta}{{}_3}{\Delta} \Delta_3 \Delta_4]}$ analogously, see Appendix \ref{sec:amp_defs} for details.

\begin{figure}[bht]
\begin{tikzpicture}[scale=1.0]
\draw (0,0) circle [radius=3];
\draw [fill=black] (-2.121,-2.121) circle [radius=0.1];
\draw [fill=black] (-2.121, 2.121) circle [radius=0.1];
\draw [fill=black] ( 2.121,-2.121) circle [radius=0.1];
\draw [fill=black] ( 2.121, 2.121) circle [radius=0.1];
\draw [fill=black] (-1, 0) circle [radius=0.1];
\draw [fill=black] ( 1, 0) circle [radius=0.1];
\draw (-2.121,-2.121) -- (-1,0) -- (-2.121, 2.121);
\draw ( 2.121, 2.121) -- ( 1,0) -- ( 2.121,-2.121);
\draw (-1,0) -- (1,0);
\draw [red] (-1.4,-0.52) -- (-1.1,0) -- (-1.4,0.52);
\node [left] at (-1.4,-0.52) {$\nabla_\mu$};
\node [left] at (-1.4, 0.52) {$\nabla^\mu$};
\node [left] at (-2.121, 2.2) {$\O_1(\bs{k}_1)$}; 
\node [left] at (-2.121,-2.2) {$\O_2(\bs{k}_2)$}; 	
\node [right] at ( 2.121,-2.2) {$\O_3(\bs{k}_3)$}; 
\node [right] at ( 2.121, 2.2) {$\O_4(\bs{k}_4)$}; 	
\node [right] at (-1.5, 1.2) {$\K_{[\Delta_1]}$};
\node [right] at (-1.5, -1.2) {$\K_{[\Delta_2]}$};
\node [left] at ( 1.5, -1.2) {$\K_{[\Delta_3]}$};
\node [left] at ( 1.5, 1.2) {$\K_{[\Delta_4]}$};
\node [above] at (0,0) {$\G_{[\Delta_x]}$};
\end{tikzpicture}
\qquad
\begin{tikzpicture}[scale=1.0]
\draw (0,0) circle [radius=3];
\draw [fill=black] (-2.121,-2.121) circle [radius=0.1];
\draw [fill=black] (-2.121, 2.121) circle [radius=0.1];
\draw [fill=black] ( 2.121,-2.121) circle [radius=0.1];
\draw [fill=black] ( 2.121, 2.121) circle [radius=0.1];
\draw [fill=black] (-1, 0) circle [radius=0.1];
\draw [fill=black] ( 1, 0) circle [radius=0.1];
\draw (-2.121,-2.121) -- (-1,0) -- (-2.121, 2.121);
\draw ( 2.121, 2.121) -- ( 1,0) -- ( 2.121,-2.121);
\draw (-1,0) -- (1,0);
\draw [red] (1.4,-0.52) -- (1.1,0) -- (1.4,0.52);
\draw [red] (-1.4,-0.52) -- (-1.1,0) -- (-1.4,0.52);
\node [left] at (-1.4,-0.52) {$\nabla_\mu$};
\node [left] at (-1.4, 0.52) {$\nabla^\mu$};
\node [right] at ( 1.4,-0.52) {$\nabla_\nu$};
\node [right] at ( 1.4, 0.52) {$\nabla^\nu$};
\node [left] at (-2.121, 2.2) {$\O_1(\bs{k}_1)$}; 
\node [left] at (-2.121,-2.2) {$\O_2(\bs{k}_2)$}; 	
\node [right] at ( 2.121,-2.2) {$\O_3(\bs{k}_3)$}; 
\node [right] at ( 2.121, 2.2) {$\O_4(\bs{k}_4)$}; 	
\node [right] at (-1.5, 1.2) {$\K_{[\Delta_1]}$};
\node [right] at (-1.5, -1.2) {$\K_{[\Delta_2]}$};
\node [left] at ( 1.5, -1.2) {$\K_{[\Delta_3]}$};
\node [left] at ( 1.5, 1.2) {$\K_{[\Delta_4]}$};
\node [above] at (0,0) {$\G_{[\Delta_x]}$};
\end{tikzpicture}
\centering
\caption{Witten diagrams representing derivative 4-point exchange amplitudes $\ino_{[\contraction[0.5ex]{}{\Delta}{{}_1}{\Delta} \Delta_1 \Delta_2; \Delta_3 \Delta_4 x \Delta_x]}$ and $\ino_{[\contraction[0.5ex]{}{\Delta}{{}_1}{\Delta} \Delta_1 \Delta_2; \contraction[0.5ex]{}{\Delta}{{}_3}{\Delta} \Delta_3 \Delta_4 x \Delta_x]}$. The additional red lines indicate how the correpsonding pairs of spacetime indices on derivatives $\nabla_\mu$ acting on the propagators are contracted.\label{fig:4ptX}}
\end{figure}
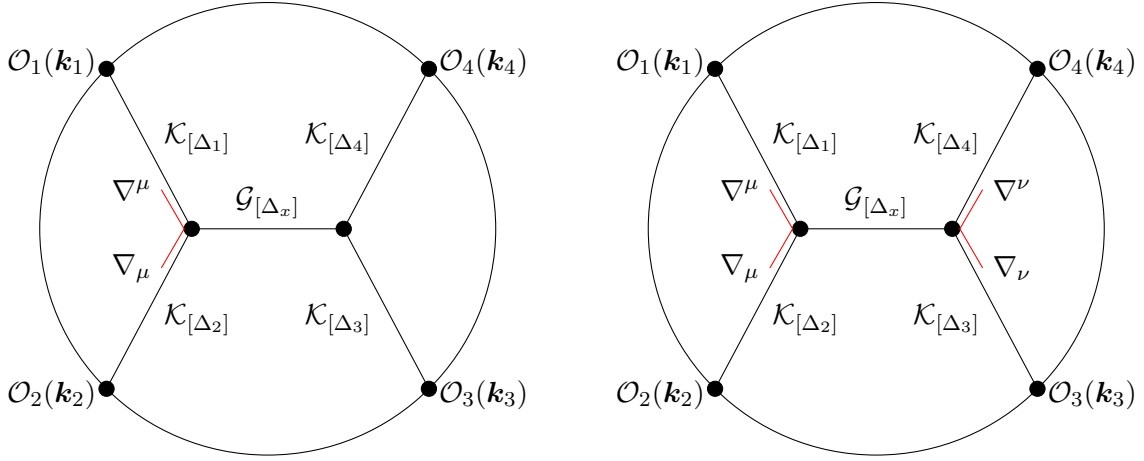

As far as the 4-point \emph{exchange} amplitudes are concerned, we have a number of different amplitudes, depending on which propagators the derivatives act. Let us define here only two exchange amplitudes, relegating all other cases to Appendix \ref{sec:amp_defs}. For the 2-derivative amplitude we take
\begin{empheq}[box=\nicebox]{align} \label{4ptX2d}
& \ino_{[\contraction[0.5ex]{}{\Delta}{{}_1}{\Delta} \Delta_1 \Delta_2; \Delta_3 \Delta_4 x \Delta_x]}(k_1, k_2, k_3, k_4, s)  \nn\\
& \qquad = \int_0^\infty \D z \, z^{-d-1} [\mathcal{D}^m \K_{[\Delta_1]}](z, \bs{k}_1) [\mathcal{D}_m \K_{[\Delta_2]}](z, \bs{k}_2) \times\nn\\
& \qquad\qquad \times \int_0^\infty \D \z \, \z^{-d-1} \G_{[\Delta_x]}(z, s; \z) \K_{[\Delta_3]}(\z, k_3) \K_{[\Delta_4]}(\z, k_4),
\end{empheq}
while for the 4-derivative amplitude
\begin{empheq}[box=\nicebox]{align} \label{4ptX4d}
& \ino_{[\contraction[0.5ex]{}{\Delta}{{}_1}{\Delta} \Delta_1 \Delta_2; \contraction[0.5ex]{}{\Delta}{{}_3}{\Delta} \Delta_3 \Delta_4 x \Delta_x]}(k_1, k_2, k_3, k_4, s)  \nn\\
& \qquad = \int_0^\infty \D z \, z^{-d-1} [\mathcal{D}^m \K_{[\Delta_1]}](z, \bs{k}_1) [\mathcal{D}_m \K_{[\Delta_2]}](z, \bs{k}_2) \times\nn\\
& \qquad\qquad \times \int_0^\infty \D \z \, \z^{-d-1} \G_{[\Delta_x]}(z, s; \z) [\mathcal{D}^n \K_{[\Delta_3]}](\z, \bs{k}_3) [\mathcal{D}_n \K_{[\Delta_4]}](\z, \bs{k}_4).
\end{empheq}
The corresponding Witten diagrams are presented in Figure \ref{fig:4ptX}.

The full set of 4-point derivative exchange amplitudes analyzed in this paper is presented in Table \ref{tab:exchange_amplitudes} below. Their precise definitions are listed in Appendix \ref{sec:amp_defs}. As we will investigate in the following section, there exists an intricate web of dependencies between various amplitudes. Thus, we will mostly concentrate on the two types of derivative exchange amplitudes defined above.

\begin{table}[thb]
\centering
\begin{tabular}{||c|c||}
\hline\hline
Derivatives & Exchange amplitudes \\ \hline\hline
0 & $\ino_{[\Delta_1 \Delta_2; \Delta_3 \Delta_4 x \Delta_x]}$ \\ \hline
2 & $\ino_{[\contraction[0.5ex]{}{\Delta}{{}_1}{\Delta} \Delta_1 \Delta_2; \Delta_3 \Delta_4 x \Delta_x]}$, $\ino_{[\Delta_1 \Delta_2; \contraction[0.5ex]{}{\Delta}{{}_3 \Delta_4 x}{\Delta} \Delta_3 \Delta_4 x \Delta_x]}$ \\ \hline
4 & $\ino_{[\contraction[0.5ex]{}{\Delta}{{}_1}{\Delta} \Delta_1 \Delta_2; \contraction[0.5ex]{}{\Delta}{{}_3}{\Delta} \Delta_3 \Delta_4 x \Delta_x]}$, $\ino_{[\contraction[0.5ex]{}{\Delta}{{}_1}{\Delta} \Delta_1 \Delta_2; \contraction[0.5ex]{}{\Delta}{{}_3 \Delta_4 x}{\Delta} \Delta_3 \Delta_4 x \Delta_x]}$, $\ino_{[\contraction[1.0ex]{}{\Delta}{{}_1 \Delta_2; \Delta_3 \Delta_4 x \,}{\Delta} \contraction[0.5ex]{\Delta_1 \Delta_2;}{\Delta}{{}_3 \Delta_4 x \!\!}{\Delta} \Delta_1 \Delta_2; \Delta_3 \Delta_4 x \Delta_x]}$ \\ \hline\hline
\end{tabular}
\caption{The list of distinct types of derivative 4-point exchange amplitudes considered in this paper and not related to each other by symmetries. For exact expressions, see Appendix \ref{sec:amp_defs}.\label{tab:exchange_amplitudes}}
\end{table}

\subsection{Regularization}

While some amplitudes are finite, others exhibit divergences. All the divergences follow from the near-boundary region of $z$ approaching $0$. From the point of view of the dual QFT, these are UV divergences. From practical point of view, all divergences in the amplitudes present themselves as divergences at the lower integration limits in all the integrals (\ref{3pt2d}, \ref{4ptC2d}, \ref{4ptX2d}, \ref{4ptX4d}). 

In order to regulate the amplitudes, we use dimensional regularization as in \cite{Bzowski:2022rlz}. We shift both the spacetime and conformal dimensions according to
\begin{empheq}[box=\nicebox]{align} \label{general}
& d \ \longmapsto \ \dreg = d + 2 u \ep, && \Delta_j \ \longmapsto \ \Dreg_j = \Delta_j + (u + v_j) \ep,
\end{empheq}
where $\ep$ is the regulator. The parameters $u$ and $v_j$ are fixed numbers, which determine the direction of the shift in the space $(d, \Delta_j)$ of dimensions. 

All regulated quantities are denoted by hats. For example, regulated propagators are denoted as $\Kreg_{[\Delta]}, \Greg_{[\Delta]}$, regulated amplitudes by $\ireg$ and so on. Once the amplitude is calculated with non-vanishing $\ep$, one can consider its series expansion around $\ep = 0$. If the expansion contains poles, it must be renormalized by the addition of suitable counterterms. We will discuss the subtleties of the procedure in the following sections.

In general, we can think about the regulated amplitudes as the amplitudes obtained by means of analytic continuation. Indeed, one can show either directly, \cite{Bzowski:2016kni}, or by means of the Mellin space, \cite{Mack:2009mi,Penedones:2010ue,Fitzpatrick:2011ia}, that the amplitudes are analytic functions of dimensions $d$ and $\Delta_j$ and there exists a non-empty, open set in which the defining integrals converge. Thus, one can use analytic continuation to obtain the unique expressions for the amplitudes for any $d$ and $\Delta_j$ except at the singularities.

The \emph{beta scheme} is the special regularization scheme in which $u = 1$ and all $v_j = 0$. In such a scheme the order of all Bessel functions involved in the regulated propagators remain unaltered. Indeed, the order of all Bessel functions in the propagators depends on the important combination of the dimensions,
\begin{align}
\beta_j = \Delta_j - \frac{d}{2}.
\end{align}
In the beta scheme
\begin{empheq}[box=\nicebox]{align} \label{beta}
& \dreg = d + 2 \ep, && \Dreg_j = \Delta_j + \ep, && \reg{\beta}_j = \beta_j
\end{empheq}
and the beta parameters remain unaltered. This simplifies calculations of the amplitudes significantly. In particular, the propagators with half-integral beta parameters still simplify to elementary functions.

In the remainder of the paper we will work mostly in the beta scheme. However, some results are better exposed in the general $(u, v_j)$-scheme \eqref{general}. In both cases we will use the same notation for the regulated amplitudes, $\ireg$, as well as other regulated quantities. We will explicitly specify when we switch between the beta scheme \eqref{beta} and the general $(u, v_j)$-scheme \eqref{general} and such an approach should not pose confusion.

\section{Identities} \label{sec:identities}

In this section we derive and list a number of useful identities between various amplitudes. Unless specified otherwise, the results hold in \emph{any regularization scheme} or, more generally, for all analytically continued amplitudes. In many cases the derivative amplitudes can be expressed in terms of non-derivative amplitudes. In some cases, most notably those involving massless AdS fields, the order of singularities in the derivative amplitudes diminishes and many amplitudes become finite.

\subsection{3-point amplitudes}

3-point derivative amplitudes $\ireg_{[\Delta_1 \contraction[0.5ex]{}{\Delta}{{}_2}{\Delta} \Delta_2 \Delta_3]}$ are expressible in terms of non-derivative amplitudes $\ireg_{[\Delta_1 \Delta_2 \Delta_3]}$. Indeed, one can integrate expression \eqref{i3D_position} by parts to rewrite it as
\begin{align} \label{i3_by_parts}
\ireg_{[\Delta_1 \contraction[0.5ex]{}{\Delta}{{}_2}{\Delta} \Delta_2 \Delta_3]} & = \frac{1}{2} \int \D^{\dreg+1} x \, \sqrt{g} \left[ \Box \reg{\K}_{[\Delta_1]} \, \reg{\K}_{[\Delta_2]} \reg{\K}_{[\Delta_3]} - \reg{\K}_{[\Delta_1]}  \Box \reg{\K}_{[\Delta_2]} \, \reg{\K}_{[\Delta_3]} - \reg{\K}_{[\Delta_1]} \reg{\K}_{[\Delta_2]} \Box \reg{\K}_{[\Delta_3]} \right].
\end{align}
Then, using the fact that the propagators obey equations of motion in the bulk,
\begin{align} \label{boxK}
\Box \Kreg_{[\Delta]} = \reg{m}_{\Delta}^2 \Kreg_{[\Delta]}.
\end{align}
where
\begin{empheq}[box=\nicebox]{align} \label{mass}
\reg{m}_{\Delta}^2 = \Dreg (\Dreg - \dreg)
\end{empheq}
we obtain
\begin{empheq}[box=\nicebox]{align} \label{3pt_calc}
\ireg_{[\Delta_1 \contraction[0.5ex]{}{\Delta}{{}_2}{\Delta} \Delta_2 \Delta_3]} = \frac{1}{2} \left( \reg{m}_{\Delta_1}^2 - \reg{m}_{\Delta_2}^2 - \reg{m}_{\Delta_3}^2 \right) \ireg_{[\Delta_1 \Delta_2 \Delta_3]}.
\end{empheq}
From this expression we see that the derivative 3-point amplitudes are at most as singular as the non-derivative ones. In special cases, however, the prefactor may become of order $\ep$, meaning that the amplitude $\ireg_{[\Delta_1 \contraction[0.5ex]{}{\Delta}{{}_2}{\Delta} \Delta_2 \Delta_3]}$ is less singular than $\ireg_{[\Delta_1 \Delta_2 \Delta_3]}$. The most important case when such a simplification occurs concerns the amplitude $\ireg_{[d \contraction[0.5ex]{}{d}{}{d} dd]}$ involving three massless fields, corresponding to marginal dual operators with $\Delta_j = d$.

\subsection{4-point contact amplitudes}

General derivative 4-point contact amplitudes cannot be easily reduced to non-derivative amplitudes. One can only apply the differential operators of the form \eqref{Dm} to the bulk-to-boundary propagators building up the amplitude \eqref{4ptC2d}. It is straightforward to calculate contact amplitudes when all $\beta_j = \Delta_j - \frac{d}{2}$ are half-integral. In such cases the Bessel functions reduce to elementary functions, \eqref{Khalfint}, and the integrals are expressible in terms of Euler's gamma functions. For general values of parameters the multiple-$K$ integrals are expressible in terms of generalized hypergeometric series: Appell series for triple-$K$ integrals, \cite{Coriano:2013jba,Bzowski:2013sza}, and Lauricella C series for quadruple-$K$ integrals, \cite{Coriano:2019sth}.

Certain combinations of 4-point 2-derivative amplitudes simplify to non-derivative amplitudes. For example, since
\begin{align} \label{nabla_dot_nabla}
\nabla f \cdot \nabla g = \frac{1}{2} \left[ \Box (f g) - f \Box g - g \Box f \right]
\end{align}
we have the identity
\begin{align} \label{sym_diff}
\ireg_{[\Delta_1 \Delta_2 \contraction[0.5ex]{}{\Delta}{{}_3}{\Delta} \Delta_3 \Delta_4]} - \ireg_{[\contraction[0.5ex]{}{\Delta}{{}_2}{\Delta} \Delta_1 \Delta_2 \Delta_3 \Delta_4]} = - \frac{1}{2} \left( \reg{m}_{\Delta_1}^2 + \reg{m}_{\Delta_2}^2 - \reg{m}_{\Delta_3}^2 - \reg{m}_{\Delta_4}^2 \right) \ireg_{[\Delta_1 \Delta_2 \Delta_3 \Delta_4]},
\end{align}
where we used \eqref{boxK}. In particular, consider the case $\Delta_3 = \Delta_1$ and $\Delta_4 = \Delta_2$. If we work in the scheme where the equalities are preserved for the regulated dimensions, $\Dreg_3 = \Dreg_1$ and $\Dreg_2 = \Dreg_4$ (equivalently $v_1 = v_3$ and $v_2 = v_4$ in \eqref{general}) the right hand side vanishes and thus we have
\begin{align} \label{idC_sym}
\ireg_{[\Delta_1 \Delta_2 \contraction[0.5ex]{}{\Delta}{{}_1}{\Delta} \Delta_1 \Delta_2]} = \ireg_{[\contraction[0.5ex]{}{\Delta}{{}_2}{\Delta} \Delta_1 \Delta_2 \Delta_1 \Delta_2]}.
\end{align}
If the regularization scheme does not preserve the equality of the regulated dimensions, the right hand side of \eqref{sym_diff} may be non-vanishing, but at most local. Here `local' refers to any expression which in position space would contain at least one Dirac's delta function. Indeed, in such a case the prefactor is of order $O(\ep)$ and the right hand side is non-vanishing only if the non-derivative amplitude on the right hand side is divergent and all divergences are local.

Similarly, using integration by parts, we have
\begin{align}
& \int \sqrt{g} \left[ \Box(f_1 f_2) f_3 f_4 + f_1 \Box(f_2 f_3) f_4 + f_1 f_2 \Box(f_3 f_4) \right] = \nn\\
& \qquad = \int \sqrt{g} \left[ f_1 f_2 f_3 \Box f_4 + f_1 f_2 \Box f_3 \, f_4 + f_1 \Box f_2 \, f_3 f_4 + \Box f_1 \, f_2 f_3 f_4 \right].
\end{align}
Combining the left hand side with \eqref{nabla_dot_nabla} we obtain the following relations,
\begin{align}
\ireg_{[\contraction[0.5ex]{}{\Delta}{{}_1}{\Delta} \Delta_1 \Delta_2 \Delta_3 \Delta_4]} +
\ireg_{[\contraction[0.5ex]{}{\Delta}{{}_1 \Delta_1}{\Delta} \Delta_1 \Delta_2 \Delta_3 \Delta_4]} + \ireg_{[\contraction[0.5ex]{}{\Delta}{{}_1 \Delta_1 \Delta_2}{\Delta} \Delta_1 \Delta_2 \Delta_3 \Delta_4]} & = - \reg{m}_{\Delta_1}^2 \ireg_{[\Delta_1 \Delta_2 \Delta_3 \Delta_4]}, \label{idC_cycl1} \\
\ireg_{[\contraction[0.5ex]{}{\Delta}{{}_1}{\Delta} \Delta_1 \Delta_2 \Delta_3 \Delta_4]} +
\ireg_{[\contraction[0.5ex]{\Delta_1}{\Delta}{{}_1}{\Delta} \Delta_1 \Delta_2 \Delta_3 \Delta_4]} + \ireg_{[\contraction[0.5ex]{}{\Delta}{{}_1 \Delta_1}{\Delta} \Delta_1 \Delta_2 \Delta_3 \Delta_4]} & = - \frac{1}{2} \left( \reg{m}_{\Delta_1}^2 + \reg{m}_{\Delta_2}^2 + \reg{m}_{\Delta_3}^2 - \reg{m}_{\Delta_4}^2 \right) \ireg_{[\Delta_1 \Delta_2 \Delta_3 \Delta_4]}. \label{idC_cycl2}
\end{align}

Note that derivative contact amplitudes depend on Mandelstam variables, see Appendix \ref{sec:momenta_conventions} for definition. This behavior is different than for non-derivative amplitudes, which depend only on the four magnitudes of the external momenta, $k_j = | \bs{k}_j |$, $j=1,2,3,4$. Which Mandelstam variable appears in the amplitude depends on how the derivatives are contracted, as presented in Table \ref{fig:Mandelstam} below.
\begin{table}[htb]
\centering
\begin{tabular}{||c|c||}
\hline\hline
Amplitude & Mandelstam variables \\ \hline\hline
$\ireg_{[\contraction[0.5ex]{}{\Delta}{{}_1}{\Delta} \Delta_1 \Delta_2 \Delta_3 \Delta_4]}$, $\ireg_{[\Delta_1 \Delta_2 \contraction[0.5ex]{}{\Delta}{{}_3}{\Delta} \Delta_3 \Delta_4]}$ & $s^2$ \\ \hline
$\ireg_{[\contraction[0.5ex]{}{\Delta}{{}_1 \Delta_1}{\Delta} \Delta_1 \Delta_2 \Delta_3 \Delta_4]}$, $\ireg_{[\Delta_1 \contraction[0.5ex]{}{\Delta}{{}_2 \Delta_3}{\Delta} \Delta_2 \Delta_3 \Delta_4]}$ & $t^2$ \\ \hline
$\ireg_{[\contraction[0.5ex]{}{\Delta}{{}_1 \Delta_1 \Delta_2}{\Delta} \Delta_1 \Delta_2 \Delta_3 \Delta_4]}$, $\ireg_{[\Delta_1 \contraction[0.5ex]{}{\Delta}{{}_2}{\Delta} \Delta_2 \Delta_3 \Delta_4]}$ & $u^2$ \\ \hline
$\ireg_{[\contraction[0.5ex]{}{\Delta}{{}_1}{\Delta} \Delta_1 \Delta_2 \contraction[0.5ex]{}{\Delta}{{}_3}{\Delta} \Delta_3 \Delta_4]}$ & $s^4, s^2$ \\ \hline\hline
\end{tabular}
\caption{Derivative 4-point contact amplitudes depend on Mandelstam variables $s$, $t$, or $u$ as defined in Appendix \ref{sec:momenta_conventions}. Which Mandelstam variables appear, and with what powers, depends on how many derivatives are present and how they are contracted.\label{fig:Mandelstam}}
\end{table}

\subsection{Crossing symmetry}

As mentioned above, derivative contact amplitudes depend on Mandelstam variables. Which Mandelstam variables appear depend on how the derivatives are contracted, see Table \ref{fig:Mandelstam}. By default we work in the $s$ channel, always trying to write amplitudes in such a way that they depend on $s$. On occasions, however, we may encounter combinations of amplitudes that require other Mandelstam variables. 

In practical applications one encounters crossing symmetric combinations of the amplitudes. Consider the amplitude $\ireg_{[\contraction[0.5ex]{}{\Delta}{{}_1}{\Delta} \Delta_1 \Delta_2 \Delta_3 \Delta_4]}$. Its crossing-symmetric versions $\ireg^T_{[\contraction[0.5ex]{}{\Delta}{{}_1}{\Delta} \Delta_1 \Delta_2 \Delta_3 \Delta_4]}$ and $\ireg^U_{[\contraction[0.5ex]{}{\Delta}{{}_1}{\Delta} \Delta_1 \Delta_2 \Delta_3 \Delta_4]}$ are defined by exchanging the suitable external momenta,
\begin{empheq}[box=\nicebox]{align} \label{tchannel}
\ireg^T_{[\contraction[0.5ex]{}{\Delta}{{}_1}{\Delta} \Delta_1 \Delta_2 \Delta_3 \Delta_4]} & = \ireg_{[\contraction[0.5ex]{}{\Delta}{{}_1}{\Delta} \Delta_1 \Delta_2 \Delta_3 \Delta_4]}(\bs{k}_2 \leftrightarrow \bs{k}_3), \\
\ireg^U_{[\contraction[0.5ex]{}{\Delta}{{}_1}{\Delta} \Delta_1 \Delta_2 \Delta_3 \Delta_4]} & = \ireg_{[\contraction[0.5ex]{}{\Delta}{{}_1}{\Delta} \Delta_1 \Delta_2 \Delta_3 \Delta_4]}(\bs{k}_1 \leftrightarrow \bs{k}_3). \label{uchannel}
\end{empheq}
As far as the dependence of $\ireg_{[\contraction[0.5ex]{}{\Delta}{{}_1}{\Delta} \Delta_1 \Delta_2 \Delta_3 \Delta_4]}$ on scalar quantities is concerned, the amplitude depends on $4$ momenta magnitudes, $k_i$, and the Mandelstam variable $s = | \bs{k}_1 + \bs{k}_2|$. Its crossing-symmetric version $\ireg^T_{[\contraction[0.5ex]{}{\Delta}{{}_1}{\Delta} \Delta_1 \Delta_2 \Delta_3 \Delta_4]}$ then depends on the Mandelstam variable $t = |\bs{k}_1 + \bs{k}_3|$. Thus, for example,
\begin{align}
\ireg^T_{[\contraction[0.5ex]{}{\Delta}{{}_1}{\Delta} \Delta_1 \Delta_2 \Delta_3 \Delta_4]}(k_i, t) & = \ireg_{[\contraction[0.5ex]{}{\Delta}{{}_1}{\Delta} \Delta_1 \Delta_2 \Delta_3 \Delta_4]}(k_1, k_3, k_2, k_4, t) = \ireg_{[\contraction[0.5ex]{}{\Delta}{{}_1}{\Delta} \Delta_1 \Delta_2 \Delta_3 \Delta_4]}(k_4, k_2, k_3, k_1, t) \nn\\
& = \ireg_{[\contraction[0.5ex]{}{\Delta}{{}_1 \Delta_3}{\Delta} \Delta_1 \Delta_3 \Delta_2 \Delta_4]}(k_j, t) = \ireg_{[\Delta_4 \contraction[0.5ex]{}{\Delta}{{}_2 \Delta_3}{\Delta}  \Delta_2 \Delta_3 \Delta_1]}(k_j, t),
\end{align}
where $k_j$ in the last line denotes the canonical order, $k_1, k_2, k_3, k_4$. Similar expressions hold for the $u$-channel. In the second line, instead of permuting the momentum magnitudes, we can permute the order of the operators. Thus, if some dimensions $\Delta_j$ are equal and the equality is preserved by the regularization scheme, equations \eqref{idC_cycl1} and \eqref{idC_cycl2} become the sums of crossing-symmetric terms. We find
\begin{align}
\ireg_{[\contraction[0.5ex]{}{\Delta}{{}_1}{\Delta} \Delta_1 \Delta \Delta \Delta]} + \ireg^T_{[\contraction[0.5ex]{}{\Delta}{{}_1}{\Delta} \Delta_1 \Delta \Delta \Delta]} + \ireg^U_{[\contraction[0.5ex]{}{\Delta}{{}_1}{\Delta} \Delta_1 \Delta \Delta \Delta]} & = - \reg{m}_{\Delta_1}^2 \ireg_{[\Delta_1 \Delta \Delta \Delta]}, \\
\ireg_{[\contraction[0.5ex]{}{\Delta}{}{\Delta} \Delta \Delta \Delta \Delta_4]} + \ireg^T_{[\contraction[0.5ex]{}{\Delta}{}{\Delta} \Delta \Delta \Delta \Delta_4]} + \ireg^U_{[\contraction[0.5ex]{}{\Delta}{}{\Delta} \Delta \Delta \Delta \Delta_4]} & = \frac{1}{2} \left( \reg{m}_{\Delta_4}^2 - 3 \reg{m}_{\Delta}^2 \right) \ireg_{[\Delta \Delta \Delta \Delta_4]}.
\end{align}
If the regularization scheme does not preserve the equality of the regulated dimensions, the terms in brackets acquire additional terms of order $O(\ep)$; in such a case one should use \eqref{idC_cycl1} and \eqref{idC_cycl2} directly. In any case, note that the right hand sides contain non-derivative amplitudes only. Thus, in highly symmetric theories, where correlators are built up by crossing-symmetric combination of amplitudes, significant simplifications can occur. We will present an example of such a theory in Section \ref{sec:symmetric}.

\subsection{Invariant variables} \label{sec:variables}

In the beta regularization scheme \eqref{beta} the amplitude simplifies considerably for half-integral $\beta = \Delta - d/2$ parameters due to \eqref{Khalfint}. Looking at the integrands of (\ref{intro_4ptC}, \ref{4ptC2d}, \ref{4ptC4d}) we see that in the beta scheme all contact amplitudes exhibit the following structure,
\begin{align} \label{iC_R}
& \ireg_{\text{any 4-point contact}} = \Gamma(2 \ep) k_T^{-2 \ep} \times \frac{W(k_1, k_2, k_3, k_4, s; \ep)}{k_T^{d-2}}
\end{align}
where $W$ is a polynomial of momenta lengths with a finite or vanishing $\ep \rightarrow 0$ limit. All contact amplitudes are either linearly divergent at $\ep = 0$ or finite. As it stands the prefactor is divergent at $\ep = 0$, but for finite amplitudes the divergence can canceled against a zero in $W$.

In this section we are interested in the momentum dependence of contact amplitudes rather than $\ep$-dependence. Consider a non-derivative 4-point contact amplitude $\ireg_{[\Delta \Delta \Delta \Delta]}$ with all dimensions $\Delta_j = \Delta$ equal and regulated in the beta scheme. Such an amplitude exhibits a large discreet symmetry group. Indeed, without derivatives, the amplitude $\ireg_{[\Delta \Delta \Delta \Delta]}(k_1, k_2, k_3, k_4)$ does not depend on Mandelstam variables and is invariant under any permutation of momenta $k_j$, $j=1,2,3,4$. Neglecting the $\ep$-dependence and concentrating on momenta, the amplitude becomes the rational function of the form
\begin{align}
\ireg_{[\Delta \Delta \Delta \Delta]}(k_1, k_2, k_3, k_4) = \frac{1}{k_T^{\dreg - 2}} P_0(k_1, k_2, k_3, k_4),
\end{align}
where $k_T = k_1 + k_2 + k_3 + k_4$ and $P_0$ is a polynomial of dimension $4 \Delta - 2d - 2$. In particular, $P_0$ is invariant under the full symmetry group $S_4$ of the 4 momenta, \textit{i.e.}, for any permutation $s \in S_4$ we have $P_0(k_1, k_2, k_3, k_4) = P_0(k_{s(1)}, k_{s(2)}, k_{s(3)}, k_{s(4)})$. Thus it can be rewritten entirely in terms of symmetric polynomials of four variables, $k_1, k_2, k_3, k_4$ which we denote by $\sigma_i = \s{i}{1234}$ with $i=1,2,3,4$. All the conventions are summarized in Appendix \ref{sec:momenta_conventions}.

Consider now a derivative amplitude $\ireg_{[\contraction[0.5ex]{}{\Delta}{}{\Delta} \Delta \Delta \Delta \Delta]}(k_1, k_2, k_3, k_4, s)$. The amplitude depends on the Mandelstam variable $s$. Furthermore, it remains invariant under swapping $k_1 \leftrightarrow k_2$ or $k_3 \leftrightarrow k_4$. Finally, equation \eqref{idC_sym} shows that the amplitude is invariant under swapping the two pairs of momenta altogether, $(k_1, k_2) \leftrightarrow (k_3, k_4)$. Those permutations generate the 8-element dihedral subgroup $D_4 \leq S_4$ of the full permutation group and the amplitude is invariant under such permutations. Thus, for half-integral $\beta = \Delta - d/2$, the polynomial $W$ in \eqref{iC_R} must exhibit the $D_4$ symmetry. 

By looking more carefully at the integrands of the contact amplitudes we can figure out on what Mandelstam combinations the amplitudes depend. Those are gathered in Table \ref{fig:Mandelstam}, from which we find the general structure of the 2- and 4-derivative contact amplitudes to be
\begin{align}
\ireg_{[\contraction[0.5ex]{}{\Delta}{}{\Delta} \Delta \Delta \Delta \Delta]} = \ireg_{[\Delta \Delta \contraction[0.5ex]{}{\Delta}{}{\Delta} \Delta \Delta]} & = \frac{1}{k_T^{\dreg - 2}} \left[ s^2 P_1 + Q_0 \right], \\
\ireg_{[\contraction[0.5ex]{}{\Delta}{}{\Delta} \Delta \Delta \contraction[0.5ex]{}{\Delta}{}{\Delta} \Delta \Delta]} & = \frac{1}{k_T^{\dreg - 2}} \left[ s^4 P_2 + s^2 Q_1 + Q_0 \right].
\end{align}
The polynomials $P_j = P_j(k_1, k_2, k_3, k_4)$ are completely symmetric, invariant under the full permutation group $S_4$, while the polynomials $Q_j = Q_j(k_1, k_2, k_3, k_4)$ are invariant under $D_4$ subgroup only.

Polynomials $Q_j$ can be neatly rewritten in terms of elementary polynomials invariant under the symmetry group $D_4$. The problem of finding the minimal set of generators of the invariant ring $\C[k_1, k_2, k_3, k_4]^{D_4}$ is the standard problem of the theory of ring invariants, see \cite{paule2008algorithms} for a simple practical introduction. Standard symmetric polynomials $\sigma_i$ on $k_i$ for $i=1,2,3,4$ are all invariant under the full $S_4$ and we can keep them. By calculating the Molien series we see that a single new invariant polynomial of degree $2$ is required. This can be taken as
\begin{empheq}[box=\nicebox]{align} \label{tau}
\tau = (k_1 + k_2) (k_3 + k_4),
\end{empheq}
clearly an invariant of $D_4$ and not an invariant of $S_4$. The five polynomials: four symmetric polynomials $\sigma_i$ and $\tau$ are not independent. The relation between them reads
\begin{align}
0 = \tau^3 - 2 \tau^2 \sigma_2 + \tau (\sigma_2^2 + \sigma_1 \sigma_3 - 4 \sigma_4 ) + \sigma_1^2 \sigma_4 - \sigma_1 \sigma_2 \sigma_3 + \sigma_3^2
\end{align}
and can be found by calculating the reduced Gr\"obner basis of the invariant ring. The relation can be used to remove all $\tau^n$ for $n \geq 3$ from any expression.

\subsection{4-point exchange amplitudes} \label{sec:identities_4ptX}

All 4-point exchange amplitudes in (\ref{4ptX2d}, \ref{4ptX4d}) as well as in (\ref{4ptX}, \ref{4ptX2da}, \ref{4ptX4da}, \ref{4ptX4db}) contain two integrals, each of which contains a product of propagators. For this reason we can still integrate by parts as in \eqref{i3_by_parts} to produce only Laplacians acting on propagators. Since in position space
\begin{align} \label{boxG}
\Box_x \Greg_{[\Delta]}(x, x') = \Box_{x'} \Greg_{[\Delta]}(x, x') = \reg{m}_{\Delta}^2 \Greg_{[\Delta]}(x, x') - \frac{\delta(x - x')}{\sqrt{g(x)}},
\end{align}
where the mass-squared is given in \eqref{mass}, this reduces all derivative 4-point exchange amplitudes to a combination of non-derivative 4-point exchange amplitudes as well as derivative and non-derivative 4-point contact amplitudes. In particular we find
\begin{empheq}[box=\nicebox]{align}
\ireg_{[\contraction[0.5ex]{}{\Delta}{{}_1}{\Delta} \Delta_1 \Delta_2; \Delta_3 \Delta_4 x \Delta_x]} & = \frac{1}{2} \left( \reg{m}_{\Delta_x}^2 - \reg{m}_{\Delta_1}^2 - \reg{m}_{\Delta_2}^2 \right) \ireg_{[\Delta_1 \Delta_2; \Delta_3 \Delta_4 x \Delta_x]} - \frac{1}{2} \ireg_{[\Delta_1 \Delta_2 \Delta_3 \Delta_4]}, \label{redi2da} \\
\ireg_{[\Delta_1 \Delta_2; \contraction[0.5ex]{}{\Delta}{{}_3}{\Delta} \Delta_3 \Delta_4 x \Delta_x]} & = \frac{1}{2} \left( \reg{m}_{\Delta_x}^2 - \reg{m}_{\Delta_3}^2 - \reg{m}_{\Delta_4}^2 \right) \ireg_{[\Delta_1 \Delta_2; \Delta_3 \Delta_4 x \Delta_x]} - \frac{1}{2} \ireg_{[\Delta_1 \Delta_2 \Delta_3 \Delta_4]}. \label{redi2db}
\end{empheq}
The exchange amplitudes with four derivatives can be written in two ways, depending on whether we start with \eqref{redi2da} or \eqref{redi2db}. We find
\begin{align}
\ireg_{[\contraction[0.5ex]{}{\Delta}{{}_1}{\Delta} \Delta_1 \Delta_2; \contraction[0.5ex]{}{\Delta}{{}_3}{\Delta} \Delta_3 \Delta_4 x \Delta_x]} & = \frac{1}{4} \left( \reg{m}_{\Delta_x}^2 - \reg{m}_{\Delta_1}^2 - \reg{m}_{\Delta_2}^2 \right)\left( \reg{m}_{\Delta_x}^2 - \reg{m}_{\Delta_3}^2 - \reg{m}_{\Delta_4}^2 \right) \ireg_{[\Delta_1 \Delta_2; \Delta_3 \Delta_4 x \Delta_x]} \nn\\
& \qquad - \frac{1}{4} \left( \reg{m}_{\Delta_x}^2 - \reg{m}_{\Delta_1}^2 - \reg{m}_{\Delta_2}^2 \right) \ireg_{[\Delta_1 \Delta_2 \Delta_3 \Delta_4]} - \frac{1}{2} \ireg_{[\Delta_1 \Delta_2 \contraction[0.5ex]{}{\Delta}{{}_3}{\Delta} \Delta_3 \Delta_4]} \nn\\
& = \frac{1}{4} \left( \reg{m}_{\Delta_x}^2 - \reg{m}_{\Delta_1}^2 - \reg{m}_{\Delta_2}^2 \right)\left( \reg{m}_{\Delta_x}^2 - \reg{m}_{\Delta_3}^2 - \reg{m}_{\Delta_4}^2 \right) \ireg_{[\Delta_1 \Delta_2; \Delta_3 \Delta_4 x \Delta_x]} \nn\\
& \qquad - \frac{1}{4} \left( \reg{m}_{\Delta_x}^2 - \reg{m}_{\Delta_3}^2 - \reg{m}_{\Delta_4}^2 \right) \ireg_{[\Delta_1 \Delta_2 \Delta_3 \Delta_4]} - \frac{1}{2} \ireg_{[\contraction[0.5ex]{}{\Delta}{{}_2}{\Delta} \Delta_1 \Delta_2 \Delta_3 \Delta_4]}.
\end{align}
All remaining 4-point exchange amplitudes can be related to $\ireg_{[\contraction[0.5ex]{}{\Delta}{{}_1}{\Delta} \Delta_1 \Delta_2; \Delta_3 \Delta_4 x \Delta_x]}$ and $\ireg_{[\contraction[0.5ex]{}{\Delta}{{}_1}{\Delta} \Delta_1 \Delta_2; \contraction[0.5ex]{}{\Delta}{{}_3}{\Delta} \Delta_3 \Delta_4 x \Delta_x]}$ by integrating by parts. For the 2-derivative amplitude $\ireg_{[\Delta_1 \Delta_2; \contraction[0.5ex]{}{\Delta}{{}_3 \Delta_4 x}{\Delta} \Delta_3 \Delta_4 x \Delta_x]}$ we find
\begin{align}
\ireg_{[\Delta_1 \Delta_2; \contraction[0.5ex]{}{\Delta}{{}_3 \Delta_4 x}{\Delta} \Delta_3 \Delta_4 x \Delta_x]} & = - \ireg_{[\Delta_1 \Delta_2; \contraction[0.5ex]{}{\Delta}{{}_3}{\Delta} \Delta_3 \Delta_4 x \Delta_x]} - \reg{m}_{\Delta_3}^2 \ireg_{[\Delta_1 \Delta_2; \Delta_3 \Delta_4 x \Delta_x]} \nn\\
& = \frac{1}{2} \left( \reg{m}_{\Delta_4}^2 - \reg{m}_{\Delta_3}^2 - \reg{m}_{\Delta_x}^2 \right) \ireg_{[\Delta_1 \Delta_2; \Delta_3 \Delta_4 x \Delta_x]} + \frac{1}{2} \ireg_{[\Delta_1 \Delta_2 \Delta_3 \Delta_4]}.
\end{align}
For the remaining two 4-derivative exchange amplitudes we obtain
\begin{align}
\ireg_{[\contraction[0.5ex]{}{\Delta}{{}_1}{\Delta} \Delta_1 \Delta_2; \contraction[0.5ex]{}{\Delta}{{}_3 \Delta_4 x}{\Delta} \Delta_3 \Delta_4 x \Delta_x} & = - \ireg_{[\contraction[0.5ex]{}{\Delta}{{}_1}{\Delta} \Delta_1 \Delta_2; \contraction[0.5ex]{}{\Delta}{{}_3}{\Delta} \Delta_3 \Delta_4 x \Delta_x]} - \reg{m}_{\Delta_3}^2 \ireg_{[\contraction[0.5ex]{}{\Delta}{{}_1}{\Delta} \Delta_1 \Delta_2;  \Delta_3 \Delta_4 x \Delta_x]}, \\
\ireg_{[\contraction[1.0ex]{}{\Delta}{{}_1 \Delta_2; \Delta_3 \Delta_4 x \,}{\Delta} \contraction[0.5ex]{\Delta_1 \Delta_2;}{\Delta}{{}_3 \Delta_4 x \!\!}{\Delta} \Delta_1 \Delta_2; \Delta_3 \Delta_4 x \Delta_x]} & = \ireg_{[\contraction[0.5ex]{}{\Delta}{{}_1}{\Delta} \Delta_1 \Delta_2; \contraction[0.5ex]{}{\Delta}{{}_3}{\Delta} \Delta_3 \Delta_4 x \Delta_x]} + \reg{m}_{\Delta_1}^2 \ireg_{[\Delta_1 \Delta_2; \contraction[0.5ex]{}{\Delta}{{}_3}{\Delta} \Delta_3 \Delta_4 x \Delta_x]} \nn\\
& \qquad + \reg{m}_{\Delta_3}^2 \ireg_{[\contraction[0.5ex]{}{\Delta}{{}_1}{\Delta} \Delta_1 \Delta_2; \Delta_3 \Delta_4 x \Delta_x]} + \reg{m}_{\Delta_1}^2 \reg{m}_{\Delta_3}^2 \ireg_{[\Delta_1 \Delta_2; \Delta_3 \Delta_4 x \Delta_x]}.
\end{align}
Alternatively, one can integrate by parts and use \eqref{boxG} directly to obtain the relations
\begin{align}
\ireg_{[\contraction[0.5ex]{}{\Delta}{{}_1}{\Delta} \Delta_1 \Delta_2; \contraction[0.5ex]{}{\Delta}{{}_3 \Delta_4 x}{\Delta} \Delta_3 \Delta_4 x \Delta_x} & = \frac{1}{4} \left( \reg{m}_{\Delta_x}^2 - \reg{m}_{\Delta_1}^2 - \reg{m}_{\Delta_2}^2 \right)\left( \reg{m}_{\Delta_4}^2 - \reg{m}_{\Delta_3}^2 - \reg{m}_{\Delta_x}^2 \right) \ireg_{[\Delta_1 \Delta_2; \Delta_3 \Delta_4 x \Delta_x]} \nn\\
& \qquad - \frac{1}{4} \left( \reg{m}_{\Delta_4}^2 + \reg{m}_{\Delta_3}^2 - \reg{m}_{\Delta_x}^2 \right) \ireg_{[\Delta_1 \Delta_2 \Delta_3 \Delta_4]} + \frac{1}{2} \ireg_{[\contraction[0.5ex]{}{\Delta}{{}_2}{\Delta} \Delta_1 \Delta_2 \Delta_3 \Delta_4]}
\end{align}
as well as
\begin{align} \label{ired1}
\ireg_{[\contraction[1.0ex]{}{\Delta}{{}_1 \Delta_2; \Delta_3 \Delta_4 x \,}{\Delta} \contraction[0.5ex]{\Delta_1 \Delta_2;}{\Delta}{{}_3 \Delta_4 x \!\!}{\Delta} \Delta_1 \Delta_2; \Delta_3 \Delta_4 x \Delta_x]} & = \frac{1}{4} \left( \reg{m}_{\Delta_2}^2 - \reg{m}_{\Delta_1}^2 - \reg{m}_{\Delta_x}^2 \right)\left( \reg{m}_{\Delta_4}^2 - \reg{m}_{\Delta_3}^2 - \reg{m}_{\Delta_x}^2 \right) \ireg_{[\Delta_1 \Delta_2; \Delta_3 \Delta_4 x \Delta_x]} \nn\\
& \qquad - \frac{1}{4} \left( \reg{m}_{\Delta_1}^2 - \reg{m}_{\Delta_2}^2 + 2 \reg{m}_{\Delta_3}^2 + \reg{m}_{\Delta_x}^2 \right) \ireg_{[\Delta_1 \Delta_2 \Delta_3 \Delta_4]} - \frac{1}{2} \ireg_{[\Delta_1 \Delta_2 \contraction[0.5ex]{}{\Delta}{{}_3}{\Delta} \Delta_3 \Delta_4]} \nn\\
& = \frac{1}{4} \left( \reg{m}_{\Delta_2}^2 - \reg{m}_{\Delta_1}^2 - \reg{m}_{\Delta_x}^2 \right)\left( \reg{m}_{\Delta_4}^2 - \reg{m}_{\Delta_3}^2 - \reg{m}_{\Delta_x}^2 \right) \ireg_{[\Delta_1 \Delta_2; \Delta_3 \Delta_4 x \Delta_x]} \nn\\
& \qquad - \frac{1}{4} \left( \reg{m}_{\Delta_3}^2 - \reg{m}_{\Delta_4}^2 + \reg{m}_{\Delta_x}^2 \right) \ireg_{[\Delta_1 \Delta_2 \Delta_3 \Delta_4]} + \frac{1}{2} \ireg_{[\contraction[0.5ex]{}{\Delta}{{}_1 \Delta_2}{\Delta} \Delta_1 \Delta_2 \Delta_3 \Delta_4]} + \frac{1}{2} \ireg_{[\contraction[0.5ex]{}{\Delta}{{}_1 \Delta_2 \Delta_3}{\Delta} \Delta_1 \Delta_2 \Delta_3 \Delta_4]}.
\end{align}
Since all derivative 4-point exchange amplitudes reduce in such a way, they are slightly less interesting than contact amplitudes. Nevertheless, further interesting simplifications occur for special values of parameters. We will investigate such cases in the following subsection.

\subsection{Cancellations in special cases} \label{sec:cancellations}

Further simplifications for 4-point exchange amplitudes occur when some of the dimensions are equal. We assume throughout this section that the regularization preserves equalities between such dimensions. The combinations of the masses in the equations above simplify in such cases. For example, we find
\begin{align}
\ireg_{[\contraction[0.5ex]{}{\Delta}{}{\Delta} \Delta \Delta'; \Delta_3 \Delta_4 x \Delta]} & = - \frac{1}{2} \reg{m}_{\Delta'}^2 \ireg_{[\Delta \Delta_2; \Delta_3 \Delta_4 x \Delta]} - \frac{1}{2} \ireg_{[\Delta \Delta' \Delta_3 \Delta_4]}, \\
\ireg_{[\Delta_1 \Delta_2; \contraction[0.5ex]{}{\Delta}{\Delta x}{\Delta} \Delta \Delta x \Delta']} & = - \frac{1}{2} \reg{m}^2_{\Delta'} \ireg_{[\Delta_1 \Delta_2; \Delta \Delta x \Delta']} + \frac{1}{2} \ireg_{[\Delta_1 \Delta_2 \Delta \Delta]}, \\
\ireg_{[\Delta_1 \Delta_2; \contraction[0.5ex]{}{\Delta}{{}_3 \Delta x}{\Delta} \Delta' \Delta x \Delta]} & = - \frac{1}{2} \reg{m}^2_{\Delta'} \ireg_{[\Delta_1 \Delta_2; \Delta' \Delta x \Delta]} + \frac{1}{2} \ireg_{[\Delta_1 \Delta_2 \Delta' \Delta]}.
\end{align}
In particular if $\Delta' = d$ corresponds to the marginal operator, the masses $\reg{m}^2_{\Delta'}$ in \eqref{mass} are of order $\ep$, as $\Dreg' - \dreg = (v'-u) \ep$ in the general scheme \eqref{general}. In such a case only the local part of the exchange amplitudes enters. Thus, up to possible local terms we have
\begin{align}
\ireg_{[\contraction[0.5ex]{}{\Delta}{}{3} \Delta d; \Delta_3 \Delta_4 x \Delta]} & = - \frac{1}{2} \ireg_{[\Delta d \Delta_3 \Delta_4]} + \text{local}, \\
\ireg_{[\Delta_1 \Delta_2; \contraction[0.5ex]{}{\Delta}{\Delta x}{d} \Delta \Delta x d]} & = \frac{1}{2} \ireg_{[\Delta_1 \Delta_2 \Delta \Delta]} + \text{local}, \\
\ireg_{[\Delta_1 \Delta_2; \contraction[0.5ex]{}{d}{\Delta x}{\Delta} d \Delta x \Delta]} & = \frac{1}{2} \ireg_{[\Delta_1 \Delta_2 d \Delta]} + \text{local}.
\end{align}
Here `local' refers to any expression which in position space would contain at least one Dirac's delta function. This explains why we should expect various amplitudes involving marginal operators to simplify significantly.

Similar simplifications occur for 4-derivative exchange amplitudes. For example, we find
\begin{align}
\ireg_{[\contraction[0.5ex]{}{\Delta}{}{\Delta} \Delta \Delta'; \contraction[0.5ex]{}{\Delta}{}{\Delta} \Delta_3 \Delta_4 x \Delta]} & = - \frac{1}{4} \reg{m}^2_{\Delta'} \left( \reg{m}^2_{\Delta} - \reg{m}^2_{\Delta_3} - \reg{m}^2_{\Delta_4} \right) \ireg_{[\Delta \Delta'; \Delta_3 \Delta_4 x \Delta]} \nn\\
& \qquad\qquad + \frac{1}{4} \reg{m}_{\Delta'}^2 \ireg_{[\Delta \Delta_2 \Delta_3 \Delta_4]} - \frac{1}{2} \ireg_{[\Delta \Delta' \contraction[0.5ex]{}{\Delta}{}{\Delta} \Delta_3 \Delta_4]}.
\end{align}
Thus, if $\Delta' = d$ corresponds to the marginal operator, the amplitude simplifies to the contact amplitude, up to local terms,
\begin{align}
\ireg_{[\contraction[0.5ex]{}{\Delta}{}{d} \Delta d; \contraction[0.5ex]{}{\Delta}{{}_3}{\Delta} \Delta_3 \Delta_4 x \Delta]} & = - \frac{1}{2} \ireg_{[\Delta d \contraction[0.5ex]{}{\Delta}{{}_3}{\Delta} \Delta_3 \Delta_4]} + \text{local}.
\end{align}
In particular all of the following exchange amplitudes become equal to some contact amplitudes, up to local terms,
\begin{align} \label{Xtolocal}
\begin{array}{c}
\ireg_{[\contraction[0.5ex]{}{\Delta}{}{d} \Delta d; \contraction[0.5ex]{}{\Delta}{{}_3}{\Delta} \Delta_3 \Delta_4 x \Delta]}, \\
\ireg_{[\contraction[0.5ex]{}{\Delta}{}{d} \Delta d; \contraction[0.5ex]{}{\Delta}{{}_3}{\Delta} \Delta_3 \Delta_4 x \Delta]}, 
\ireg_{[\contraction[0.5ex]{}{\Delta}{{}_1}{\Delta} \Delta_1 \Delta_2; \contraction[0.5ex]{}{d}{\Delta x}{\Delta} d \Delta x \Delta]},
\ireg_{[\contraction[0.5ex]{}{\Delta}{{}_1}{\Delta} \Delta_1 \Delta_2; \contraction[0.5ex]{}{\Delta}{\Delta x}{d} \Delta \Delta x d]},
\ireg_{[\contraction[0.5ex]{}{\Delta}{}{d} \Delta d; \contraction[0.5ex]{}{\Delta}{{}_3 \Delta_4 x}{\Delta} \Delta_3 \Delta_4 x \Delta]}, \\
\ireg_{[\contraction[1.0ex]{}{\Delta}{{}_1 \Delta_2; d \Delta x \,}{\Delta} \contraction[0.5ex]{\Delta_1 \Delta_2;}{d}{\Delta x \!\!}{\Delta} \Delta_1 \Delta_2; d \Delta x \Delta]}, \ireg_{[\contraction[1.0ex]{}{\Delta}{{}_1 \Delta_2; \Delta d x \,}{\Delta} \contraction[0.5ex]{\Delta_1 \Delta_2;}{\Delta}{d x \!\!}{\Delta} \Delta_1 \Delta_2; \Delta d x \Delta]}, \ireg_{[\contraction[1.0ex]{}{\Delta}{d; \Delta_3 \Delta_4 x \,}{\Delta} \contraction[0.5ex]{\Delta d;}{\Delta}{{}_3 \Delta_4 x \!\!}{\Delta} \Delta d; \Delta_3 \Delta_4 x \Delta]}, \ireg_{[\contraction[1.0ex]{}{\Delta}{\Delta; \Delta_3 \Delta_4 x \,}{d} \contraction[0.5ex]{\Delta \Delta;}{\Delta}{{}_3 \Delta_4 x \!\!}{d} \Delta \Delta; \Delta_3 \Delta_4 x d]}.
\end{array}
\end{align}
We refer to the attached Mathematica package \verb|HandbooK| for the evaluation of such amplitudes.

\section{From action to amplitudes} \label{sec:action_to_amplitudes}

In physical context amplitudes are building blocks of correlators. By means of AdS/CFT each boundary correlator is expressed in terms of combinations of bulk amplitudes. Thus, we specify the physical theory by specifying the bulk AdS action. Only then the relation between actual correlation functions and the amplitudes is derived.

\subsection{Amplitudes are correlators}

In \cite{Bzowski:2022rlz} we introduced the \emph{asymmetric theory}. This is the bulk theory in which a single given Witten diagram represents the entire correlation function. As long as 3- and 4-point non-derivative amplitudes are concerned, the bulk theory contains 5 scalar fields $\Phi_j$ with $j=1,2,3,4$ or $j=x$. The bulk action reads
\begin{align} \label{Sasym_0d}
S^{\text{asym}} & = \frac{1}{2} \int \D^{d+1} x \sqrt{g} \sum_{j =1,2,3,4,x} \left[ \partial_\mu \Phi_j \partial^\mu \Phi_j + m^2_{\Delta_j} \Phi_j^2 \right] \nn\\
& \qquad + \int \D^{d+1} x \sqrt{g} \left[ \lambda_{12x} \Phi_1 \Phi_2 \Phi_x + \lambda_{34x} \Phi_x \Phi_3 \Phi_4 - \lambda_{1234} \Phi_1 \Phi_2 \Phi_3 \Phi_4 \right],
\end{align}
where $\lambda_{12x}, \lambda_{34x}, \lambda_{1234}$ are arbitrary AdS couplings. We refer to this theory as the \emph{asymmetric theory}, since the resulting correlators have as few discrete symmetries as possible. The aim of the asymmetric theory is to express 3- and 4-point correlators by means of single amplitudes. The non-vanishing 3-point functions are
\begin{align} \label{asym3}
\lla \O_i(\bs{k}_1) \O_j(\bs{k}_2) \O_x(\bs{k}_3) \rra & = \lambda_{ijx} \, \ino_{[\Delta_i \Delta_j \Delta_x]}(k_1, k_2, k_3),
\end{align}
where $(ij) = (12)$ or $(34)$. The double brackets on the left indicate that in momentum space the momentum-conserving delta function has been dropped, see \eqref{double_brackets}.

Among the non-vanishing 4-point functions, the most important is $\< \O_1 \O_2 \O_3 \O_4 \>$.  This contains two contributions: a single exchange diagram and a single contact diagram,
\begin{align} \label{asym4}
& \lla \O_1(\bs{k}_1) \O_2(\bs{k}_2) \O_3(\bs{k}_3) \O_4(\bs{k}_4) \rra 
 \nn\\& \qquad\qquad 
=
 \lambda_{12x} \lambda_{34x} \, \ino_{[\Delta_1 \Delta_2; \Delta_3 \Delta_4 x \Delta_x]} + \lambda_{1234} \, \ino_{[\Delta_1 \Delta_2 \Delta_3 \Delta_4]}.
\end{align}
If we want a bulk theory where the contact diagram $\ino_{[\Delta_1 \Delta_2 \Delta_3 \Delta_4]}$ is a correlator on its own, it therefore suffices to consider a bulk action with  $\lambda_{12x}=0$ or  $\lambda_{34x}=0$. On the other hand, if we want the exchange diagram to be a correlator on its own, then we need $\lambda_{1234}=0$.

We conclude that in the asymmetric theory we can always choose the bulk fields $\Phi_j$ in such a way that certain 3- and 4-point functions are expressed in terms of single non-derivative amplitudes. Thus, any 3- or 4-point scalar amplitude must obey all the properties of the actual correlation function it represents.

\subsection{Asymmetric derivative theory} \label{sec:asymmetric_theory}

In this paper we are interested in derivative amplitudes. We want to construct \emph{asymmetric derivative theory} in which 3- and 4-point functions can be realized by single derivative scalar 3- and 4-point amplitudes. Consider the action
\begin{align} \label{Sasym}
S^{\text{asym}} = S^{\text{asym}}_{\text{free}} + S_{\text{asym}}^{(3)} + S_{\text{asym}}^{(4)}
\end{align}
where
\begin{align}
S^{\text{asym}}_{\text{free}} & = \frac{1}{2} \int \D^{d+1} x \, \sqrt{g} \sum_{j=1,2,3,4,x} \left[ \contraction{}{\Phi}{{}_j}{\Phi} \Phi_j \Phi_j + m_j^2 \Phi_j^2 \right], \\
S_{\text{asym}}^{(3)} & = \int \D^{d+1} x \, \sqrt{g} \left[ \lambda_{12x} \Phi_1 \Phi_2 \Phi_x + \lambda_{\contraction[0.5ex]{}{1}{}{2} 1 2 x} \contraction{}{\Phi}{{}_1}{\Phi} \Phi_1 \Phi_2 \Phi_x + \lambda_{\contraction[0.5ex]{}{1}{2}{x} 1 2 x} \contraction{}{\Phi}{{}_i \Phi_2}{\Phi} \Phi_1 \Phi_2 \Phi_x \right.\nn\\
& \qquad\qquad \left. + \, \lambda_{34x} \Phi_3 \Phi_4 \Phi_x + \lambda_{\contraction[0.5ex]{}{3}{}{4} 3 4 x} \contraction{}{\Phi}{{}_3}{\Phi} \Phi_3 \Phi_4 \Phi_x + \lambda_{\contraction[0.5ex]{}{3}{4}{x} 3 4 x} \contraction{}{\Phi}{{}_i \Phi_4}{\Phi} \Phi_3 \Phi_4 \Phi_x \right], \label{S3asym} \\
S_{\text{asym}}^{(4)} & = - \int \D^{d+1} x \, \sqrt{g} \left[ \lambda_{1234} \Phi_1 \Phi_2 \Phi_3 \Phi_4 + \lambda_{\contraction[0.5ex]{}{1}{2}{3} 1 2 3 4} \contraction{}{\Phi}{{}_1 \Phi_2}{\Phi} \Phi_1 \Phi_2 \Phi_3 \Phi_4 + \lambda_{\contraction[0.5ex]{}{1}{}{2} \contraction[0.5ex]{12}{3}{}{4} 1234} \contraction{}{\Phi}{{}_1}{\Phi} \Phi_1 \Phi_2 \contraction{}{\Phi}{{}_3}{\Phi} \Phi_3 \Phi_4 \right].
\end{align}
Contractions acting on fields indicate contracted derivatives, \textit{e.g.},
\begin{align}
\contraction{}{\Phi}{{}_1}{\Phi} \Phi_1 \Phi_1 = \nabla_\mu \Phi_1 \nabla^\mu \Phi_1
\end{align}
is the usual kinetic term and so on. Contractions within the coupling constants, \textit{e.g.}, $\lambda_{\contraction[0.5ex]{}{1}{}{2} 1 2 x}$, simply indicate which term they multiply. Equations of motion of the asymmetric theory read
\begin{align} \label{eqofmo1}
& (-\Box_{AdS} + \reg{m}^2_{\Delta_{1}}) \Phi_{1} = - \lambda_{12x} \Phi_2 \Phi_x + \lambda_{1234} \Phi_2 \Phi_3 \Phi_4 \nn\\
& \qquad + \lambda_{\contraction[0.5ex]{}{1}{}{2} 1 2 x} \partial_\mu \left( \partial^{\mu} \Phi_2 \, \Phi_x \right) + \lambda_{\contraction[0.5ex]{}{1}{2}{x} 1 2 x} \partial_\mu \left( \Phi_2 \partial^{\mu} \Phi_x \right) \nn\\
& \qquad - \lambda_{\contraction[0.5ex]{}{1}{2}{3} 1 2 3 4} \partial_\mu \left( \partial^{\mu} \Phi_2 \, \Phi_3 \Phi_4 \right) - \lambda_{\contraction[0.5ex]{}{1}{}{2} 1 2 \contraction[0.5ex]{}{3}{}{4} 3 4} \partial_\mu \left( \partial^{\mu} \Phi_2 \partial_{\nu} \Phi_3 \partial^{\nu} \Phi_4 \right), \\
& (-\Box_{AdS} + \reg{m}^2_{\Delta_{2}}) \Phi_{2} = - \lambda_{12x} \Phi_1 \Phi_x + \lambda_{1234} \Phi_1 \Phi_3 \Phi_4 \nn\\
& \qquad + \lambda_{\contraction[0.5ex]{}{1}{}{2} 1 2 x} \partial_\mu \left( \partial^{\mu} \Phi_1 \, \Phi_x \right) - \lambda_{\contraction[0.5ex]{}{1}{2}{x} 1 2 x} \partial_\mu \Phi_1 \partial^{\mu} \Phi_x \nn\\
& \qquad + \lambda_{\contraction[0.5ex]{}{1}{2}{3} 1 2 3 4} \partial_\mu \Phi_1 \, \partial^{\mu} \Phi_3 \, \Phi_4 - \lambda_{\contraction[0.5ex]{}{1}{}{2} 1 2 \contraction[0.5ex]{}{3}{}{4} 3 4} \partial_\mu \left( \partial^{\mu} \Phi_1 \partial_{\nu} \Phi_3 \partial^{\nu} \Phi_4 \right), \\
& (-\Box_{AdS} + \reg{m}^2_{\Delta_{x}}) \Phi_{x} = - \lambda_{12x} \Phi_1 \Phi_2 - \lambda_{34x} \Phi_3 \Phi_4  \nn\\
& \qquad - \lambda_{\contraction[0.5ex]{}{1}{}{2} 1 2 x} \partial_\mu \Phi_1 \partial^{\mu} \Phi_2 + \lambda_{\contraction[0.5ex]{}{1}{2}{x} 1 2 x} \partial_\mu \left( \partial^{\mu} \Phi_1 \, \Phi_2 \right) \nn\\
& \qquad - \lambda_{\contraction[0.5ex]{}{3}{}{4} 3 4 x} \partial_\mu \Phi_3 \partial^{\mu} \Phi_4 + \lambda_{\contraction[0.5ex]{}{3}{4}{x} 3 4 x} \partial_\mu \left( \partial^{\mu} \Phi_3 \, \Phi_4 \right). \label{eqofmox}
\end{align}
Since the action \eqref{Sasym} is symmetric under exchanging fields $\Phi_1, \Phi_2$ with $\Phi_3, \Phi_4$, equations of motion for $\Phi_3$ and $\Phi_4$ are equal to those of $\Phi_1$ and $\Phi_2$ with $(12) \leftrightarrow (34)$.

We can now calculate correlators perturbatively with the conventions summarized in Appendix \ref{sec:conventions}. There are two non-vanishing 3-point functions. The first one is
\begin{align} \label{asym_3pt}
\lla \O_1(\bs{k}_1) \O_2(\bs{k}_2) \O_x(\bs{k}_3) \rra & = \lambda_{12x} \ireg_{[\Delta_1 \Delta_2 \Delta_x]} + \lambda_{\contraction[0.5ex]{}{1}{}{2} 1 2 x} \ireg_{[\contraction[0.5ex]{}{\Delta}{{}_1}{\Delta} \Delta_1 \Delta_2 \Delta_x]} + \lambda_{\contraction[0.5ex]{}{1}{2}{x} 1 2 x} \ireg_{[\contraction[0.5ex]{}{\Delta}{{}_1 \Delta_2}{\Delta} \Delta_1 \Delta_2 \Delta_x]}.
\end{align}
The other non-vanishing correlator is $\lla \O_3 \O_4 \O_x \rra$, which can be obtained by exchanging $(12) \leftrightarrow (34)$ in the expression above. Similarly, the 4-point function $\lla \O_1 \O_2 \O_3 \O_4 \rra$ reads
\begin{align} \label{asym_4pt}
& \lla \O_1(\bs{k}_1) \O_2(\bs{k}_2) \O_3(\bs{k}_3) \O_4(\bs{k}_4) \rra = \lambda_{12x} \lambda_{34x} \ireg_{[\Delta_1 \Delta_2; \Delta_3 \Delta_4 x \Delta_x]} + \lambda_{1234} \ireg_{[\Delta_1 \Delta_2 \Delta_3 \Delta_4]} \nn\\
& \qquad + \lambda_{12x} \lambda_{\contraction[0.5ex]{}{3}{}{4} 3 4 x} \ireg_{[\Delta_1 \Delta_2; \contraction[0.5ex]{}{\Delta}{{}_3}{\Delta} \Delta_3 \Delta_4 x \Delta_x]} + \lambda_{\contraction[0.5ex]{}{1}{}{2} 1 2 x} \lambda_{3 4 x} \ireg_{[\contraction[0.5ex]{}{\Delta}{{}_1}{\Delta} \Delta_1 \Delta_2; \Delta_3 \Delta_4 x \Delta_x]} \nn\\
& \qquad + \lambda_{12x} \lambda_{\contraction[0.5ex]{}{3}{4}{x} 3 4 x} \ireg_{[\Delta_1 \Delta_2; \contraction[0.5ex]{}{\Delta}{{}_3 \Delta_4 x}{\Delta} \Delta_3 \Delta_4 x \Delta_x]} + \lambda_{\contraction[0.5ex]{}{1}{2}{x} 1 2 x} \lambda_{34x} \ireg_{[\contraction[0.5ex]{}{\Delta}{{}_1 \Delta_2; \Delta_3 \Delta_4 x}{\Delta} \Delta_1 \Delta_2; \Delta_3 \Delta_4 x \Delta_x]} \nn\\
& \qquad + \lambda_{\contraction[0.5ex]{}{1}{}{2} 1 2 x} \lambda_{\contraction[0.5ex]{}{3}{}{4} 3 4 x} \ireg_{[\contraction[0.5ex]{}{\Delta}{{}_1}{\Delta} \Delta_1 \Delta_2; \contraction[0.5ex]{}{\Delta}{{}_3}{\Delta} \Delta_3 \Delta_4 x \Delta_x]} + \lambda_{\contraction[0.5ex]{}{1}{2}{x} 1 2 x} \lambda_{\contraction[0.5ex]{}{3}{4}{x} 3 4 x} \ireg_{[\contraction[1.0ex]{}{\Delta}{{}_1 \Delta_2; \Delta_3 \Delta_4 x \,}{\Delta} \contraction[0.5ex]{\Delta_1 \Delta_2;}{\Delta}{{}_3 \Delta_4 x \!\!}{\Delta} \Delta_1 \Delta_2; \Delta_3 \Delta_4 x \Delta_x]} \nn\\
& \qquad + \lambda_{\contraction[0.5ex]{}{1}{}{2} 1 2 x} \lambda_{\contraction[0.5ex]{}{3}{4}{x} 3 4 x} \ireg_{[\contraction[0.5ex]{}{\Delta}{{}_1}{\Delta} \Delta_1 \Delta_2; \contraction[0.5ex]{}{\Delta}{{}_3 \Delta_4 x}{\Delta} \Delta_3 \Delta_4 x \Delta_x} + \lambda_{\contraction[0.5ex]{}{1}{2}{x} 1 2 x} \lambda_{\contraction[0.5ex]{}{3}{}{4} 3 4 x} \ireg_{[\contraction[1.0ex]{}{\Delta}{{}_1 \Delta_2; \Delta_3 \Delta_4 x}{\Delta} \Delta_1 \Delta_2; \contraction[0.5ex]{}{\Delta}{{}_3}{\Delta} \Delta_3 \Delta_4 x \Delta_x} \nn\\
& \qquad + \lambda_{\contraction[0.5ex]{}{1}{2}{3} 1 2 3 4} \ireg_{[\contraction[0.5ex]{}{\Delta}{{}_1 \Delta_2}{\Delta} \Delta_1 \Delta_2 \Delta_3 \Delta_4]} + \lambda_{\contraction[0.5ex]{}{1}{}{2} 1 2 \contraction[0.5ex]{}{3}{}{4} 3 4} \ireg_{[\contraction[0.5ex]{}{\Delta}{{}_1}{\Delta} \Delta_1 \Delta_2 \contraction[0.5ex]{}{\Delta}{{}_3}{\Delta} \Delta_3 \Delta_4]}
\end{align}
This expression looks a little daunting, mostly due to the large number of couplings present. The point here is that by choosing specific values of the couplings we can make the 4-point function equal to a single derivative or non-derivative amplitude. For example, by taking $\lambda_{12x} = \lambda_{\contraction[0.5ex]{}{3}{}{4} 3 4 x} = 1$ with all remaining couplings set to zero the 4-point function becomes equal to the amplitude $\ireg_{[\Delta_1 \Delta_2; \contraction[0.5ex]{}{\Delta}{{}_3}{\Delta} \Delta_3 \Delta_4 x \Delta_x]}$. Thus, all amplitudes discussed in this paper are physical in the sense that they represent a valid correlator in the asymmetric theory. Amplitudes are correlators.

\subsection{Alternative derivation of the amplitude identities}

There is an important consequence of the fact that the asymmetric theory realizes each amplitude as a boundary correlator. If we can prove some statement in the asymmetric theory for arbitrary values of the coupling constants, it must hold amplitude by amplitude. Such statements will then encompass relations between all the amplitudes present in \eqref{asym_3pt} and \eqref{asym_4pt}, proving a number of relations all at once. Since the manipulations of the action are generally simpler than the manipulations of the amplitudes, such an approach can simplify and unify the identities between various amplitudes.

Let us begin with 3-point functions. By integrating the action $S_{\text{asym}}^{(3)}$ in \eqref{S3asym} by parts we obtain
\begin{align} \label{S3asym_int}
S_{\text{asym}}^{(3)} & = \int \D^{d+1} x \, \sqrt{g} \left[ \lambda_{12x} \Phi_1 \Phi_2 \Phi_x + \lambda_{34x} \Phi_3 \Phi_4 \Phi_x \right] \nn\\
& \qquad + \frac{1}{2} \int \D^{d+1} x \, \sqrt{g} \left[ \lambda_{\contraction[0.5ex]{}{1}{}{2} 1 2 x} \left( \Phi_1 \Phi_2 \Box \Phi_x - \Phi_1 \Box \Phi_2 \Phi_x - \Box \Phi_1 \Phi_2 \Phi_x \right) + \right.\nn\\
& \qquad\qquad\qquad \left. \lambda_{\contraction[0.5ex]{}{1}{2}{x} 1 2 x} \left( \Phi_1 \Box \Phi_2 \Phi_x - \Box \Phi_1 \Phi_2 \Phi_x - \Phi_1 \Phi_2 \Box \Phi_x \right) + (12) \leftrightarrow (34) \right].
\end{align}
Since the boundary correlation functions in AdS/CFT are obtained from the on-shell action, we can substitute to this expression the bulk equations of motion \eqref{eqofmo1} - \eqref{eqofmox}. As long as we are interested in 3-point functions, we can limit our attention to the leading terms in the couplings. Thus, by substituting the zero-th order solutions, $\Box \Phi_j = m_j^2 \Phi_j$, we obtain the equivalent action
\begin{align}
S'^{(3)}_{\text{asym}} & = \left[ \lambda_{12x} + \frac{1}{2} \lambda_{\contraction[0.5ex]{}{1}{}{2} 1 2 x} \left( \reg{m}_{\Delta_x}^2 - \reg{m}_{\Delta_1}^2 - \reg{m}_{\Delta_2}^2 \right) + \frac{1}{2} \lambda_{\contraction[0.5ex]{}{1}{2}{x} 1 2 x} \left( \reg{m}_{\Delta_2}^2 - \reg{m}_{\Delta_1}^2 - \reg{m}_{\Delta_x}^2 \right)  \right] \times\nn\\
& \qquad\qquad \times \int \D^{d+1} x \, \sqrt{g} \, \Phi_1 \Phi_2 \Phi_x + \left[ (12) \leftrightarrow (34) \right],
\end{align}
This theory contains only two non-derivative vertices, $\Phi_1 \Phi_2 \Phi_x$ and $\Phi_3 \Phi_4 \Phi_x$. Nevertheless, as long as 3-point functions are concerned, it reproduces the same correlators as $S_{\text{asym}}^{(3)}$ in \eqref{S3asym}. Now we find
\begin{align}
\lla \O_1(\bs{k}_1) \O_2(\bs{k}_2) \O_x(\bs{k}_3) \rra & = \ireg_{[\Delta_1 \Delta_2 \Delta_x]} \times \left[ \lambda_{12x} + \frac{\lambda_{\contraction[0.5ex]{}{1}{}{2} 1 2 x}}{2} \left( - \reg{m}_{\Delta_1}^2 - \reg{m}_{\Delta_2}^2 + \reg{m}_{\Delta_x}^2 \right) \right.\nn\\
& \qquad\qquad\qquad\qquad \left. + \frac{\lambda_{\contraction[0.5ex]{}{1}{2}{x} 1 2 x}}{2} \left( - \reg{m}_{\Delta_1}^2 + \reg{m}_{\Delta_2}^2 - \reg{m}_{\Delta_x}^2 \right) \right].
\end{align}
and we can compare this expression with \eqref{asym_3pt}. By comparing the two expressions coupling by coupling, we arrive at the identity \eqref{3pt_calc}.

We can use the same logic to derive the identities obeyed by the 4-point amplitudes. In this case we must keep terms up to quadratic order in the couplings. After the equations of motion \eqref{eqofmo1} - \eqref{eqofmox} are substituted to \eqref{S3asym_int}, a number of new 4-point contact vertices is produced. However, as we are interested only in the correlator $\< \O_1 \O_2 \O_3 \O_4 \>$, it is sufficient to keep only those 4-point contact vertices that contain exactly the 4 four dual fields, $\Phi_1, \Phi_2, \Phi_3, \Phi_4$. This means that we can immediately drop all terms in \eqref{S3asym_int} with no derivatives on $\Phi_x$. All in all we find
\begin{align}
S'^{(4)}_{\text{asym}} & = - \int \D^{d+1} x \, \sqrt{g} \, \Phi_1 \Phi_2 \Phi_3 \Phi_4 \left[ \lambda_{1234} + \frac{1}{2} \left( \lambda_{12x} \lambda_{\contraction[0.5ex]{}{3}{4}{x} 3 4 x} - \lambda_{12x} \lambda_{\contraction[0.5ex]{}{3}{}{4} 3 4 x} + \lambda_{\contraction[0.5ex]{}{1}{2}{x} 1 2 x} \lambda_{34x} - \lambda_{\contraction[0.5ex]{}{1}{}{2} 1 2 x} \lambda_{34x} \right) \right] \nn\\
& \qquad - \int \D^{d+1} x \, \sqrt{g} \, \contraction{}{\Phi}{{}_1 \Phi_2}{\Phi} \Phi_1 \Phi_2 \Phi_3 \Phi_4 \left[ \lambda_{\contraction[0.5ex]{}{1}{2}{3} 1 2 3 4} + \frac{1}{2} \left( \lambda_{\contraction[0.5ex]{}{1}{}{2} 12x} \lambda_{\contraction[0.5ex]{}{3}{4}{x} 34x} - \lambda_{\contraction[0.5ex]{}{1}{2}{x} 12x} \lambda_{\contraction[0.5ex]{}{3}{4}{x} 34x} + \lambda_{\contraction[0.5ex]{}{1}{2}{x} 12x} \lambda_{\contraction[0.5ex]{}{3}{}{4} 34x} - \lambda_{\contraction[0.5ex]{}{1}{2}{x} 12x} \lambda_{\contraction[0.5ex]{}{3}{4}{x} 34x} \right) \right] \nn\\
& \qquad - \frac{1}{2} \int \D^{d+1} x \, \sqrt{g} \, \contraction{\Phi_1}{\Phi}{{}_2}{\Phi} \Phi_1 \Phi_2 \Phi_3 \Phi_4 \left[ \lambda_{\contraction[0.5ex]{}{1}{}{2} 12x} \lambda_{\contraction[0.5ex]{}{3}{4}{x} 34x} - \lambda_{\contraction[0.5ex]{}{1}{2}{x} 12x} \lambda_{\contraction[0.5ex]{}{3}{4}{x} 34x} \right] \nn\\
& \qquad - \frac{1}{2} \int \D^{d+1} x \, \sqrt{g} \, \contraction{}{\Phi}{{}_1 \Phi_2 \Phi_3}{\Phi} \Phi_1 \Phi_2 \Phi_3 \Phi_4 \left[ \lambda_{\contraction[0.5ex]{}{1}{2}{x} 12x} \lambda_{\contraction[0.5ex]{}{3}{}{4} 34x} - \lambda_{\contraction[0.5ex]{}{1}{2}{x} 12x} \lambda_{\contraction[0.5ex]{}{3}{4}{x} 34x} \right] \nn\\
& \qquad - \int \D^{d+1} x \, \sqrt{g} \lambda_{\contraction[0.5ex]{}{1}{}{2} \contraction[0.5ex]{12}{3}{}{4} 1234} \contraction{}{\Phi}{{}_1}{\Phi} \Phi_1 \Phi_2 \contraction{}{\Phi}{{}_3}{\Phi} \Phi_3 \Phi_4 + \text{ discarded 4-point terms}.
\end{align}
Now, one can derive the 4-point correlator $\< \O_1 \O_2 \O_3 \O_4 \>$ by using the action
\begin{align}
S'_{\text{asym}} = S^{\text{asym}}_{\text{free}} + S'^{(3)}_{\text{asym}} + S'^{(4)}_{\text{asym}}.
\end{align}
By comparing the coefficients of various combinations of the couplings between the derived 4-point function and the expression \eqref{asym_4pt} one independently verifies the entire set of the identities between the amplitudes derived in Section \ref{sec:identities_4ptX}.

We treat the described approach as a check on our results from the previous section, but it may be very useful for \textit{en masse} analysis of relations between higher-point amplitudes in the future.

\section{Warm-up: 3-point derivative amplitudes} \label{sec:3pt}

In this section we discuss 3-point derivative amplitudes. As far as the \emph{regulated} expressions are concerned, the 3-point derivative amplitudes trivialize due to the identity \eqref{3pt_calc},
\begin{align} \label{3pt_calc_2}
& \ireg_{[\Delta_1 \contraction[0.5ex]{}{\Delta}{{}_2}{\Delta} \Delta_2 \Delta_3]} = \frac{1}{2} \left( \reg{m}_{\Delta_1}^2 - \reg{m}_{\Delta_2}^2 - \reg{m}_{\Delta_3}^2 \right) \ireg_{[\Delta_1 \Delta_2 \Delta_3]}, && \reg{m}_{\Delta}^2 = \Dreg (\Dreg - \dreg),
\end{align}
valid in any regularization scheme. Nevertheless, when it comes to \emph{renormalized} amplitudes this identity may fail, if the prefactor vanishes. At the first glance this would suggest that the renormalized amplitude vanishes, but, as well will see, such a conclusion is not correct. 

Furthermore, we can use the 3-point functions to showcase several features that occur in derivative 4-point functions as well. The key observation is that some amplitudes can be scheme-dependent, even though they remain finite. This behavior is in stark contrast with the non-derivative amplitudes analyzed in \cite{Bzowski:2022rlz}, where scheme-dependence was always the consequence of divergences present in the regulated amplitude. As we will see, for derivative amplitudes this is not the case. In particular, renormalization for such amplitudes is essential to consider.

\subsection{Divergences and scheme-dependence} \label{sec:scheme_dependence}

Before formulating and addressing the issues raised above, let us recall from \cite{Bzowski:2015yxv,Bzowski:2015pba,Bzowski:2022rlz} how divergences and scheme-dependence can be extracted from the near-boundary analysis of the triple-$K$ integrals near $z = 0$. Consider a regulated contact amplitude $\ireg(\ep; u, v_j)$ of the form
\begin{align}
\ireg(\ep; u, v_j) = \int_0^{\infty} \D z \, \reg{\mathcal{I}}(\ep; u, v_j; z),
\end{align}
where $\reg{\mathcal{I}}$ is the integrand. The amplitude is regulated in an arbitrary $(u, v_j)$-scheme \eqref{general}, which means that the integrand $\reg{\mathcal{I}}$ depends on the regulator $\ep$ and the parameters $u$ and $v_j$.

First, let us recall that in dimensional regularization the only source of possible divergences and scheme-dependence are terms of order $z^{-1 + O(\ep)}$ in the integrand $\reg{\mathcal{I}}$. In \cite{Bzowski:2015yxv} we showed that all divergent and scheme-dependent terms can obtained by the following procedure. First, series expand $\reg{\mathcal{I}}$ around $z = 0$ and keep only terms of order $z^{-1 + O(\ep)}$. The exact form of such terms depends on the regularization scheme and certain singularity conditions satisfied by the amplitude; please, refer to \cite{Bzowski:2015yxv,Bzowski:2015pba,Bzowski:2022rlz} for details. In general, we define $\mathcal{I}^{\text{div}}$ as the sum of all terms of order $z^{-1 + O(\ep)}$, \textit{i.e.},
\begin{align} \label{Idiv}
\mathcal{I}^{\text{div}}(\ep; u, v_j; z) = \sum_n c_n(\ep; u, v_j) z^{-1 + a_n(u, v_j) \ep},
\end{align}
where $a_n$ and $c_n$ are some scheme-dependent constants with $a_n$ $\ep$-independent and $c_n$ having a finite $\ep \rightarrow 0$ limit\footnote{This assumes non-integral $\beta_j$'s, but the procedure for extraction of divergences and scheme-dependent terms works for integral $\beta_j$'s as well.}. The sum is over all, necessarily finite and usually small number of terms of order $z^{-1 + O(\ep)}$ in the power expansion. This can be integrated from $z = 0$ to $z = \mu^{-1}$, where $\mu > 0$ is an arbitrary cut-off,
\begin{align} \label{idiv}
\idiv(\ep; u, v_j) = \int_0^{\mu^{-1}} \D z \, \mathcal{I}^{\text{div}}(\ep; u, v_j; z) = \sum_n \frac{c_n(\ep; u, v_j) \mu^{-a_n(u, v_j) \ep}}{a_n(u, v_j) \ep}.
\end{align}
It turns out that when series expanded in $\ep$, this expression yields all divergent terms in the amplitude, \textit{i.e.},
\begin{align}
\ireg(\ep; u, v_j) = \idiv(\ep; u, v_j) + O(\ep^0).
\end{align}
Furthermore, assume that we know the regulated amplitude in some specific regularization scheme with given $u = \bar{u}$ and $v_j = \bar{v}_j$. Then, we can evaluate the amplitude in any other regularization scheme with $u$ and $v_j$, up to terms of order $O(\ep)$, by means of the scheme-changing formula
\begin{align}
\ireg(\ep; u, v_j) = \ireg(\ep; \bar{u}, \bar{v}_j) + \left[ \idiv(\ep; u, v_j) - \idiv(\ep; \bar{u}, \bar{v}_j) \right] + O(\ep).
\end{align}

Usually, one can calculate an amplitude with ease in the beta scheme \eqref{beta}, where $\bar{u} = 1$ and $\bar{v}_j = 0$. Using expression above one can express the amplitude in any other regularization scheme. For at most linearly divergent amplitudes, which is the case for contact amplitudes with non-integral $\beta_j$'s, the formula simplifies. To write it explicitly, let us define $\ireg|_{\ep^n}$ to represent the terms of order $\ep^n$ in a regulated amplitude $\ireg$, so that up to linear order a linearly divergent amplitude is given by
\begin{align}
\ireg(\ep; u, v_j) = \frac{\ireg(u, v_j)|_{\ep^{-1}}}{\ep} + \ireg(u, v_j)|_{\ep^0} + O(\ep).
\end{align}
With this notation in place, in an arbitrary $(u, v_j)$-scheme the expansion coefficients are given by
\begin{align}
\ireg(u, v_j)|_{\ep^{-1}} & = \idiv \left(u,v_j; 1 \right) |_{\ep^{-1}}, \label{ischM1} \\
\ireg(u, v_j)|_{\ep^{0}} & = \ireg(1, 0)|_{\ep^0} + \idiv \left(u,v_j; 1 \right)|_{\ep^0}, \label{isch0}
\end{align}
In \cite{Bzowski:2022rlz} this formula was extended to subleading orders in $\ep$ as well.

\subsection{Finite amplitudes} \label{sec:finite_amps}

First let us recall that all non-derivative 3-point amplitudes, $\ireg_{[222]}, \ireg_{[322]}, \ireg_{[332]}, \ireg_{[333]}$ are linearly divergent at $\ep = 0$. On the other hand two derivative amplitudes, $\ireg_{[2\contraction[0.5ex]{}{3}{}{2} 32]}$ and $\ireg_{[3\contraction[0.5ex]{}{3}{}{3} 33]}$, are finite, see Table \ref{fig:deriv_3pt} below. This would suggest that such amplitudes are scheme-independent (\textit{i.e.}, they do not depend on the regularization parameters $u$ and $v_j$) and can be calculated by setting $\epsilon = 0$ before the integration. As we will discuss now, this is not the case.

\begin{table}[ht]
\centering
\begin{tabular}{||c|c|c||}
\hline\hline
Degree of divergence & No-derivative amplitudes & Derivative amplitudes\\ \hline\hline
0 & none & $\ireg_{[2\contraction[0.5ex]{}{3}{}{2} 32]}, \ireg_{[3\contraction[0.5ex]{}{3}{}{3} 33]}$ \\ \hline
1 & $\ireg_{[222]}, \ireg_{[322]}, \ireg_{[332]}, \ireg_{[333]}$ & $\ireg_{[2\contraction[0.5ex]{}{2}{}{2} 22]}, \ireg_{[3\contraction[0.5ex]{}{2}{}{2} 22]}, \ireg_{[3\contraction[0.5ex]{}{3}{}{2} 32]}, \ireg_{[2\contraction[0.5ex]{}{3}{}{3} 33]}$ \\ \hline\hline
\end{tabular}
\caption{Degrees of divergence of 3-point amplitudes.\label{fig:deriv_3pt}}
\end{table}

Let us consider the two finite derivative amplitudes, $\ireg_{[2\contraction[0.5ex]{}{3}{}{2} 32]}$ and $\ireg_{[3\contraction[0.5ex]{}{3}{}{3} 33]}$, regulated in an arbitrary $(u, v_j)$-scheme, \eqref{general}. By using \eqref{3pt_calc_2} we immediately find,
\begin{align}
\ireg_{[\contraction[0.5ex]{3}{3}{}{3} 333]} & = \frac{1}{2} k_1^2 + (2 u + v_1 - v_2 - v_3) \left[ \frac{k_2^2}{2(2u - v_1 + v_2 - v_3)} + \frac{k_3^2}{2(2u - v_1 - v_2 + v_3)} \right] + O(\ep), \label{i3DD} \\
\ireg_{[\contraction[0.5ex]{2}{3}{}{2} 232]} & = - \frac{6 u + v_1 - 3 v_2 - v_3}{2} \left[ \frac{k_1}{2u + v_1 - v_2 - v_3} + \frac{k_3}{2u - v_1 - v_2 + v_3} \right]  + O(\ep). \label{i2Dd}
\end{align}
Despite the fact that the amplitudes are finite, they remain scheme-dependent. This behavior is different than what we found for non-derivative 3- and 4-point amplitudes in \cite{Bzowski:2015yxv,Bzowski:2015pba,Bzowski:2022rlz}. There, scheme-dependence was the consequence of the divergence: different regularizations yielded slightly different expressions, an expected phenomenon.

What is more, with the amplitudes  $\iuv_{[2\contraction[0.5ex]{}{3}{}{2} 32]}$ and $\iuv_{[3\contraction[0.5ex]{}{3}{}{3} 33]}$ finite, we could try substituting $\ep = 0$ before the integration. For the amplitude $\ireg_{[\contraction[0.5ex]{2}{3}{}{2} 232]}$ the $\epsilon \rightarrow 0$ limit of the integrand can be taken before the integration. We will denote the result by $\ino^0_{[2\contraction[0.5ex]{}{3}{}{2} 32]} $ and it reads
\begin{align} \label{i0_2Dd}
\ino^0_{[2\contraction[0.5ex]{}{3}{}{2} 32]} & = \frac{1}{2} \int_0^{\infty} \D z \, e^{-k_t z} \left[ ( k_3^2 - k_1^2 - k_2^2 ) + z k_t k_2 (k_2 + k_3 - k_1) \right] \nn\\
& = \frac{1}{2} ( -k_1 + k_3 ).
\end{align}
On the other hand the $\epsilon \rightarrow 0$ limit of the integrand of $\ireg_{[\contraction[0.5ex]{3}{3}{}{3} 333]}$ exihibits a quadratic pole at $z = 0$. Thus, its integral is divergent, $\ino^0_{[3\contraction[0.5ex]{}{3}{}{3} 33]} = \infty$. All in all, we encountered two new phenomena here:
\begin{enumerate}
\item Some amplitudes, despite being fintie, remain scheme-dependent.
\item For some finite amplitudes the regulator $\ep$ cannot be removed before integration.
\end{enumerate}

\subsubsection{Mathematical explanation}

First, let us concentrate on the scheme-dependence of the finite amplitudes \eqref{i3DD} and \eqref{i2Dd}. Scheme-dependence enters amplitudes through the divergent part $\idiv \left(u,v_j; 1 \right)|_{\ep^0}$ in \eqref{isch0}. For non-derivative amplitudes considered in \cite{Bzowski:2022rlz} the  scheme-dependence is always the result of the divergence. This happens because the sum of coefficients $c_n(\ep; u, v_j)$ in \eqref{Idiv} are in such cases non-vanishing for $\ep = 0$. Thus, $\idiv(\ep; u, v_j)$ are indeed divergent with scheme-dependent finite piece.

On the other hand, for derivative amplitudes the coefficients $c_n$ can be of order $O(\ep)$. Consider such $c_n$, \textit{i.e.}, say, $c_n = c(u, v_j) \ep + O(\ep^2)$. Such a term now enters $\idiv$ in \eqref{idiv}. With the corresponding $a_n(u, v_j)$ denoted simply by $a(u, v_j)$, the contribution to $\idiv$ reads
\begin{align}
\idiv \supseteq \int_0^{\mu^{-1}} \D z \, c \ep \, z^{-1 + a \ep} = \frac{c(u, v_j)}{a(u, v_j)} \mu^{-a(u, v_j) \ep}
\end{align}
and is finite at $\ep = 0$. In such a case $\idiv$ remains finite but scheme-dependent.

The reason why terms with $c_n = O(\ep)$ may appear in \eqref{Idiv} for derivative amplitudes is the presence of $z$-derivatives acting on propagators in \eqref{3pt2d}. Imagine that the series expansion about $z = 0$ of the propagator on which the derivative acts contains terms of the form $C z^{A \ep}$, for some $A$ and $C$. When the derivative hits, it produces terms $C A \ep z^{-1 + A \ep}$. In other words the derivative produces the term $c_n = C A \ep$ in the expansion. This is what distinguishes the derivative amplitudes from non-derivative ones.

Finally we can address the issue of when the $\ep \rightarrow 0$ limit can be taken before the integration and why the result does not have to match any $(u, v_j)$-scheme. In general, the series expansion of the integrand $\reg{\mathcal{I}}$ around $z = 0$ contains a finite number of terms of the negative order, \textit{i.e.}, terms of the form $z^{-a + O(\ep)}$ for $a \geq 1$. If such terms are present in the expansion, the integral of the limit $\lim_{\ep \rightarrow 0} \reg{\mathcal{I}}$ may diverge. To be precise the integral diverges if there exists a coefficient of some term of the form $z^{-a + O(\ep)}$ for $a \geq 1$ that is non-vanishing at $\ep = 0$. This observation also explains why the results obtained by taking the limit first do not have to match any $(u, v_j)$-regularization scheme. Indeed, for the terms of order $z^{-1 + O(\ep)}$ we have
\begin{align}
\frac{\mu^{-a \ep}}{a} = \lim_{\epsilon \rightarrow 0} \int_0^{\mu^{-1}} \D z\,\ep z^{-1 + a \ep} \ \neq \ \int_0^{\mu^{-1}} \D z\, \lim_{\epsilon \rightarrow 0} \ep z^{-1 + a \ep} = 0.
\end{align}
Thus, with non-vanishing $\ep$, the terms of the form $c \ep z^{-1 + a \ep}$ in the integrand yield a finite, scheme-dependent contribution to the amplitude. On the other hand, if the $\ep \rightarrow 0$ limit is taken before the integration, such terms simply vanish.

Let us finally see what is the situation for our two examples, $\ireg_{[2\contraction[0.5ex]{}{3}{}{2} 32]}$ and $\ireg_{[3\contraction[0.5ex]{}{3}{}{3} 33]}$. Without loss of generality let us work here in the beta scheme, \eqref{beta}, so the expressions are significantly shorter. The series expansions of the integrands $\reg{\mathcal{I}}_{[2\contraction[0.5ex]{}{3}{}{2} 32]}$ and $\reg{\mathcal{I}}_{[3\contraction[0.5ex]{}{3}{}{3} 33]}$ around $z = 0$ read
\begin{align}
\reg{\mathcal{I}}_{[2\contraction[0.5ex]{}{3}{}{2} 32]} & = \ep (1 + \ep) z^{- 2 + \ep} - \ep \left[ (1 + \ep) k_1 + (2 + \ep) k_3 \right] z^{-1 + \ep} + O(z^0), \\
\reg{\mathcal{I}}_{[3\contraction[0.5ex]{}{3}{}{3} 33]} & = \ep^2 z^{-4 + \ep} - \frac{1}{2} \left[ ( s^2 - k_2^2 - k_3^2) + O(\ep) \right] z^{-2 + \ep} + \frac{\ep}{3} \left[ 3 (k_2^3 + k_3^3) + O(\ep) \right] z^{-1 + \ep} + O(\ep^0).
\end{align}
Notice that both expansions contain terms of the form $z^{-a + O(\ep)}$ for $a \geq 1$. However, when the limit $\ep \rightarrow 0$ is taken before the integration, all divergent terms in $\reg{\mathcal{I}}_{[2\contraction[0.5ex]{}{3}{}{2} 32]}$ vanish. Thus, the resulting integral is convergent with the result given in \eqref{i0_2Dd}. On the other hand the divergent term of order $z^{-2}$ remains, when the $\ep \rightarrow 0$ limit of $\mathcal{I}_{[3\contraction[0.5ex]{}{3}{}{3} 33]}$ is taken. Thus, the unregulated integral diverges. Finally, we can see that the coefficients of the terms of order $z^{-1 + \ep}$ are indeed of order $O(\ep)$. Thus, when integrated before taking the $\ep \rightarrow 0$ limit, they produce finite contributions to the amplitudes.

\subsubsection{Physical explanation}

Physically, the scheme-dependence of any amplitude, finite or not, is related to the existence of counterterms. As argued in Section \ref{sec:asymmetric_theory} we can think about any amplitude as representing a full correlation function in the asymmetric theory. Consider the 3-point function $\< \O_1 \O_2 \O_3 \>$ of three boundary operators of dimensions $\Delta_1, \Delta_2, \Delta_3$ sourced by three boundary sources $\phi_1, \phi_2, \phi_3$. The 3-point function is then identified with the 3-point amplitude in the asymmetric theory.

The existence of a local boundary counterterm in dimensional regularization is determined by whether the dimensions of the operators ($\Delta_j$) and their sources ($d - \Delta_j$) can be arranged in such a way that they sum to $d$. The counterterm can also contain $2r$ derivatives for a non-negative integer $r$ and takes the schematic form
\begin{align} \label{Sct_form}
S_{ct} \sim \int \D^{d} \bs{x} \, X_1 X_2 X_3 \partial^{2r}, \qquad X_j = \phi_j \text{ or } \O_j.
\end{align}
The derivatives must be distributed between $X_j$'s and contracted. Such a term exists if and only if the dimensions are such that
\begin{align} \label{sch_cond}
x_1 + x_2 + x_3 + 2 r = d, \qquad x_j = d - \Delta_j \text{ or } \Delta_j.
\end{align}
The choice of the source, $X_j = \phi_j$, in the counterterm corresponds to $x_j = d - \Delta_j$, the dimension of the source. The choice of the operator in the counterterm, $X_j = \O_j$, corresponds to $x_j = \Delta_j$.

Most importantly for our discussion here is that the form and the existence of a local counterterm is independent of the details of the bulk theory. The form of the counterterms \eqref{Sct_form} does not depend on the specifics of the bulk interactions, but only on the boundary data. Thus, whenever the condition \eqref{sch_cond} is satisfied, a local contribution from \eqref{Sct_form} to the correlator is to be expected. Thus, we should expect that even finite amplitudes remain scheme-dependent when the condition is satisfied.

We should also point out that in a physical bulk theory we can expect all kinds of interactions, with and without derivatives. Thus, a 3-point function of generic boundary operators will contain contributions from, \textit{a priori}, infinite number of couplings containing higher derivatives,
\begin{align}
& \< \O_{[\Delta_1]} \O_{[\Delta_2]} \O_{[\Delta_3]} \> = \lambda_{[\Delta_1 \Delta_2 \Delta_3]} \ino_{[\Delta_1 \Delta_2 \Delta_3]} \nn\\
& \qquad + \lambda_{[\contraction[0.5ex]{}{\Delta}{{}_1}{\Delta} \Delta_1 \Delta_2 \Delta_3]} \ino_{[\contraction[0.5ex]{}{\Delta}{{}_1}{\Delta} \Delta_1 \Delta_2 \Delta_3]} + \lambda_{[\contraction[0.5ex]{\Delta_1}{\Delta}{{}_2}{\Delta} \Delta_1 \Delta_2 \Delta_3]} \ino_{[\contraction[0.5ex]{\Delta_1}{\Delta}{{}_2}{\Delta} \Delta_1 \Delta_2 \Delta_3]} + \lambda_{[\contraction[0.5ex]{}{\Delta}{{}_1 \Delta_2}{\Delta} \Delta_1 \Delta_2 \Delta_3]} \ino_{[\contraction[0.5ex]{}{\Delta}{{}_1 \Delta_2}{\Delta} \Delta_1 \Delta_2 \Delta_3]} \nn\\
& \qquad + \text{higher-derivative amplitudes}.
\end{align}
It is the regulated correlation function that is scheme-dependent and eventually gets renormalized. Thus, from the physics point of view we can always think about scheme-dependence of any derivative amplitude as the artifact of the scheme-dependence of the corresponding non-derivative amplitude.

\subsection{Renormalization}

\subsubsection{Symmetries}

As discussed above, the form of available counterterms does not depend on the details of the bulk theory. Thus, as far as the form of counterterms go, we could copy here equations (4.25) -- (4.28) of \cite{Bzowski:2022rlz}. The problem with such an approach is that the counterterms there assumed certain symmetry properties of the renormalized amplitudes, which may not hold for derivative amplitudes. For example, it was assumed there that the renormalized amplitude $\iren_{[333]}$ should be completely symmetric under any permutation of momenta. This forced the counterterm to prevent the symmetry. Hence, only a single counterterm constant, $\act_{[333]}$, is present in the action (4.28) of \cite{Bzowski:2022rlz} and it reads
\begin{align}
S_{[333]}^{\text{ct}\,(3)} & = \frac{1}{3} \Gamma(\ep) \mu^{-\ep} \act_{[333]}  \int \D^{3 + 2 \ep} \bs{x} \, ( \Src_{1} \Src_{2} \Op_{3} + \Src_{2} \Src_{3} \Op_{1} + \Src_{3} \Src_{1} \Op_{2} ), \label{oldS3asymCt333}
\end{align}
where $\act_{[333]} = 1 + \ep \, \act_{[333]}^{(1)} + O(\ep^2)$. We have three boundary operators here, $\O^{[3]}_j$, and their three sources, $\phi^{[0]}_j$, for $j = 1,2,3$. The numbers in brackets indicate unregulated dimension of those objects, according to the rules listed in Appendix \ref{sec:QFT_conventions}. The contribution to the 3-point amplitude from such a counterterm reads
\begin{align}
- \frac{1}{3} \Gamma(\ep) \mu^{-\ep} \act_{[333]} \left[ k_1^3 + k_2^3 + k_3^3 \right].
\end{align}
As we can see, the contribution is completely symmetric in momenta, which renders the renormalized amplitude $\iren_{[333]}$ completely symmetric as well,
\begin{align}
\iren_{[333]} & = -\frac{1}{3} (k_1^3 + k_2^3 + k_3^3) \left[ \log \left( \frac{k_t}{\mu} \right) + \act_{[333]}^{(1)} - \tfrac{4}{3} \right] + \frac{1}{3} \left[ k_1^2 k_2 + 5 \text{ perms.} - k_1 k_2 k_3 \right],
\end{align}
where $k_t = k_1 + k_2 + k_3$. On the other hand, if the amplitude represents a 3-point function $\<\O_1 \O_2 \O_3 \>$ of 3 \emph{different} operators all of dimension $3$, there is no reason to impose the total symmetry of the renormalized 3-point amplitude. In particular, there is no reason to assume the total symmetry of the counterterm for the derivative amplitudes such as $\iren_{[3\contraction[0.5ex]{}{3}{}{3} 33]}$.

Due to this reasoning we will correct the analysis of counterterms in \cite{Bzowski:2022rlz} to include all possible counterterms, regardless of the symmetry properties of the underlying amplitude. For this reason, we must consider the counterterm \eqref{oldS3asymCt333} with 3 different renormalization constants, $\act_{[333], j}$, $j=1,2,3$, one for each term there. The counterterm action to consider reads
\begin{align}
S_{[333]}^{\text{ct}\,(3)} & = \frac{1}{3} \Gamma(\ep) \mu^{-\ep}  \int \D^{3 + 2 \ep} \bs{x} \, \left[  \act_{[333], 1} \Src_{2} \Src_{3} \Op_{1} + \act_{[333], 2} \Src_{3} \Src_{1} \Op_{2} + \act_{[333], 3} \Src_{1} \Src_{2} \Op_{3} \right] \label{newS3asymCt333}
\end{align}
where $\act_{[333], j} = 1 + \ep \, \act_{[333], j}^{(1)} + O(\ep^2)$ for $j = 1,2,3$. The renormalized amplitude thus reads
\begin{align}
\iren_{[333]} & = -\frac{1}{3} (k_1^3 + k_2^3 + k_3^3) \log \left( \frac{k_t}{\mu} \right) + \frac{1}{3} \left[ k_1^2 k_2 + 5 \text{ perms.} - k_1 k_2 k_3 \right] \nn\\
& \qquad\qquad - \frac{1}{3} \left[ \left( \act_{[333], 1}^{(1)} - \tfrac{4}{3} \right) k_1^3 + \left( \act_{[333], 2}^{(1)}  - \tfrac{4}{3} \right) k_2^3 + \left( \act_{[333], 3}^{(1)}  - \tfrac{4}{3} \right) k_3^3 \right]
\end{align}
and it is fully symmetric only if the constant $\act_{[333], j}^{(1)}$ are all equal.

\subsubsection{Renormalization} \label{sec:3pt_renorm}

With the above discussion in mind, we can renormalize all 3-point functions by using equations (4.25) -- (4.28) of \cite{Bzowski:2022rlz}. We modify them in order to incorporate the maximal number of counterterm constants, not restricted by imposing any unnecessary symmetries. We have
\begin{align} \label{S3asymCt222}
S_{[222]}^{\text{ct}\,(3)} & = \int \D^{3 + 2\ep} \bs{x} \, \left[\mathfrak{s}_{[222]} \, \src_{1} \src_{2} \src_{3} \right], \\
S_{[322]}^{\text{ct}\,(3)} & = \int \D^{3 + 2 \ep} \bs{x} \, \left[ \mathfrak{s}_{[322], 3} \Src_{1} \src_{2} \op_{3} + \mathfrak{s}_{[322], 2} \Src_{1} \src_{3} \op_{2} \right],  \label{S3asymCt322} \\
S_{[332]}^{\text{ct}\,(3)} & = \int \D^{3 + 2 \ep} \bs{x} \, \left[ \mathfrak{s}_{[332]} \, \src_{3} \partial_\mu \Src_{1} \partial^\mu \Src_{2} \right], \\
S_{[333]}^{\text{ct}\,(3)} & =  \int \D^{3 + 2 \ep} \bs{x} \, \left[ \mathfrak{s}_{[333], 1} \Src_{2} \Src_{3} \Op_{1} + \mathfrak{s}_{[333], 2} \Src_{3} \Src_{1} \Op_{2} + \mathfrak{s}_{[333], 3} \Src_{1} \Src_{2} \Op_{3} \right]. \label{S3asymCt333}
\end{align}
The values of the counterterm constants, however, depend on the amplitude we consider. With no derivatives we recover relations (3.19) -- (3.20) of \cite{Bzowski:2022rlz},
where the values of the constants are
\begin{align}
& \mathfrak{s}_{[222]} = - \Gamma(\ep) \mu^{-\ep} \act_{[222]}, && \mathfrak{s}_{[322], j} = \Gamma(\ep) \mu^{-\ep} \act_{[322], j}, \label{s222ands322} \\
& \mathfrak{s}_{[332]} = \Gamma(\ep) \mu^{-\ep} \act_{[332]}, && \mathfrak{s}_{[333], j} = \frac{1}{3} \Gamma(\ep) \mu^{-\ep} \act_{[333], j}. \label{s332}
\end{align} 
Analogously, for the four divergent derivative amplitudes from Table \ref{fig:deriv_3pt} we need
\begin{align}
& \mathfrak{s}_{[2\contraction[0.5ex]{}{2}{}{2} 22]} = - \Gamma(\ep) \mu^{-\ep} \act_{[2 \contraction[0.5ex]{}{2}{}{2} 22]}, && \mathfrak{s}_{[3\contraction[0.5ex]{}{2}{}{2} 22], j} = 2 \Gamma(\ep) \mu^{-\ep} \act_{[3\contraction[0.5ex]{}{2}{}{2} 22], j}, \label{s2ddands3dd} \\
& \mathfrak{s}_{[2\contraction[0.5ex]{}{3}{}{3} 33]} = - \Gamma(\ep) \mu^{-\ep} \act_{[2\contraction[0.5ex]{}{3}{}{3} 33]}, && \mathfrak{s}_{[3\contraction[0.5ex]{}{3}{}{3} 32]} = \Gamma(\ep) \mu^{-\ep} \act_{[3\contraction[0.5ex]{}{3}{}{3} 32]}. \label{s2DDands3Dd}
\end{align} 
These counterterms are linearly divergent and all the counterterm constants are of the form $\act_j = 1 + \act_j^{(1)} \ep + O(\ep^2)$ with the leading term fixed to cancel the divergence and the subleading term unfixed and scheme-dependent. When the contribution from the counterterms is added to the regulated amplitudes, we obtain finite renormalized expressions. For example, we find
\begin{align}
\iren_{[2\contraction[0.5ex]{}{2}{}{2} 22]} & = - \log \left( \frac{k_t}{\mu} \right) - \act_{[2\contraction[0.5ex]{}{2}{}{2} 22]} + \frac{3}{2}.
\end{align}

Let us now concentrate on the two finite derivative amplitudes, $\ireg_{[2\contraction[0.5ex]{}{3}{}{2} 32]}$ and $\ireg_{[3\contraction[0.5ex]{}{3}{}{2} 33]}$. Their regulated expressions are given by \eqref{i3DD} and \eqref{i2Dd} and, as noted there, they are completely local. Thus, we can add local finite terms to the action, which yield the renormalized amplitudes equal to
\begin{align}
\iren_{[3\contraction[0.5ex]{}{3}{}{3} 33]} & = \act_{[3\contraction[0.5ex]{}{3}{}{3} 33], 1}^{(1)} k_1^3 + \act_{[3\contraction[0.5ex]{}{3}{}{3} 33], 2}^{(1)} k_2^3 + \act_{[3\contraction[0.5ex]{}{3}{}{3} 33], 3}^{(1)} k_3^3, \label{iren3DD} \\
\iren_{[2\contraction[0.5ex]{}{3}{}{2} 32]} & = \act_{[2\contraction[0.5ex]{}{3}{}{2} 32], 1}^{(1)} k_1 + \act_{[2\contraction[0.5ex]{}{3}{}{2} 32], 3}^{(1)} k_3, \label{iren2Dd}
\end{align}
with arbitrary constants $\act_{[2\contraction[0.5ex]{}{3}{}{2} 32], j}^{(1)}$ and $\act_{[3\contraction[0.5ex]{}{3}{}{2} 33], j}^{(1)}$. In order to bring the renormalized amplitudes to this form, the values of the constants $\mathfrak{s}_{[2\contraction[0.5ex]{}{3}{}{2} 32], j}$ and  $\mathfrak{s}_{[3\contraction[0.5ex]{}{3}{}{3} 33], j}$ in \eqref{S3asymCt322} and \eqref{S3asymCt333} must be parameterized as
\begin{align} \label{s3deriv}
& \mathfrak{s}_{[2\contraction[0.5ex]{}{3}{}{2} 32], j} = \mu^{-\ep} \left( \frac{3}{2} + \act_{[2\contraction[0.5ex]{}{3}{}{2} 32], j}^{(1)} + O(\ep) \right), && \mathfrak{s}_{[3\contraction[0.5ex]{}{3}{}{3} 33], j} = \mu^{-\ep} \left( \frac{1}{2} - \act_{[3\contraction[0.5ex]{}{3}{}{3} 33], j}^{(1)} + O(\ep) \right).
\end{align}
In particular, by choosing specific values of the renormalization constants we can set the two amplitudes, $\iren_{[2\contraction[0.5ex]{}{3}{}{2} 32]}$ and $\iren_{[3\contraction[0.5ex]{}{3}{}{3} 33]}$, to vanish,
\begin{align} \label{ren3_zero}
& \iren_{[3\contraction[0.5ex]{}{3}{}{3} 33]} \doteq 0, && \iren_{[2\contraction[0.5ex]{}{3}{}{2} 32]} \doteq 0.
\end{align}
We use $\doteq$ to denote the fact that we fixed some or all scheme-dependence and the equality holds in a specific renormalization scheme only. As we shall see next, the choice of the renormalization scheme for 3-point amplitudes will now influence the renormalized 4-point exchange amplitudes.

\section{4-point contact amplitudes} \label{sec:4ptC}

In this section we consider 4-point derivative contact amplitudes for scalar fields of dimensions $\Delta_j = 2$ or $3$ and in $d = 3$ spacetime dimensions. 

\subsection{Divergences and analytic structure} \label{sec:divs_4ptC}

Since we consider scalar fields of dimensions $\Delta_j = 2$ or $3$ and in $d = 3$ spacetime dimensions, the orders of the Bessel functions $\beta_j = \Delta_j - d/2$ in the propagators are equal to either $1/2$ or $3/2$. Since in the beta scheme \eqref{beta} the orders of the Bessel functions do not regulate, the 4-point contact amplitudes trivialize to elementary integrals by means of the relation \eqref{Khalfint}. Looking at the integrands of (\ref{intro_4ptC}, \ref{4ptC2d}, \ref{4ptC4d}) we see that all contact amplitudes exhibit the structure given in \eqref{iC_R}, with $W$ being a polynomial of momenta with a finite or vanishing $\ep \rightarrow 0$ limit. All contact amplitudes for non-integral $\beta_j$'s are either linearly divergent at $\ep = 0$ or finite. Thus, when series expanded around $\ep = 0$, the divergent contact amplitudes contain a scale-violating logarithm, $\log(k_T)$, from the expansion of $k_T^{-2 \ep}$. On the other hand all finite contact amplitudes are rational functions of momenta. 

\begin{table}[ht]
\centering
\begin{tabular}{||c||c|c|c|c|c||}
\hline \hline
Amplitude & $\ireg_{[2222]}$ & $\ireg_{[3222]}$ & $\ireg_{[3322]}$ & $\ireg_{[3332]}$ & $\ireg_{[3333]}$ \\ \hline
Divergence & \indep & $\one$ & $\one$ & $\one$ & $\one$ \\ \hline\hline
\end{tabular}

\vspace{0.5cm}

\begin{tabular}{||c||c|c|c||}
\hline \hline
Amplitude & $\ireg_{[\contraction[0.5ex]{}{2}{}{2} 2222]}$ & 
$\ireg_{[\contraction[0.5ex]{}{2}{}{2} 2232]}$ & 
$\ireg_{[\contraction[0.5ex]{}{2}{}{2} 2233]}$ \\ \hline 
Divergence & \indep & $\one$ & $\one$ \\ \hline \hline
Amplitude & $\ireg_{[\contraction[0.5ex]{}{3}{}{2} 3222]}$ & 
$\ireg_{[\contraction[0.5ex]{}{3}{}{2} 3232]}$ & 
$\ireg_{[\contraction[0.5ex]{}{3}{}{2} 3233]}$ \\ \hline
Divergence & \dep & \dep & $\one$ \\ \hline\hline
Amplitude & $\ireg_{[\contraction[0.5ex]{}{3}{}{3} 3322]}$ &
$\ireg_{[\contraction[0.5ex]{}{3}{}{3} 3332]}$ &
$\ireg_{[\contraction[0.5ex]{}{3}{}{3} 3333]}$ \\ \hline
Divergence & \indep & $\one$ & \dep \\ \hline\hline
\end{tabular}

\vspace{0.5cm}

\begin{tabular}{||c||c|c|c|c|c|c||}
\hline \hline
Amplitude &
$\ireg_{[\contraction[0.5ex]{}{2}{}{2} 22 \contraction[0.5ex]{}{2}{}{2} 22]}$ &
$\ireg_{[\contraction[0.5ex]{}{3}{}{2} 32 \contraction[0.5ex]{}{2}{}{2} 22]}$ &
$\ireg_{[\contraction[0.5ex]{}{3}{}{3} 33 \contraction[0.5ex]{}{2}{}{2} 22]}$ &
$\ireg_{[\contraction[0.5ex]{}{3}{}{2} 32 \contraction[0.5ex]{}{3}{}{2} 32]}$ &
$\ireg_{[\contraction[0.5ex]{}{3}{}{3} 33 \contraction[0.5ex]{}{3}{}{2} 32]}$ &
$\ireg_{[\contraction[0.5ex]{}{3}{}{3} 33 \contraction[0.5ex]{}{3}{}{3} 33]}$ \\ \hline
Divergence & \indep & \dep & \indep & \indep & \dep & \indep \\ \hline\hline
\end{tabular}
\caption{Degrees of divergences of various 0-, 2-, and 4-derivative 4-point contact amplitudes. The `Wick contractions' indicate on which propagators the derivatives act and how they are contracted. For finite amplitudes ($\text{Div} = 0$) we indicate by `dep' if the finite part is scheme-dependent (\textit{i.e.}, $u$ and $v_j$-dependent) and by `indep' if it is scheme-independent. See equations \eqref{intro_4ptC}, \eqref{4ptC2d} and \eqref{4ptC4d} for definitions of the derivative contact amplitudes.\label{fig:divsC}}
\end{table}

The degrees of divergences of the contact amplitudes analyzed in this paper are listed in Table \ref{fig:divsC}. As we can see the higher number of derivatives, the lower divergence of the integrals. In particular all 4-derivative 4-point contact amplitudes considered in this paper are finite on their own. Nevertheless, similarly to the case of 3-point amplitudes discussed in Section \ref{sec:finite_amps}, some finite 4-point amplitudes are scheme-independent, while some retain the memory of the regularization scheme used.

\subsection{Examples}

Let us present here a few examples of contact derivative 4-point amplitudes. The symbols used in the following expressions are gathered in the Appendix \ref{sec:conventions}. The complete list is available in the \verb|HandbooK| Mathematica package attached to this paper. In Section \ref{sec:package} I will show how to access all the amplitudes stored in the package.

\subsubsection[\texorpdfstring{${[22 \contraction[0.5ex]{}{2}{}{2} 22]}$}{[22dd]}]{Amplitude $\bs{\ino_{[22 \contraction[0.5ex]{}{2}{}{2} 22]}}$}

The simplest derivative 4-point amplitude is $\ino_{[22 \contraction[0.5ex]{}{2}{}{2} 22]}$. The amplitude is finite and scheme-independent as no counterterms exist due to the dimensional reasons. It reads
\begin{align}
\iren_{[22 \contraction[0.5ex]{}{2}{}{2} 22]} = \ifin_{[22 \contraction[0.5ex]{}{2}{}{2} 22]} & = - \frac{s^2}{k_T^3} + \frac{1}{k_T} - \frac{\tau}{k_T^3}.
\end{align}
This amplitude is also maximally symmetric for a derivative amplitude. Its symmetry group is $D_4 \leq S_4$ and thus it can be rewritten in terms of 4 symmetric polynomials plus an additional polynomial $\tau = (k_1 + k_2)(k_3 + k_4)$. See Section \ref{sec:variables} for the discussion of the invariant polynomials and variables.

\subsubsection[\texorpdfstring{${[33 \contraction[0.5ex]{}{3}{}{3} 33]}$}{[33DD]}]{Amplitude $\bs{\ino_{[33 \contraction[0.5ex]{}{3}{}{3} 33]}}$} 

Let us now consider another maximally symmetric amplitude, $\ino_{[33 \contraction[0.5ex]{}{3}{}{3} 33]}$. In the beta scheme the amplitude reads
\begin{align}
\ireg_{[33 \contraction[0.5ex]{}{3}{}{3} 33]} & = (1 + 2 \ep) \Gamma(1 + \ep) k_T^{-2 \ep} \left\{ s^2 \left[ - \frac{2(1+\ep) \sigma_4}{k_T^3} - \frac{\sigma_3}{k_T^2} - \frac{\sigma_2}{(1 + 2\ep) k_T} + \frac{k_T}{1-4\ep^2} \right] \right.\nn\\
& \qquad - \frac{2(1+\ep) \tau \sigma_4}{k_T^3} - \frac{\tau \sigma_3}{k_T^2} + \frac{\tau(\tau - 2\sigma_2) + 2(-1 + 2 \ep + \ep^2) \sigma_4}{(1 + 2\ep) k_T} \nn\\
& \qquad \left. + \frac{2+\ep}{1 + 2 \ep} \sigma_3 - \frac{\ep k_T \sigma_2}{1 - 4 \ep^2} + \frac{\ep k_T^3}{(1 + 2 \ep)(-1 + 2\ep)(-3 + 2\ep)} \right\}.
\end{align}
This expression is valid to all orders in $\ep$. The symmetry group of this amplitude is the dihedral subgroup $D_4 \leq S_4$. As discusses in Section \ref{sec:variables} such amplitudes can be expressed in terms of 4 symmetric polynomials $\sigma_j$ of the 4 momenta magnitudes $k_j$ for $j = 1,2,3,4$ as well as the additional polynomial $\tau = (k_1 + k_2)(k_3 + k_4)$.

We can use the methods described in Section \ref{sec:scheme_dependence} to see if the amplitude is scheme-dependent. Just as was the case for the 3-point functions, the amplitude turns out to be scheme-dependent, despite being finite. In a general $(u,v_j)$-scheme \eqref{general} the amplitude reads
\begin{align} \label{iDD33}
\iuv_{[33 \contraction[0.5ex]{}{3}{}{3} 33]} & = s^2 \left[ - \frac{\sigma_4}{k_T^3} - \frac{\sigma_3}{2 k_T^2} - \frac{\sigma_2}{2 k_T} + \frac{k_T}{2} \right] \nn\\
& \qquad - \frac{\tau (2 \sigma_4 + k_T \sigma_3)}{2 k_T^3} + \frac{\tau(\tau - 2 \sigma_2) - 2 \sigma_4}{2 k_T} + \sigma_3 \nn\\
& \qquad + k_3^3 \left( \frac{u-v_4}{2u - v_T + 2 v_3} - \frac{1}{2} \right) + k_4^3 \left( \frac{u-v_3}{2u - v_T + 2 v_4} - \frac{1}{2} \right) + O(\ep).
\end{align}
The terms in the last line are scheme-dependent. Note that the scheme-dependent terms can also break the $D_4$ symmetry of the beta scheme-regulated correlator.

The scheme-dependence can be explained by the existence of local terms, which can be added to the action. If the amplitude represents the 4-point function $\< \O_1 \O_2 \O_3 \O_4 \>$ of four operators $\O_j$ of dimension $3$ sourced by four sources $\phi_j$ for $j = 1,2,3,4$, then we can form the local term
\begin{align} \label{Sloc3333}
S_{\text{loc}} = \int \D^{3} \bs{x} \left[ \mathfrak{a}_1  \O_1 \phi_2 \phi_3 \phi_4 + \mathfrak{a}_2 \phi_1 \O_2 \phi_3 \phi_4 + \mathfrak{a}_3 \phi_1 \phi_2 \O_3 \phi_4 + \mathfrak{a}_4 \phi_1 \phi_2 \phi_3 \O_4 \right],
\end{align}
where $\act_j$ are 4 undetermined constants. In the spirit of the paper these constants should be denoted as $\act_{[33 \contraction[0.5ex]{}{3}{}{3} 33], j}^{(1)}$, but we will omit most of their indices for clarity. The contribution of the action to the amplitude is $\sum_{j=1}^4 \mathfrak{a}_j k_j^3$. Thus, in the renormalized amplitude we can set these combinations of momenta to whatever values suits us. The renormalized amplitude can be written as
\begin{align}
\iren_{[33 \contraction[0.5ex]{}{3}{}{3} 33]} & = s^2 \left[ - \frac{\sigma_4}{k_T^3} - \frac{\sigma_3}{2 k_T^2} - \frac{\sigma_2}{2 k_T} + \frac{k_T}{2} \right] \nn\\
& \qquad - \frac{\tau (2 \sigma_4 + k_T \sigma_3)}{2 k_T^3} + \frac{\tau(\tau - 2 \sigma_2) - 2 \sigma_4}{2 k_T} + \sigma_3 \nn\\
& \qquad + \mathfrak{a}'_1 k_1^3 + \mathfrak{a}'_2 k_2^3 + \mathfrak{a}'_3 k_3^3 + \mathfrak{a}'_4 k_4^3.
\end{align}
The constants $\mathfrak{a}'_j$ are adjusted in such a way that the local, scheme-dependent terms in \eqref{iDD33} are removed in favor of the renormalization constants,
\begin{align}
\mathfrak{a}_{1,2} & = \mathfrak{a}'_{1,2}, \\
\mathfrak{a}_3 & = \mathfrak{a}'_3 - \left( \frac{u-v_4}{2u - v_T + 2 v_3} - \frac{1}{2} \right), \\
\mathfrak{a}_4 & = \mathfrak{a}'_4 - \left( \frac{u-v_3}{2u - v_T + 2 v_4} - \frac{1}{2} \right).
\end{align}

\textit{A priori} the four renormalization constants $\act'_j$ for $j=1,2,3,4$ can be chosen independently. This is a valid choice even though it breaks the dihedral symmetry of the amplitude, unless all $\act'_j$ are equal. Indeed, if the amplitude represented the 4-point function of four \emph{different} operators of dimension 3, there would be no reason to impose the symmetry. Finally, there is no obstacle for the arbitrary choice of the renormalization constants from the point of view of conformal symmetry either. Each expression $k_j^3$ for $j=1,2,3,4$ satisfies conformal Ward identities separately and thus can represent a valid, albeit local, 4-point amplitude.

It is also worth mentioning that despite the amplitude being finite, the $\ep \rightarrow 0$ limit cannot be taken before the integration in \eqref{4ptC2d}. Indeed, the integrand exhibits a quadratic pole at $z = 0$ and therefore the integral diverges without the regulator.

\subsubsection[\texorpdfstring{${[22 \contraction[0.5ex]{}{3}{}{3} 33]}$}{[22DD]}]{Amplitude $\bs{\ino_{[22 \contraction[0.5ex]{}{3}{}{3} 33]}}$} \label{sec:i22DD}

Next, let us consider another interesting amplitude, $\ireg_{[22 \contraction[0.5ex]{}{3}{}{3} 33]}$. This amplitude is finite and it turns out to be scheme-independent, as its finite piece does not depend on $u$ or $v_j$. In any regularization scheme we find
\begin{align} \label{i22DD}
\iuv_{[22 \contraction[0.5ex]{}{3}{}{3} 33]} & = - s^2 \left[ \frac{k_3 k_4}{k_T^3} + \frac{k_3 + k_4}{2 k_T^2} + \frac{1}{2 k_T} \right] \nn\\
& \qquad + \frac{k_3 k_4 (k_3 + k_4)^2}{k_T^3} + \frac{(k_3 + k_4)(k_3^2 + k_4^2)}{2 k_T^2} + \frac{k_3^2 + k_4^2}{2 k_T} + O(\ep).
\end{align}
In this particular case we can also take $\ep \rightarrow 0$ in the integrand and the resulting integral converges. The finiteness of the amplitude does not, however, imply that the \emph{renormalized} amplitude is scheme-independent. The existence of local terms does not depend on the details of the bulk interactions, but only on the dimensions of the sources and operators in the boundary. We should consider local terms
\begin{align}
S_{\text{loc}} = \int \D^3 \bs{x} \left[ \mathfrak{a}_1 \O_1 \phi_2 \phi_3 \phi_4 + \mathfrak{a}_2 \phi_1 \O_2 \phi_3 \phi_4 \right],
\end{align}
which lead to local contribution $\mathfrak{a}_1 k_1 + \mathfrak{a}_2 k_2$. The renormalized amplitude thus reads
\begin{align}
\iren_{[22 \contraction[0.5ex]{}{3}{}{3} 33]} = \ifin_{[22 \contraction[0.5ex]{}{3}{}{3} 33]} + \mathfrak{a}_1 k_1 + \mathfrak{a}_2 k_2,
\end{align}
where $\ifin_{[22 \contraction[0.5ex]{}{3}{}{3} 33]}$ denotes the finite, explicitly written right hand side of \eqref{i22DD}. Just as before, the renormalization constants $\act_1$ and $\act_2$ can be selected independently, thus breaking the $k_1 \leftrightarrow k_2$ symmetry of the regulated amplitude. Again, this is a valid result and it can represent the 4-point function involving two different operators $\O_1$ and $\O_2$, both of dimension $2$.

\subsubsection[\texorpdfstring{${[23 \contraction[0.5ex]{}{2}{}{2} 22]}$}{[23dd]}]{Amplitude $\bs{\ino_{[23 \contraction[0.5ex]{}{2}{}{2} 22]}}$}

Let us point out that not all derivative 4-point contact amplitudes are finite. For example, in a general $(u, v_j)$-scheme \eqref{general} we find
\begin{align}
\iuv_{[23 \contraction[0.5ex]{}{2}{}{2} 22]} & = \frac{1}{(2 u - v_T) \ep} + \left[ - \log k_T - \gamma_E + 1 + \frac{v_1 + v_2}{2u - v_T} - s^2 \left( \frac{k_2}{k_T^3} + \frac{1}{2 k_T^2} \right) \right.\nn\\
& \qquad \left. + \, \frac{k_2 (k_3 + k_4)^2}{k_T^3} + \frac{(k_3 + k_4)(-2k_2 + k_3 + k_4)}{2 k_T^2} + \frac{k_2 - k_3 - k_4}{k_T} \right],
\end{align}
where $v_T = v_1 + v_2 + v_3 + v_4$. This amplitude does require a genuine counterterm to make it finite. With the counterterm action
\begin{align} \label{Sct23dd}
S_{\text{ct}} & = \left[ \frac{1}{(2u - v_T)\ep} + 1 + \frac{v_1 + v_2}{2u - v_T} - \gamma_E - \mathfrak{a}_{[23 \contraction[0.5ex]{}{2}{}{2} 22]}^{(1)} + O(\ep) \right] \times\nn\\
& \qquad\qquad \times \int \D^{3 + 2 u \ep} \bs{x} \, \phi_1 \phi_2 \phi_3 \phi_4 \, \mu^{-(2u - v_T) \ep}
\end{align}
the renormalized amplitude becomes
\begin{align}
\iren_{[23 \contraction[0.5ex]{}{2}{}{2} 22]} & =  - \log \left( \frac{k_T}{\mu} \right) + \mathfrak{a}_{[23 \contraction[0.5ex]{}{2}{}{2} 22]}^{(1)} - s^2 \left( \frac{k_2}{k_T^3} + \frac{1}{2 k_T^2} \right)\nn\\
& \qquad + \frac{k_2 (k_3 + k_4)^2}{k_T^3} + \frac{(k_3 + k_4)(-2k_2 + k_3 + k_4)}{2 k_T^2} + \frac{k_2 - k_3 - k_4}{k_T}.
\end{align}
The counterterm constant can always be chosen in such a way that the entire contact contribution to the renormalized amplitude remains adjustable by $\mathfrak{a}_{[23 \contraction[0.5ex]{}{2}{}{2} 22]}^{(1)}$. This is the reason for the specific choice of the term in the brackets in the counterterm action \eqref{Sct23dd} to make the renormalized amplitude as simple as possible.

\subsubsection[\texorpdfstring{${[\contraction[0.5ex]{}{2}{}{2} 22 \contraction[0.5ex]{}{2}{}{2} 22]}$ and ${[\contraction[0.5ex]{}{3}{}{3} 33 \contraction[0.5ex]{}{3}{}{3} 33]}$}{[dddd] and [DDDD]}]{Amplitudes $\bs{\ino_{[\contraction[0.5ex]{}{2}{}{2} 22 \contraction[0.5ex]{}{2}{}{2} 22]}}$ and $\bs{\ino_{[\contraction[0.5ex]{}{3}{}{3} 33 \contraction[0.5ex]{}{3}{}{3} 33]}}$} 

Finally, let us present examples of a 4-derivative contact diagrams. We find
\begin{align}
\iuv_{[\contraction[0.5ex]{}{2}{}{2} 22 \contraction[0.5ex]{}{2}{}{2} 22]} & = \frac{6 s^4}{k_T^5} + s^2 \left[ \frac{12 \tau}{k_T^5} - \frac{5}{k_T^3} \right] + \frac{6 \tau^2}{k_T^5} - \frac{3 \tau}{k_T^3} + \frac{1}{k_T} + O(\ep)
\end{align}
and
\begin{align}
\iuv_{[\contraction[0.5ex]{}{3}{}{3} 33 \contraction[0.5ex]{}{3}{}{3} 33]} & = s^4 \left[ \frac{6 \sigma_4}{k_T^5} + \frac{3 \sigma_3}{2 k_T^4} + \frac{\sigma_2}{2 k_T^3} + \frac{1}{2 k_T} \right] + \nn\\
& \qquad + s^2 \left[ \frac{12 \tau \sigma_4}{k_T^5} + \frac{3 \tau \sigma_3}{k_T^4} - \frac{\tau^2 - 2 \tau \sigma_2 + \sigma_4}{k_T^3} - \frac{3 \sigma_3}{2 k_T^2} + \frac{\sigma_2}{2 k_T} - \frac{1}{2} k_T \right] \nn\\
& \qquad + \frac{6 \tau^2 \sigma_4}{k_T^5} + \frac{3 \tau^2 \sigma_3}{2 k_T^4} + \frac{ -\tfrac{1}{2} \tau^2 \sigma_2 + \tau \sigma_2^2 + \sigma_3^2 - 3 \tau \sigma_4}{k_T^3} \nn\\
& \qquad\qquad\qquad - \frac{\sigma_2 \sigma_3}{k_T^2} + \frac{- \tfrac{1}{2} \tau^2 + \tau \sigma_2 + 2 \sigma_4}{k_T} + \sigma_3 + O(\ep).
\end{align}
Both amplitudes are scheme-independent. Nevertheless, as for the 2-derivative amplitude $\ireg_{[\contraction[0.5ex]{}{3}{}{3} 33  33]}$ or the non-derivative amplitude $\ireg_{[3333]}$ a finite local contribution $\sum_{j=1} \act_j k_j^3$ can be added to the renormalized amplitude $\iren_{[\contraction[0.5ex]{}{3}{}{3} 33 \contraction[0.5ex]{}{3}{}{3} 33]}$. Such a contribution follows from the local action \eqref{Sloc3333}, which can always be included.

\section{4-point exchange amplitudes} \label{sec:4ptX}

In this section we will discuss derivative 4-point exchange amplitudes. All such amplitudes reduce to combinations of non-derivative 4-point exchange amplitudes plus derivative and non-derivative contact 4-point amplitudes. The precise formulas are listed in Section \ref{sec:identities_4ptX}. This may suggest that they are less interesting as their properties stem from the analysis of the simpler amplitudes. Nevertheless, in many cases interesting cancellations described in Section \ref{sec:cancellations} occur, making the 4-point exchange derivative amplitudes interesting to discuss.

Let us recall that we only deal here with amplitudes in $d = 3$ spacetime dimensions involving operators of dimensions $\Delta_j = 2$ or $3$ for all external legs $j=1,2,3,4$ as well as the internal lines denoted by $j = x$.

\subsection{Divergences and analytical structure}

\begin{table}[ht]
\centering
\begin{tabular}{||c||c|c||c|c||c|c||}
\hline \hline
& \multicolumn{2}{|c||}{Contact} & \multicolumn{2}{|c||}{$\Delta_x = 2$} & \multicolumn{2}{|c||}{$\Delta_x = 3$} \\ \hline
Amplitude & Div & Tran & Div & Tran & Div & Tran \\ \hline \hline
$\ireg_{[\contraction[0.5ex]{}{2}{}{2} 22;22x\Delta_x]}$ & \indep & rational & \indep & $\Li_2$ & \indep & $\Li_2$ \\ \hline
$\ireg_{[\contraction[0.5ex]{}{2}{}{2} 22;32x\Delta_x]}$ & $\one$ & $\log$ & $\two$ & $\Li_2$ & $\one$ & $\Li_2$ \\ \hline
$\ireg_{[\contraction[0.5ex]{}{2}{}{2} 22;33x\Delta_x]}$ & $\one$ & $\log$ & $\one$ & $\Li_2$ & $\two$ & $\Li_2$ \\ \hline \hline
$\ireg_{[\contraction[0.5ex]{}{3}{}{2} 32;22x\Delta_x]}$ & \dep & rational & $\one$ & $\log$ & \dep & $\Li_2$ \\ \hline
$\ireg_{[\contraction[0.5ex]{}{3}{}{2} 32;32x\Delta_x]}$ & \dep & rational & $\one$ & $\log$ & \dep & $\Li_2$ \\ \hline
$\ireg_{[\contraction[0.5ex]{}{3}{}{2} 32;33x\Delta_x]}$ & $\one$ & $\log$ & $\one$ & $\log$ & $\two$ & $\Li_2$ \\ \hline \hline
$\ireg_{[\contraction[0.5ex]{}{3}{}{3} 33;22x\Delta_x]}$ & \indep & rational & \indep & $\Li_2$ & \dep & $\log$ \\ \hline
$\ireg_{[\contraction[0.5ex]{}{3}{}{3} 33;32x\Delta_x]}$ & $\one$ & $\log$ & $\two$ & $\Li_2$ & $\one$ & $\log$ \\ \hline
$\ireg_{[\contraction[0.5ex]{}{3}{}{3} 33;33x\Delta_x]}$ & \dep & rational & \dep & $\Li_2$ & $\one$ & $\log$ \\ \hline \hline
\end{tabular}
\caption{Degrees of divergence and transcendence of 4-derivative 4-point AdS amplitudes. The `Wick contractions' indicate on which propagators the derivatives act and how they are contracted. Degree of transcendence indicates whether the function is rational in momenta (label `rational'), contains logarithms (`log'), or dilogarithms ($\Li_2$). For finite amplitudes ($\text{Div} = 0$) we indicate by `dep' if the finite part is scheme-dependent (\textit{i.e.}, $u$ and $v_j$-dependent) and by `indep' if it is scheme-independent.\label{fig:divs4pt2d}}
\end{table}

\begin{table}[ht]
\centering
\begin{tabular}{||c||c|c||c|c||c|c||}
\hline \hline
& \multicolumn{2}{|c||}{Contact} & \multicolumn{2}{|c||}{$\Delta_x = 2$} & \multicolumn{2}{|c||}{$\Delta_x = 3$} \\ \hline
Amplitude & Div & Tran & Div & Tran & Div & Tran \\ \hline \hline
$\ireg_{[\contraction[0.5ex]{}{2}{}{2} 22; \contraction[0.5ex]{}{2}{}{2} 22x\Delta_x]}$ & \indep & rational & \indep & $\Li_2$ & \indep & $\Li_2$ \\ \hline
$\ireg_{[\contraction[0.5ex]{}{3}{}{2} 32; \contraction[0.5ex]{}{2}{}{2} 22x\Delta_x]}$ & \dep & rational & $\one$ & $\log$ & \dep & $\Li_2$ \\ \hline
$\ireg_{[\contraction[0.5ex]{}{3}{}{3} 33; \contraction[0.5ex]{}{2}{}{2} 22x\Delta_x]}$ & \indep & rational & \indep & $\Li_2$ & \dep & $\log$ \\ \hline
$\ireg_{[\contraction[0.5ex]{}{3}{}{2} 32; \contraction[0.5ex]{}{3}{}{2} 32x\Delta_x]}$ & \indep & rational & \dep & rational & \indep & $\Li_2$ \\ \hline
$\ireg_{[\contraction[0.5ex]{}{3}{}{3} 33; \contraction[0.5ex]{}{3}{}{2} 32x\Delta_x]}$ & \dep & rational & $\one$ & $\log$ & \dep & $\log$ \\ \hline
$\ireg_{[\contraction[0.5ex]{}{3}{}{3} 33; \contraction[0.5ex]{}{3}{}{3} 33x\Delta_x]}$ & \indep & rational & \indep & $\Li_2$ & \indep & rational \\ \hline \hline
\end{tabular}
\caption{Degrees of divergence and transcendence of 4-derivative 4-point AdS amplitudes. The `Wick contractions' indicate on which propagators the derivatives act and how they are contracted. Degree of transcendence indicates whether the function is rational in momenta (label `rational'), contains logarithms (`log'), or dilogarithms (`$\Li_2$'). For finite amplitudes ($\text{Div} = 0$) we indicate by `dep' if the finite part is scheme-dependent (\textit{i.e.}, $u$ and $v_j$-dependent) and by `indep' if it is scheme-independent.\label{fig:divs4pt4d}}
\end{table}

\begin{table}[ht]
\centering
\begin{tabular}{||c||c|c||c|c||c|c||}
\hline \hline
& \multicolumn{2}{|c||}{Contact} & \multicolumn{2}{|c||}{$\Delta_x = 2$} & \multicolumn{2}{|c||}{$\Delta_x = 3$} \\ \hline
Amplitude & Div & Tran & Div & Tran & Div & Tran \\ \hline \hline
$\ireg_{[22;22x\Delta_x]}$ & \indep & rational & \indep & $\Li_2$ & \indep & $\Li_2$ \\ \hline
$\ireg_{[32;22x\Delta_x]}$ & $\one$ & $\log$ & $\two$ & $\Li_2$ & $\one$ & $\Li_2$ \\ \hline
$\ireg_{[33;22x\Delta_x]}$ & $\one$ & $\log$ & $\one$ & $\Li_2$ & $\two$ & $\Li_2$ \\ \hline
$\ireg_{[32;32x\Delta_x]}$ & $\one$ & $\log$ & $\two$ & $\Li_2$ & $\one$ & $\Li_2$ \\ \hline
$\ireg_{[33;32x\Delta_x]}$ & $\one$ & $\log$ & $\two$ & $\Li_2$ & $\two$ & $\Li_2$ \\ \hline
$\ireg_{[33;33x\Delta_x]}$ & $\one$ & $\log$ & $\one$ & $\Li_2$ & $\two$ & $\Li_2$ \\ \hline \hline
\end{tabular}
\caption{Degrees of divergence and transcendence of non-derivative 4-point AdS amplitudes for comparison with Tables \ref{fig:divs4pt2d} and \ref{fig:divs4pt4d}.\label{fig:divs4pt0d}}
\end{table}

In Section \ref{sec:divs_4ptC} we discussed divergences and analytical structure of the 4-point contact amplitudes. In these cases the analytical structure followed from the degree of divergence: finite contact amplitudes were rational, while divergent ones contained logarithms. For the derivative exchange amplitudes the situation is more interesting: the analytical structure does not have to correspond to the divergences. Whether finite or not, three different analytical structures may occur in the exchange amplitudes: they can be rational functions, contain logarithms, or dilogarithms. For example, a finite non-derivative amplitude $\ino_{[22;22x2]}$ is given by the well-known formula, see \textit{e.g.}, \cite{Boos:1987bg,Davydychev:1992xr,tHooft:1978jhc,Bzowski:2015yxv},
\begin{empheq}[box=\nicebox]{align} \label{ifin2222x2}
\ifin_{[22,22x2]} & = - \frac{1}{2 s} \left[ \Li_2 \left( \frac{\m{34}}{k_T} \right) + \Li_2 \left( \frac{\m{12}}{k_T} \right) + \log \left( \frac{\p{12}}{k_T} \right) \log \left( \frac{\p{34}}{k_T} \right) - \frac{\pi^2}{6} \right].
\end{empheq}
As long as non-derivative amplitudes are concerned, they all contain the dilogarithms in combinations presented above. On the other hand the situation is more involved for derivative amplitudes. Due to the competition between various terms in equations of Section \ref{sec:identities_4ptX}, the derivative amplitudes may lose their dilogarithms if the prefactors are suitable. Thus, we may encounter exchange 4-point amplitudes without dilogarithms. On occasion the amplitudes may even become rational functions of momenta. 

The degree of divergences and the analytic structures are presented in Tables \ref{fig:divs4pt2d} and \ref{fig:divs4pt4d}. In Table \ref{fig:divs4pt0d} we also present divergences and the analytic structures of non-derivative amplitudes for comparison. We also include the contact diagrams in the tables for completeness. 

In the following subsection we will present a few examples of the contact and exchange amplitudes. The complete list is available in the \verb|HandbooK| Mathematica package attached to this paper. In Section \ref{sec:package} I will show how to access all the amplitudes stored in the package.

\subsection{Examples}

\subsubsection[\texorpdfstring{${[22; \contraction[0.5ex]{}{2}{}{2} 22x2]}$}{[22;ddx2]}]{Amplitude $\bs{\ino_{[22; \contraction[0.5ex]{}{2}{}{2} 22x2]}}$}

The simplest derivative amplitude reads
\begin{align}
\ireg_{[22; \contraction[0.5ex]{}{2}{}{2} 22x2]} & = \ifin_{[22;22x2]} - \frac{1}{2 k_T} + O(\ep).
\end{align}
The first term is the non-derivative amplitude given by \eqref{ifin2222x2}. It encodes the structure of dilogarithms appearing in a typical 4-point exchange amplitude. The second term in the expression above is proportional to the contact 4-point function, as can be seen in equation \eqref{redi2db}.

\subsubsection[\texorpdfstring{${[22; \contraction[0.5ex]{}{3}{}{3} 33x2]}$}{[22;DDx2]}]{Amplitude $\bs{\ino_{[22; \contraction[0.5ex]{}{3}{}{3} 33x2]}}$}

Let us now have a look at two simple amplitudes involving the derivative coupling of the massless field, $\ireg_{[22; \contraction[0.5ex]{}{3}{}{3} 33x2]}$ and $\ireg_{[22; \contraction[0.5ex]{}{3}{}{3} 33x3]}$. The first of these reads
\begin{align} \label{i22DDx2}
\ireg_{[22; \contraction[0.5ex]{}{3}{}{3} 33x2]} & = \frac{1}{2} ( -s^2 + k_3^2 + k_4^2 ) \, \ifin_{[22;22x2]} - \frac{1}{2} (k_3 + k_4) \log \left( \frac{k_1 + k_2 + s}{k_T} \right) - \frac{k_3 k_4}{k_T} + O(\ep).
\end{align}
As we can see the amplitude is finite and contains dilogarithms through $\ifin_{[22;22x2]}$ given in \eqref{ifin2222x2}. Despite its finiteness, local terms can be added to the amplitude by means of local terms added to the action. Consider the local action
\begin{align}
S_{\text{loc}} = \int \D^3 \bs{x} \, \left[ \mathfrak{a}_1 \O_1 \phi_2 \phi_3 \phi_4 + \mathfrak{a}_2 \phi_1 \O_2 \phi_3 \phi_4 \right],
\end{align}
which leads to the contribution $\mathfrak{a}_1 k_1 + \mathfrak{a}_2 k_2$. Thus, the renormalized amplitude reads
\begin{align}
\iren_{[22; \contraction[0.5ex]{}{3}{}{3} 33x2]} = \ifin_{[22; \contraction[0.5ex]{}{3}{}{3} 33x2]} + \mathfrak{a}_1 k_1 + \mathfrak{a}_2 k_2,
\end{align}
where $\ifin_{[22; \contraction[0.5ex]{}{3}{}{3} 33x2]}$ denotes the finite, explicitly written terms in \eqref{i22DDx2}.

\subsubsection[\texorpdfstring{${[22; \contraction[0.5ex]{}{3}{}{3} 33x3]}$}{[22;DDx3]}]{Amplitude $\bs{\ino_{[22; \contraction[0.5ex]{}{3}{}{3} 33x3]}}$} \label{sec:i22DDx3}

Let us now compare the amplitude $\ireg_{[22; \contraction[0.5ex]{}{3}{}{3} 33x3]}$ to $\ireg_{[22; \contraction[0.5ex]{}{3}{}{3} 33x2]}$ discussed above. It reads
\begin{align}
\ireg_{[22; \contraction[0.5ex]{}{3}{}{3} 33x3]} & = \frac{1}{2} (k_1 + k_2) \left[ \log \left( \frac{k_1 + k_2 + s}{k_T} \right) - \frac{1}{3} \right] - \frac{k_3 k_4}{2 k_T} + \frac{1}{2} (-s + k_3 + k_4) + O(\ep).
\end{align}
As we can see the amplitude is finite, but it does not contain dilogarithms at all; it only contain logarithms and it remains scale-invariant. 

As far as renormalization is concerned, local terms of the form $\mathfrak{a}_1 k_1 + \mathfrak{a}_2 k_2$ can be added to the action here as well. In this particular case, however, additional local term exists. Indeed, renormalization of the non-derivative amplitude $\ireg_{[3\contraction[0.5ex]{}{3}{}{3} 33]}$ in Section \ref{sec:3pt_renorm} resulted in the renormalized 3-point amplitude \eqref{iren3DD} with the value of the constants $\mathfrak{s}_{[3\contraction[0.5ex]{}{3}{}{3} 33], j}$ given by \eqref{s3deriv}. Thus, here we must consider the contribution from the following terms,
\begin{align}
S_{\text{loc}} & = \mu^{-\ep} \left( \frac{1}{2} - \act_{[3\contraction[0.5ex]{}{3}{}{3} 33], x}(\ep) \right) \int \D^{3 + 2 \ep} \bs{x} \, \O_x \phi_3 \phi_4 + \nn\\
& \qquad + \mu^{-2 \ep} \int \D^{3 + 2 \ep} \bs{x} \, \left[ \mathfrak{a}_1(\ep) \O_1 \phi_2 \phi_3 \phi_4 + \mathfrak{a}_2(\ep) \phi_1 \O_2 \phi_3 \phi_4 \right].
\end{align}
Here $\act_{[3\contraction[0.5ex]{}{3}{}{3} 33], x}(\ep) = \act_{[3\contraction[0.5ex]{}{3}{}{3} 33], x}^{(1)} + \ep \, \act_{[3\contraction[0.5ex]{}{3}{}{3} 33], x}^{(2)} + O(\ep^2)$ and we can think of $\act_{[3\contraction[0.5ex]{}{3}{}{3} 33], x}^{(1)}$ as already fixed by renormalization of $\iren_{[3\contraction[0.5ex]{}{3}{}{3} 33]}$. The $\ep$-dependence of the remaining renormalization constants $\mathfrak{a}_1$ and $\mathfrak{a}_2$ is yet to be determined. 

The contribution from the counterterm $S_{\text{loc}}$ to the 4-point function reads
\begin{align}
- \mu^{-\ep} \left( \frac{1}{2} - \act_{[3\contraction[0.5ex]{}{3}{}{3} 33], x}(\ep) \right) \ireg_{[223]}(k_1, k_2, s) + \mu^{-2 \ep} \left( \mathfrak{a}_1(\ep) \ireg_{[22]}(k_1) + \mathfrak{a}_2(\ep) \ireg_{[22]}(k_2) \right).
\end{align}
Since the regulated amplitude $\ireg_{[223]}$ is linearly divergent, the first term exhibits a $1/\ep$ singularity. Thus, to cancel the divergence, now we have to choose $\mathfrak{a}_1$ and $\mathfrak{a}_2$ to be divergent as well. One finds that we have to choose
\begin{align}
& \mathfrak{a}_{1,2} = \frac{\act_{[3\contraction[0.5ex]{}{3}{}{3} 33], x}^{(1)} - \frac{1}{2}}{\ep} + \mathfrak{a}_{1,2}^{(0)} + O(\ep)
\end{align}
for undetermined constants $\mathfrak{a}_{1}^{(0)}$ and $\mathfrak{a}_{2}^{(0)}$. Thus, the renormalized amplitude $\iren_{[22; \contraction[0.5ex]{}{3}{}{3} 33x3]}$ reads
\begin{align}
\iren_{[22; \contraction[0.5ex]{}{3}{}{3} 33x3]} & = - \frac{k_3 k_4}{2 k_T} + \frac{1}{2} (k_3 + k_4) - \frac{1}{2} (k_1 + k_2) \log \left( \frac{k_T}{\mu} \right) \nn\\
& \qquad + \act_{[3\contraction[0.5ex]{}{3}{}{3} 33], x}^{(1)} (k_1 + k_2) \log \left( \frac{k_1 + k_2 + s}{\mu} \right) - \act_{[3\contraction[0.5ex]{}{3}{}{3} 33], x}^{(1)} \, s \nn\\
& \qquad + (k_1 + k_2) \left[ \frac{1}{3} + \gamma_E \left( \act_{[3\contraction[0.5ex]{}{3}{}{3} 33], x}^{(1)} - \frac{1}{2} \right) - \act_{[3\contraction[0.5ex]{}{3}{}{3} 33], x}^{(1)} - \act_{[3\contraction[0.5ex]{}{3}{}{3} 33], x}^{(2)} \right] \nn\\
& \qquad + \mathfrak{a}_{1}^{(0)} k_1 + \mathfrak{a}_{2}^{(0)} k_2.
\end{align}
The amplitude depends on the renormalization scale $\mu$ and 4 scheme-dependent constants, $\act_{[3\contraction[0.5ex]{}{3}{}{3} 33], x}^{(1)}$, $\act_{[3\contraction[0.5ex]{}{3}{}{3} 33], x}^{(2)}$ and $\mathfrak{a}_1^{(0)}$, $\mathfrak{a}_2^{(0)}$. Given the values of $\act_{[3\contraction[0.5ex]{}{3}{}{3} 33], x}^{(1)}$ and $\act_{[3\contraction[0.5ex]{}{3}{}{3} 33], x}^{(2)}$ one can always choose $\mathfrak{a}_{1}^{(0)}$ and $\mathfrak{a}_{2}^{(0)}$ in such a way that the two last lines cancel each other. The remainder of the 4-point exchange amplitude depends on the renormalization of $\ireg_{[3\contraction[0.5ex]{}{3}{}{3} 33]}$. There are two natural choices to consider. If we set $\act_{[3\contraction[0.5ex]{}{3}{}{3} 33], x}^{(1)} = 0$ as we did in \eqref{ren3_zero}, the 3-point function vanishes entirely and we obtain
\begin{align}
\lla \O_1 \O_2 \O_x \rra & \doteq \act_{[3\contraction[0.5ex]{}{3}{}{3} 33], 1}^{(1)} k_1^3 + \act_{[3\contraction[0.5ex]{}{3}{}{3} 33], 2}^{(1)} k_2^3 + 0 \times s^3, \\
\iren_{[22; \contraction[0.5ex]{}{3}{}{3} 33x3]} & \doteq - \frac{k_3 k_4}{2 k_T} + \frac{1}{2} (k_3 + k_4) - \frac{1}{2} (k_1 + k_2) \log \left( \frac{k_T}{\mu} \right). \label{iren22DDx3s1}
\end{align}
The fact that those expressions are valid only in a special renormalization scheme is denoted by the dot over the equal sign. The 4-point amplitude becomes scale-violating due to the presence of the renormalization scale $\mu$, but it does not depend on the Mandelstam variable $s$. 

Another interesting choice is $\act_{[3\contraction[0.5ex]{}{3}{}{3} 33], x}^{(1)} = \frac{1}{2}$. In such a case the $s^3$ term in the 3-point amplitude does not vanish, but we find
\begin{align}
\lla \O_1 \O_2 \O_x \rra & \doteq \act_{[3\contraction[0.5ex]{}{3}{}{3} 33], 1}^{(1)} k_1^3 + \act_{[3\contraction[0.5ex]{}{3}{}{3} 33], 2}^{(1)} k_2^3 + \frac{1}{2} s^3, \\
\iren_{[22; \contraction[0.5ex]{}{3}{}{3} 33x3]} & \doteq - \frac{k_3 k_4}{2 k_T} + \frac{1}{2} (k_3 + k_4) + \frac{1}{2} (k_1 + k_2) \log \left( \frac{k_1 + k_2 + s}{k_T} \right) - \frac{s}{2}. \label{iren22DDx3s2}
\end{align}
The 4-point amplitude now is completely scale-invariant, but becomes $s$-dependent.

This is a new behavior that, to the best of my knowledge, was not observed for renormalized correlators before. In one choice of scheme in \eqref{iren22DDx3s1} the amplitude is scale-violating, as it depends on the renormalization scale $\mu$, but remains $s$-independent. With another choice of scheme in \eqref{iren22DDx3s2} the amplitude becomes scale-invariant but it does depend on the Mandelstam variable $s$.

\subsubsection[\texorpdfstring{${[33; \contraction[0.5ex]{}{3}{}{3} 33x2]}$}{[33;DDx2]}]{Amplitude $\bs{\ino_{[33; \contraction[0.5ex]{}{3}{}{3} 33x2]}}$}

Next, let us analyze two more 2-derivative amplitudes involving the derivatives acting on the massless scalar. The regulated amplitude $\ireg_{[33; \contraction[0.5ex]{}{3}{}{3} 33x2]}$ is finite, but keeps the dilogarithmic terms,
\begin{align}
\ireg_{[33; \contraction[0.5ex]{}{3}{}{3} 33x2]} & = \frac{1}{4} ( -s^2 + k_1^2 + k_2^2 )( -s^2 + k_3^2 + k_4^2 ) \, \ireg_{[22;22x2]} \nn\\
& \qquad + \frac{1}{4} (-s^2 + k_1^2 + k_2^2) (k_3 + k_4) \log \left( \frac{k_1 + k_2 + s}{k_T} \right) \nn\\
& \qquad\qquad\qquad + \frac{1}{4} (-s^2 + k_3^2 + k_4^2) (k_1 + k_2) \log \left( \frac{k_3 + k_4 + s}{k_T} \right) \nn\\
& \qquad - \frac{1}{4} s^2 k_T + \frac{1}{4} s \tau - \frac{\sigma_4}{2 k_T}.
\end{align}
The renormalization on this amplitude is not particularly exciting. The local term \eqref{Sloc3333} can be added to the action leading to scheme-dependent terms proportional to $k_j^3$, $j=1,2,3,4$. We obtain
\begin{align}
\iren_{[33; \contraction[0.5ex]{}{3}{}{3} 33x2]} = \ifin_{[33; \contraction[0.5ex]{}{3}{}{3} 33x2]} + \sum_{j=1}^4 \mathfrak{a}_j k_j^3.
\end{align}
where $\mathfrak{a}_j$ are the scheme-dependent constants.

\subsubsection[\texorpdfstring{${[33; \contraction[0.5ex]{}{3}{}{3} 33x3]}$}{[33;DDx3]}]{Amplitude $\bs{\ino_{[33; \contraction[0.5ex]{}{3}{}{3} 33x3]}}$}

With comparison to its sister amplitude, $\ireg_{[33; \contraction[0.5ex]{}{3}{}{3} 33x2]}$, this amplitude does not contain any dilogarithms, but remains linearly divergent,
\begin{align} \label{i33DDx3}
\ireg_{[33; \contraction[0.5ex]{}{3}{}{3} 33x3]} & = \frac{s^3}{6} \left[ \frac{1}{\ep} - 2 \log k_T - 2 \gamma_E + 3 \right] \nn\\
& \qquad - \frac{1}{6} (s^3 + k_1^3 + k_2^3) \log \left( \frac{k_1 + k_2 + s}{k_T} \right) - \frac{1}{6} (s^3 + k_3^3 + k_4^3) \log \left( \frac{k_3 + k_4 + s}{k_T} \right) \nn\\
& \qquad + \frac{1}{6} s^2 k_T + \frac{1}{6} s ( \tau - 3 \sigma_2 + k_T^2 ) - \frac{\sigma_4}{2 k_T} + \left( \frac{2}{3} \sigma_3 - \frac{1}{6} k_T ( \tau + \sigma_2 ) + \frac{1}{18} k_T^3 \right).
\end{align}
When it comes to renormalization, here we also have to consider contributions from the renormalization of the 3-point amplitudes $\ireg_{[333]}$ and $\ireg_{[\contraction[0.5ex]{}{3}{}{3} 333]}$. In total, the most general local contribution we have we consider comes from the action,
\begin{align} \label{Sct33DDx3}
S_{\text{ct}} & = \mu^{-\ep} \int \D^{3 + 2 \ep} \bs{x} \left[ \frac{1}{3} \Gamma(\ep) \mathfrak{a}_{[333], x} \phi_1 \phi_2 \O_x + \left( \frac{1}{2} - \act_{[3\contraction[0.5ex]{}{3}{}{2} 33], x} \right) \phi_3 \phi_4 \O_x \right] \nn\\
& \qquad + \mu^{-2 \ep}\int \D^{3 + 2 \ep} \bs{x} \left[ \mathfrak{a}_1 \O_1 \phi_2 \phi_3 \phi_4 + \mathfrak{a}_2 \phi_1 \O_2 \phi_3 \phi_4 + \mathfrak{a}_3 \phi_1 \phi_2 \O_3 \phi_4 + \mathfrak{a}_4 \phi_1 \phi_2 \phi_3 \O_4 \right].
\end{align}
The constants $\mathfrak{a}_{[333], x}$ and $\act_{[3\contraction[0.5ex]{}{3}{}{2} 33], x}$ are partially fixed by the renormalization of 3-point amplitudes $\ireg_{[333]}$ and $\ireg_{[\contraction[0.5ex]{}{3}{}{3} 333]}$ respectively. By calculating the contribution from the counterterm action to the 4-point amplitude the finiteness of the renormalized amplitude now requires that
\begin{align}
& \mathfrak{a}_{1,2} = \frac{1}{3 \ep} \left( \frac{1}{2} - \act_{[3\contraction[0.5ex]{}{3}{}{2} 33], x}^{(1)} \right) + \mathfrak{a}_{1,2}^{(0)} + O(\ep), && \mathfrak{a}_{3,4} = \frac{1}{6 \ep} + \mathfrak{a}_{3,4}^{(0)} + O(\ep).
\end{align}
This yields the finite renormalized amplitude
\begin{align}
\iren_{[33; \contraction[0.5ex]{}{3}{}{3} 33x3]} & = \frac{1}{6} (k_1^3 + k_2^3 + k_3^3 + k_4^3) \log \left( \frac{k_T}{\mu} \right) - \frac{\sigma_4}{2 k_T} + \frac{2 \sigma_3}{3} - \frac{k_T \sigma_2}{6} \nn\\
& \qquad - \frac{1}{3} \act_{[3\contraction[0.5ex]{}{3}{}{2} 33], x}^{(1)} \left\{ (s^3 + k_1^3 + k_2^3) \left[ \log \left( \frac{s+k_1+k_2}{\mu} \right) + \gamma_E - \frac{4}{3} \right] \right.\nn\\
& \qquad\qquad\qquad \left. - \, \s{1}{12s} \s{2}{12s} + 4 \s{3}{12s} + ( \act_{[333], x}^{(1)} - \gamma_E ) \, s^3 \right\} \nn\\
& \qquad + \frac{1}{6} (k_1^3 + k_2^3) \left( \gamma_E - 1 + 2 \act_{[3\contraction[0.5ex]{}{3}{}{2} 33], x}^{(2)} \right) + (k_3^3 + k_4^3) \left( - \frac{2}{9} + \frac{1}{3} \gamma_E - \frac{1}{6} \act_{[333]}^{(1)} \right) \nn\\
& \qquad + \act_1^{(0)} k_1^3 + \act_2^{(0)} k_2^3 + \act_3^{(0)} k_3^3 + \act_4^{(0)} k_4^3.
\end{align}
As we can see only the terms present in the first line are scheme-independent. Just as before, we can clearly arrange the constants $\act_j^{(0)}$ for $j=1,2,3,4$ in such a way that the last two lines cancel. Furthermore, with the 3-point renormalization constant $\act_{[3\contraction[0.5ex]{}{3}{}{2} 33], x}^{(1)}$ vanishing, the 4-point amplitude simplifies to
\begin{align}
\iren_{[33; \contraction[0.5ex]{}{3}{}{3} 33x3]} & \doteq \frac{1}{6} (k_1^3 + k_2^3 + k_3^3 + k_4^3) \log \left( \frac{k_T}{\mu} \right) - \frac{\sigma_4}{2 k_T} + \frac{2 \sigma_3}{3} - \frac{k_T \sigma_2}{6}.
\end{align}
The dot over the equal sign indicates that a special renormalization scheme was selected to obtain this result.

As we can see the renormalized amplitude is much simpler than what the regulated amplitude \eqref{i33DDx3} suggests. The terms containing logarithms $\log [ (k_1 + k_2 + s)/ k_T ]$ and $\log [ (k_3 + k_4 + s) / k_T ]$ as well as $s^3$ can be completely removed from the renormalized amplitude.

\subsubsection[\texorpdfstring{${[\contraction[0.5ex]{}{3}{}{3} 33; \contraction[0.5ex]{}{3}{}{3} 33x2]}$}{[DD;DDx2]}]{Amplitude $\bs{\ino_{[\contraction[0.5ex]{}{3}{}{3} 33; \contraction[0.5ex]{}{3}{}{3} 33x2]}}$}

Finally, let us look at two examples of the 4-derivative amplitudes. We find the following amplitude regulated in the beta scheme,
\begin{align}
\ireg_{[\contraction[0.5ex]{}{3}{}{3} 33; \contraction[0.5ex]{}{3}{}{3} 33x2]} & = \frac{1}{4} (-s^2 + k_1^2 + k_2^2) (-s^2 + k_3^2 + k_4^2) \ifin_{[22;22x2]} \nn\\
& \qquad - \frac{1}{4} (-s^2 + k_1^2 + k_2^2) (k_3 + k_4) \log \left( \frac{s + k_1 + k_2}{k_T} \right) \nn\\
& \qquad\qquad\qquad - \frac{1}{4} (-s^2 + k_3^2 + k_4^2) (k_1 + k_2) \log \left( \frac{s + k_3 + k_4}{k_T} \right) \nn\\
& \qquad + \frac{s^2}{4} \left[ \frac{2 \sigma_4}{k_T^3} + \frac{\sigma_3}{k_T^2} + \frac{\sigma_2}{k_T} \right] - \frac{1}{4} s \tau \nn\\
& \qquad + \frac{\tau \sigma_4}{2 k_T^3} + \frac{\tau \sigma_3}{4 k_T^2} + \frac{-\tau^2 + 2 \tau \sigma_2 + 4 \sigma_4}{4 k_T} - \frac{\sigma_3}{2} + O(\ep).
\end{align}
The amplitude is finite, although it does contain the dilogarithms through $\ifin_{[22;22x2]}$. Just as was the case for $\ireg_{[33; \contraction[0.5ex]{}{3}{}{3} 33x2]}$, a finite local term \eqref{Sloc3333} can be added to the action. The effect of the addition is the same as before and the renormalized 4-point amplitude reads
\begin{align}
\iren_{[\contraction[0.5ex]{}{3}{}{3} 33; \contraction[0.5ex]{}{3}{}{3} 33x2]} = \ifin_{[\contraction[0.5ex]{}{3}{}{3} 33; \contraction[0.5ex]{}{3}{}{3} 33x2]} + \sum_{j=1}^4 \mathfrak{a}_j k_j^3,
\end{align}
where $\ifin_{[\contraction[0.5ex]{}{3}{}{3} 33; \contraction[0.5ex]{}{3}{}{3} 33x2]}$ denotes the finite, explicitly written part of $\ireg_{[\contraction[0.5ex]{}{3}{}{3} 33; \contraction[0.5ex]{}{3}{}{3} 33x2]}$.

\subsubsection[\texorpdfstring{${[\contraction[0.5ex]{}{3}{}{3} 33; \contraction[0.5ex]{}{3}{}{3} 33x3]}$}{[DD;DDx3]}]{Amplitude $\bs{\ino_{[\contraction[0.5ex]{}{3}{}{3} 33; \contraction[0.5ex]{}{3}{}{3} 33x3]}}$}

Similarly to the case of $\ireg_{[33; \contraction[0.5ex]{}{3}{}{3} 33x3]}$ we can expect interesting complications for this amplitude. Its regulated version is finite and reads
\begin{align}
\ireg_{[\contraction[0.5ex]{}{3}{}{3} 33; \contraction[0.5ex]{}{3}{}{3} 33x3]} & = \frac{s^3}{4} + \frac{s^2}{4} \left[ \frac{2 \sigma_4}{k_T^3} + \frac{\sigma_3}{k_T^2} + \frac{\sigma_2}{k_T} - k_T \right] \nn\\
& \qquad + \frac{\tau \sigma_4}{2 k_T^3} + \frac{\tau \sigma_3}{4 k_T^2} + \frac{-\tau^2 + 2 \tau \sigma_2 + 2 \sigma_4}{4 k_T} - \frac{\sigma_3}{2}.
\end{align}
Notice that the first term $s^3/4$ is local, so we may expect that it is removable by suitable local terms. Indeed, the local term contributing to this amplitude has the form similar to \eqref{Sct33DDx3}, but with both renormalization constants in the first line corresponding to the derivative 3-point function,
\begin{align}
S_{\text{ct}} & = \mu^{-\ep} \int \D^{3 + 2 \ep} \bs{x} \left[ \left( \frac{1}{2} - \act_{[3\contraction[0.5ex]{}{3}{}{2} 33], x} \right) \phi_1 \phi_2 \O_x + \left( \frac{1}{2} - \act'_{[3\contraction[0.5ex]{}{3}{}{2} 33], x} \right) \phi_3 \phi_4 \O_x \right] \nn\\
& \qquad + \mu^{-2 \ep}\int \D^{3 + 2 \ep} \bs{x} \left[ \mathfrak{a}_1 \O_1 \phi_2 \phi_3 \phi_4 + \mathfrak{a}_2 \phi_1 \O_2 \phi_3 \phi_4 + \mathfrak{a}_3 \phi_1 \phi_2 \O_3 \phi_4 + \mathfrak{a}_4 \phi_1 \phi_2 \phi_3 \O_4 \right].
\end{align}
Formally, the two interaction vertices $\phi_1 \phi_2 \O_x$ and $\phi_3 \phi_4 \O_x$ can be renormalized differently. For this reason we introduced two renormalization constants, $\act_{[3\contraction[0.5ex]{}{3}{}{2} 33], x}$ and $\act'_{[3\contraction[0.5ex]{}{3}{}{2} 33], x}$. This time, however, all contributions are finite, since both renormalization constants as well as the 3-point function $\ireg_{[3\contraction[0.5ex]{}{3}{}{2} 33]}$ are finite. The renormalized amplitude becomes
\begin{align}
\iren_{[\contraction[0.5ex]{}{3}{}{3} 33; \contraction[0.5ex]{}{3}{}{3} 33x3]} & =
\frac{s^2}{4} \left[ \frac{2 \sigma_4}{k_T^3} + \frac{\sigma_3}{k_T^2} + \frac{\sigma_2}{k_T} - k_T \right] + \frac{\tau \sigma_4}{2 k_T^3} + \frac{\tau \sigma_3}{4 k_T^2} + \frac{-\tau^2 + 2 \tau \sigma_2 + 2 \sigma_4}{4 k_T} - \frac{\sigma_3}{2} \nn\\
& \qquad + \act_{[3\contraction[0.5ex]{}{3}{}{2} 33], x} \act'_{[3\contraction[0.5ex]{}{3}{}{2} 33], x} s^3 + \frac{1}{2} \left( \act'_{[3\contraction[0.5ex]{}{3}{}{2} 33], x} - \frac{1}{2} \right) (k_1^3 + k_2^3) + \frac{1}{2} \left( \act_{[3\contraction[0.5ex]{}{3}{}{2} 33], x} - \frac{1}{2} \right) (k_3^3 + k_4^3) \nn\\
& \qquad + \act_{1} k_1^3 + \act_{2} k_2^3 + \act_{3} k_3^3 + \act_{4} k_4^3.
\end{align}
Only local terms $k_j^3$ and $s^3$ can be removed from the amplitude by the choice of a scheme. The first line remains scheme-independent.

\section{Symmetric theory} \label{sec:symmetric}

In section 4.5 of \cite{Bzowski:2022rlz} we analyzed \emph{the symmetric theory} containing only two fields $\Phi_{[2]}$ and $\Phi_{[3]}$ dual to operators of dimension $2$ and $3$ respectively. We considered all possible non-derivative couplings between the fields and derived correlation functions within the theory. The theory is symmetric in the sense that its correlation functions usually exhibit discreet crossing symmetries.

In the context of derivative theories the attention is shifted towards theories containing massless bulk field $\mf$ with derivative interactions. The motivation for such theory usually stems from the effective theory of inflation, \cite{Cheung:2007st,Weinberg:2008hq}. Thus, let us consider here the symmetric theory with two fields: the massless field $\mf$ dual to the marginal operator $X_{[3]}$ and another field $\Phi_{[\Delta]}$ dual to the operator $\O_{[\Delta]}$ of dimension $\Delta$ with $\Delta = 2$ or $3$. The bulk action reads
\begin{align}
S^{\text{sym}} & = S^{\text{sym}}_{\text{free}} + S^{(3)}_{\text{sym}} + S^{(4)}_{\text{sym}},
\end{align}
where
\begin{align}
S^{\text{sym}}_{\text{free}} & = \frac{1}{2} \int \D^{4 + 2 \ep} x \, \sqrt{g} \left[ \partial_\mu \Phi_{[\Delta]} \partial^\mu \Phi_{[\Delta]} + \partial_\mu \mf \partial^\mu \mf + \reg{m}_{\Delta}^2 \Phi_{[\Delta]}^2 - \ep (3 + \ep) \mf^2 \right], \\
S^{(3)}_{\text{sym}} & = \int \D^{4 + 2 \ep} x \sqrt{g} \left[ \frac{1}{6} \lambda_{[\Delta \Delta \Delta]} \Phi_{[\Delta]}^3 + \frac{1}{2} \lambda_{[\contraction[0.5ex]{}{3}{}{3} 33 \Delta]} \Phi_{[\Delta]} \partial_\mu \mf \partial^\mu \mf \right] \nn\\
S^{(4)}_{\text{sym}} & = - \int \D^{4 + 2 \ep} x \sqrt{g} \left[ \frac{1}{24} \lambda_{[\Delta \Delta \Delta \Delta]} \Phi_{[\Delta]}^4 + \frac{1}{4} \lambda_{[\contraction[0.5ex]{}{3}{}{3} 33 \Delta \Delta]} \Phi_{[\Delta]}^2 \partial_\mu \mf \partial^\mu \mf + \frac{1}{8} \lambda_{[\contraction[0.5ex]{}{3}{}{3} 33 \contraction[0.5ex]{}{3}{}{3} 33]} ( \partial_\mu \mf \partial^\mu \mf )^2 \right].
\end{align}
Equations of motion for the symmetric theory read
\begin{align}
& (-\Box_{AdS} + \reg{m}^2_{\Delta}) \Phi_{[\Delta]} = - \frac{1}{2} \lambda_{[\Delta \Delta \Delta]} \Phi_{[\Delta]}^2 - \frac{1}{2} \lambda_{[\contraction[0.5ex]{}{3}{}{3} 33 \Delta]} \partial_\mu \chi \partial^\mu \chi \nn\\
& \qquad + \frac{1}{6} \lambda_{[\Delta \Delta \Delta \Delta]} \Phi_{[\Delta]}^3 + \frac{1}{2} \lambda_{[\contraction[0.5ex]{}{3}{}{3} 33 \Delta \Delta]} \Phi_{[\Delta]} \partial_\mu \chi \partial^\mu \chi, \\
& (-\Box_{AdS} + \reg{m}^2_{0}) \chi = \lambda_{[\contraction[0.5ex]{}{3}{}{3} 33 \Delta]} \partial_\mu \left( \Phi_{[\Delta]} \partial^\mu \chi \right) - \frac{1}{2} \lambda_{[\contraction[0.5ex]{}{3}{}{3} 33 \Delta \Delta]} \partial_\mu \left( \Phi_{[\Delta]}^2 \partial^\mu \chi \right) \nn\\
& \qquad - \frac{1}{2} \lambda_{[\contraction[0.5ex]{}{3}{}{3} 33  \contraction[0.5ex]{}{3}{}{3} 33]} \partial_\mu \left( \partial^\mu \chi \partial_\nu \chi \partial^\nu \chi \right).
\end{align}
The boundary correlators follow from perturbative solutions to the equations of motion. As far as the 3-point functions are concerned we find two non-vanishing correlators,
\begin{align}
\lla \O_{[\Delta]}(\bs{k}_1) \O_{[\Delta]}(\bs{k}_2) \O_{[\Delta]}(\bs{k}_3) \rra & = \lambda_{[\Delta \Delta \Delta]} \, \ino_{[\Delta \Delta \Delta]}(k_1, k_2, k_3), \\
\lla X_{[3]}(\bs{k}_1) X_{[3]}(\bs{k}_2) \O_{[\Delta]}(\bs{k}_3) \rra & = \lambda_{[\contraction[0.5ex]{}{3}{}{3} 33 \Delta]} \, \ino_{[\contraction[0.5ex]{}{3}{}{3} 33 \Delta]}(k_1, k_2, k_3),
\end{align}
with $\< X \O \O \> = \< X X X \> = 0$. For the non-vanishing 4-point functions we find
\begin{align}
& \lla \O_{[\Delta]}(\bs{k}_1) \O_{[\Delta]}(\bs{k}_2) \O_{[\Delta]}(\bs{k}_3) \O_{[\Delta]}(\bs{k}_4) \rra = \lambda_{[\Delta \Delta \Delta]}^2 \left[ \ino_{[\Delta \Delta; \Delta \Delta x \Delta]} + \ino_{[\Delta \Delta; \Delta \Delta x \Delta]}^T + \ino_{[\Delta \Delta; \Delta \Delta x \Delta]}^U \right] \nn\\
& \qquad\qquad + \lambda_{[\Delta \Delta \Delta \Delta]} \ino_{[\Delta \Delta \Delta \Delta]}, \\
& \lla X_{[3]}(\bs{k}_1) X_{[3]}(\bs{k}_2) \O_{[\Delta]}(\bs{k}_3) \O_{[\Delta]}(\bs{k}_4) \rra = \lambda_{[\Delta \Delta \Delta]} \lambda_{[\contraction[0.5ex]{}{3}{}{3} 33 \Delta]} \ino_{[\contraction[0.5ex]{}{3}{}{3} 33; \Delta \Delta x \Delta]} \nn\\
& \qquad\qquad + \lambda_{[\contraction[0.5ex]{}{3}{}{3} 33 \Delta]}^2 \left[ \ino_{[\contraction[1.0ex]{}{3}{\Delta; \Delta 3 x \,}{3} \contraction[0.5ex]{3 \Delta;}{3}{\Delta x \!\!}{\Delta} 3 \Delta; 3 \Delta x 3]}^T + \ino_{[\contraction[1.0ex]{}{3}{\Delta; 3 \Delta x \,}{3} \contraction[0.5ex]{3 \Delta; \Delta}{3}{x \!\!}{\Delta} 3 \Delta; \Delta 3 x 3]}^U \right] + \lambda_{[\contraction[0.5ex]{}{3}{}{3} 33 \Delta \Delta]} \ino_{[\contraction[0.5ex]{}{3}{}{3} 33 \Delta \Delta]}, \label{XXOO} \\
& \lla X_{[3]}(\bs{k}_1) X_{[3]}(\bs{k}_2) X_{[3]}(\bs{k}_3) X_{[3]}(\bs{k}_4) \rra =\lambda_{[\contraction[0.5ex]{}{3}{}{3} 33 \Delta]}^2 \left[ \ino_{[\contraction[0.5ex]{}{3}{}{3} 33; \contraction[0.5ex]{}{3}{}{3} 33 x \Delta]} + \ino_{[\contraction[0.5ex]{}{3}{}{3} 33; \contraction[0.5ex]{}{3}{}{3} 33 x \Delta]}^T + \ino_{[\contraction[0.5ex]{}{3}{}{3} 33; \contraction[0.5ex]{}{3}{}{3} 33 x \Delta]}^U \right] \nn\\
& \qquad\qquad + \lambda_{[\contraction[0.5ex]{}{3}{}{3} 33 \contraction[0.5ex]{}{3}{}{3} 33]} \left[ \ino_{[\contraction[0.5ex]{}{3}{}{3} 33 \contraction[0.5ex]{}{3}{}{3} 33]} + \ino_{[\contraction[0.5ex]{}{3}{}{3} 33 \contraction[0.5ex]{}{3}{}{3} 33]}^T + \ino_{[\contraction[0.5ex]{}{3}{}{3} 33 \contraction[0.5ex]{}{3}{}{3} 33]}^U \right]
\end{align}
with the remaining correlators $\< X \O \O \O \> = \< X X X \O \> = 0$. By the subscripts $T$ and $U$ we denote the amplitudes in the $t$- and $u$-channels defined as follows. For any 4-point amplitude $\ino(\bs{k}_j)$ depending on 4 external momenta $\bs{k}_j$ for $j = 1,2,3,4$ its $t$- and $u$-channel amplitudes are defined as in \eqref{tchannel} and \eqref{uchannel},
\begin{align}
& \ino^T = \ino(\bs{k}_2 \leftrightarrow \bs{k}_3), && \ino^U = \ino(\bs{k}_1 \leftrightarrow \bs{k}_3).
\end{align}
If $\ino$ depends on the Mandelstam variable $s$, then $\ino^T$ depends on $t$ while $\ino^U$ on $u$.

Let us now study specific correlators in the symmetric theory for $\Delta = 2$ or $3$. First, let us look at $\< X X \O \O \>$. Weirdly enough, the two amplitudes there, $\ino_{[\contraction[1.0ex]{}{3}{\Delta; \Delta 3 x \,}{3} \contraction[0.5ex]{3 \Delta;}{3}{\Delta x \!\!}{\Delta} 3 \Delta; 3 \Delta x 3]}$ and $\ino_{[\contraction[1.0ex]{}{3}{\Delta; 3 \Delta x \,}{3} \contraction[0.5ex]{3 \Delta; \Delta}{3}{x \!\!}{\Delta} 3 \Delta; \Delta 3 x 3]}$, are absent from the list \eqref{Xtolocal} and do not simplify for general $\Delta$. The simplification occurs for $\Delta = 3$ though. Indeed, by plugging exact expressions to \eqref{XXOO} we find
\begin{align}
\ireg_{[\contraction[1.0ex]{}{3}{3; 33 x \,}{3} \contraction[0.5ex]{33;}{3}{3 x \!\!}{3} 33; 33 x 3]}^T + \ireg_{[\contraction[1.0ex]{}{3}{3; 33 x \,}{3} \contraction[0.5ex]{33;3}{3}{x \!\!}{3} 33; 33 x 3]}^U & = \frac{1}{2} \ireg_{[33; \contraction[0.5ex]{}{3}{}{3} 33]} + \frac{1}{4} \left[ t^3 + u^3 + 3 (k_1^3 + k_2^3 + k_3^3 + k_4^3) \right].
\end{align}
This is easy to obtain from equation \eqref{ired1}. Similarly we find
\begin{align}
\ireg_{[\contraction[0.5ex]{}{3}{}{3} 33; \contraction[0.5ex]{}{3}{}{3} 33 x 3]} + \ireg_{[\contraction[0.5ex]{}{3}{}{3} 33; \contraction[0.5ex]{}{3}{}{3} 33 x 3]}^T + \ireg_{[\contraction[0.5ex]{}{3}{}{3} 33; \contraction[0.5ex]{}{3}{}{3} 33 x 3]}^U = \frac{1}{4} \left[ s^3 + t^3 + u^3 - k_1^3 - k_2^3 - k_3^3 - k_4^3 \right]
\end{align}

Finally, renormalized versions of these equations can be considered. Using the renormalized amplitudes we find,
\begin{align}
& \iren_{[\contraction[0.5ex]{}{3}{}{3} 33; \contraction[0.5ex]{}{3}{}{3} 33 x 3]} + ( \iren_{[\contraction[0.5ex]{}{3}{}{3} 33; \contraction[0.5ex]{}{3}{}{3} 33 x 3]} )^T + ( \iren_{[\contraction[0.5ex]{}{3}{}{3} 33; \contraction[0.5ex]{}{3}{}{3} 33 x 3]}) ^U = \act^2_{[3\contraction[0.5ex]{}{3}{}{2} 33], x} \left[ s^3 + t^3 + u^3 \right] \nn\\
& \qquad\qquad + \left( \frac{3}{2} \act_{[3\contraction[0.5ex]{}{3}{}{2} 33], x} - 1 + 3 \act_{[\contraction[0.5ex]{}{3}{}{3} 33; \contraction[0.5ex]{}{3}{}{3} 33 x 3]} \right) \left[ k_1^3 + k_2^3 + k_3^3 + k_4^3 \right].
\end{align}
This implies that the amplitude is local, scheme-dependent and can be made completely vanishing by choosing local terms.

\section{Mathematica package} \label{sec:package}

All our results and their derivations can be found in the Mathematica notebooks included in the arXiv submission of this paper. Explicit expressions for the regulated and renormalized amplitudes are stored in the Mathematica package \verb|HandbooK.wl|, whose contents are described 
in Section \ref{sec:package_in} below. The package is accompanied by a number of  notebooks where details of the remaining calculations described in this paper can be found. We summarize the contents of these notebooks in Section \ref{sec:notebooks}.

\subsection{The package} \label{sec:package_in}

The package \verb|HandbooK.wl| is described in detail in our previous paper \cite{Bzowski:2022rlz}. For this paper we used the conventions and functions already defined in the package as described in \cite{Bzowski:2022rlz} and extended them to include derivative amplitudes.

The package does not contain a dedicated installer and is loaded instead through Mathematica's \verb|Get| command:

\includegraphics*[scale=0.9, trim=2.0cm 24.8cm 0cm 2.3cm]{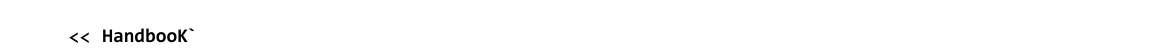}

All the regulated and renormalized 3- and 4-point amplitudes can be accessed through the commands listed in Table \ref{fig:commands} below

\begin{table}[htb]
\centering
\begin{tabular}{||c||c|c||}
\hline\hline
Amplitude & Regulated & Renormalized \\ \hline\hline
$\ino_{[\Delta_1 \Delta_2 \Delta_3]}$ & \verb|iReg3pt[|$\dreg, \{ \Dreg_1, \Dreg_2, \Dreg_3 \}$\verb|]| & \verb|iRen3pt[|$d, \{ \Delta_1, \Delta_2, \Delta_3 \}$\verb|]| \\ \hline
$\ino_{[\Delta_1 \Delta_2 \Delta_3 \Delta_4]}$ & \verb|iReg4ptC[|$\dreg, \{ \Dreg_1, \Dreg_2, \Dreg_3, \Dreg_4 \}$\verb|]| & \verb|iRen4ptC[|$d, \{ \Delta_1, \Delta_2, \Delta_3, \Delta_4 \}$\verb|]| \\ \hline
$\ino_{[\Delta_1 \Delta_2; \Delta_3 \Delta_4 x \Delta_x]}$ & \verb|iReg4ptX[|$\dreg, \{ \Dreg_1, \Dreg_2, \Dreg_3, \Dreg_4, \Dreg_x \}$\verb|]| & \verb|iRen4ptX[|$d, \{ \Delta_1, \Delta_2, \Delta_3, \Delta_4, \Delta_x \}$\verb|]| \\ \hline\hline
\end{tabular}
\caption{Mathematica commands for the access to regulated and renormalized 3- and 4-point functions in the package.\label{fig:commands}}
\end{table}

For example, the amplitude $\ireg_{[222]}$ regulated in the beta scheme with $u = 1$ and $v_j = 0$ can be obtained by the command

\includegraphics*[scale=0.9, trim=2.0cm 23.8cm 0cm 2.3cm]{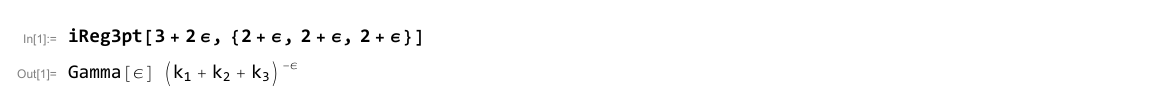}

By default, the regulator is represented by $\ep$ while the momentum magnitudes are $k_1, k_2, k_3, k_4$ and $s$ for the Mandelstam variable. If no regulator is specified, the beta scheme \eqref{beta} is assumed by default. Thus, for example, \verb|iReg3pt[3, {2,2,2}]| is equivalent to \verb|iReg3pt[|$3 + 2 \ep, \{ 2 + \ep, 2 + \ep, 2 + \ep \}$\verb|]| evaluated above. 

Additional options can be passed to the commands in Table \ref{fig:commands}. Option \verb|ExpansionOrder| determines to what order in the regulator the amplitude is expanded, the option \verb|Regulator| can be used to change the symbol for the regulator from $\ep$ to something else, and the option \verb|Momenta| can be used to change the default symbols $k_j$ and $s$ for the external momenta. For the detailed description and more examples see Section 7 of \cite{Bzowski:2022rlz}.

In order to access the derivative amplitudes, use option \verb|Derivatives|. The argument of the option is either a pair of numbers or a list of such pairs. The numbers $1,2,3,4$ within each pair indicate on which propagators the derivatives act and how they are contracted. For example, the amplitude $\ireg_{[\contraction[0.5ex]{}{2}{}{2} 222]}$ regulated in an arbitrary $(u, v_j)$-scheme \eqref{general} can be accessed by the following command

\includegraphics*[scale=0.9, trim=2.0cm 24.8cm 0cm 2.3cm]{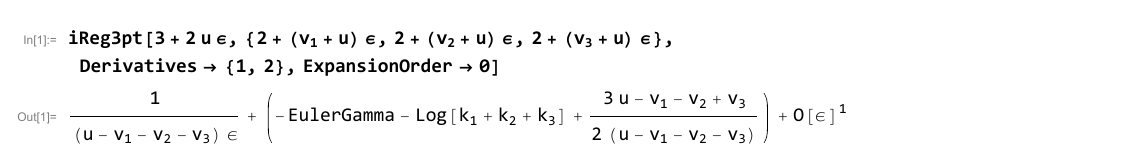}

Here we obtain the amplitude regulated in an arbitrary $(u, v_j)$-scheme and expanded to order $O(\ep^0)$ in the regulator, as indicated by the option \verb|ExpansionOrder|.

Similarly, we can obtain renormalized amplitudes. We can indicate 4-derivative amplitudes by passing a list of pairs of numbers indicating contractions. For example, we get the 4-derivative renormalized amplitude $\iren_{[\contraction[0.5ex]{}{3}{}{2} 32 \contraction[0.5ex]{}{3}{}{2} 32]}$ by calling

\includegraphics*[scale=0.9, trim=2.0cm 19.3cm 0cm 2.3cm]{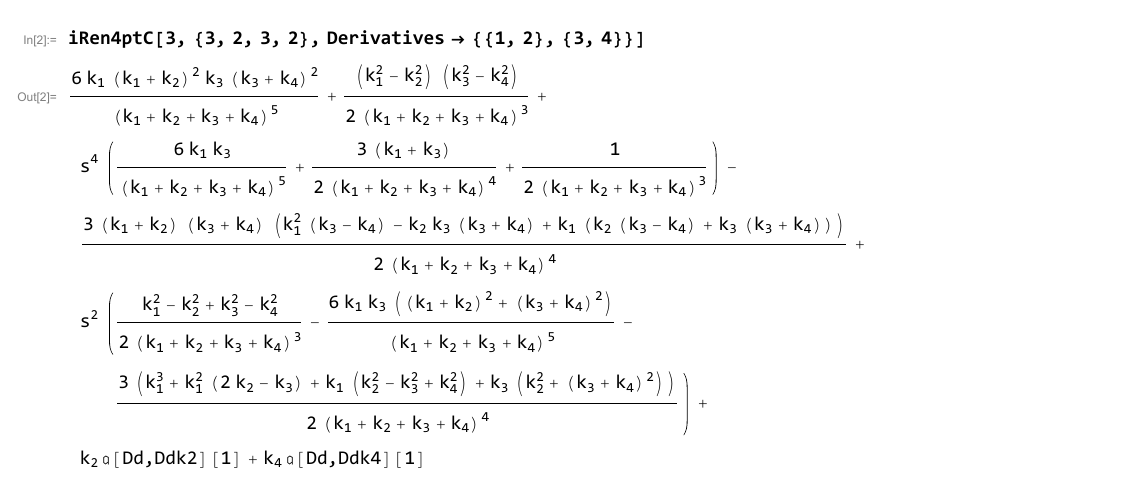}

The two constants in the last line are scheme-dependent constants. The symbol for the constants can be changed by means of the option \verb|RenormalizationConstant| passed to \verb|iRen4ptC|. See Section 7 of \cite{Bzowski:2022rlz} for details.

\begin{table}[thb]
\centering
\begin{tabular}{||c||c||}
\hline\hline
Amplitudes structure& \verb|Derivatives ->| \\ \hline\hline
$[\contraction[0.5ex]{}{\Delta}{{}_1}{\Delta} \Delta_1 \Delta_2 \Delta_3], [\contraction[0.5ex]{}{\Delta}{{}_1}{\Delta} \Delta_1 \Delta_2 \Delta_3 \Delta_4], [\contraction[0.5ex]{}{\Delta}{{}_1}{\Delta} \Delta_1 \Delta_2; \Delta_3 \Delta_4 x \Delta_x]$ &  \verb|{1,2}| \\ \hline
$[\Delta_1 \Delta_2; \contraction[0.5ex]{}{\Delta}{{}_3 \Delta_4 x}{\Delta} \Delta_3 \Delta_4 x \Delta_x]$ & \verb|{3,"x"}| \\ \hline
$[\contraction[0.5ex]{}{\Delta}{{}_1}{\Delta} \Delta_1 \Delta_2 \contraction[0.5ex]{}{\Delta}{{}_3}{\Delta} \Delta_3 \Delta_3], [\contraction[0.5ex]{}{\Delta}{{}_1}{\Delta} \Delta_1 \Delta_2; \contraction[0.5ex]{}{\Delta}{{}_3}{\Delta} \Delta_3 \Delta_4 x \Delta_x]$ & \verb|{{1,2}, {3,4}}| \\ \hline
$[\contraction[0.5ex]{}{\Delta}{{}_1}{\Delta} \Delta_1 \Delta_2; \contraction[0.5ex]{}{\Delta}{{}_3 \Delta_4 x}{\Delta} \Delta_3 \Delta_4 x \Delta_x]$ & \verb|{{1,2}, {3,"x"}}| \\ \hline
$[\contraction[1.0ex]{}{\Delta}{{}_1 \Delta_2; \Delta_3 \Delta_4 x \,}{\Delta} \contraction[0.5ex]{\Delta_1 \Delta_2;}{\Delta}{{}_3 \Delta_4 x \!\!}{\Delta} \Delta_1 \Delta_2; \Delta_3 \Delta_4 x \Delta_x]$ & \verb|{{1,"x"}, {3,"x"}}| \\ \hline\hline
\end{tabular}
\caption{The format of the `Derivatives' option passed to the 3- and 4-point functions in Table \ref{fig:commands}. To indicate the exchange, bulk-to-bulk propagator, one can use ``x" or ``X" or number $5$.\label{fig:derivatives}}
\end{table}

Finally, we can obtain exchange derivative amplitudes. In order to indicate the derivatives acting on the exchange, bulk-to-bulk propagator, use ``x" or ``X" or number $5$. For example, to obtain the amplitude $\ireg_{[\contraction[1.0ex]{}{3}{32;22x\!\!}{2} \contraction[0.5ex]{32;}{2}{2x\!}{2} 32;22x2]}$ we call

\includegraphics*[scale=0.9, trim=2.0cm 22.3cm 0cm 2.3cm]{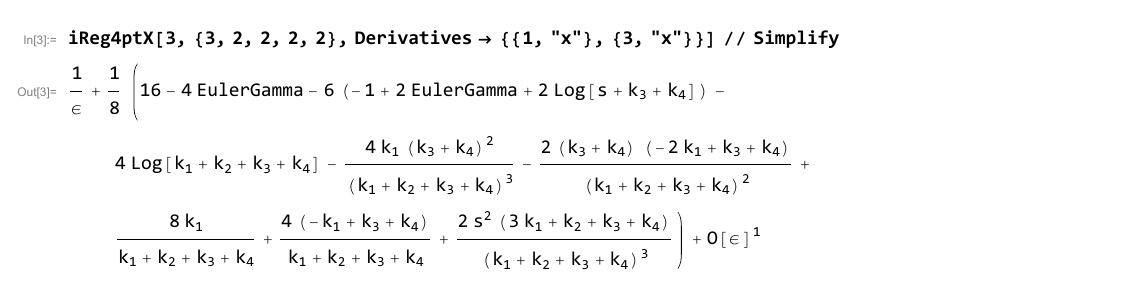}

Since no regulator was used in the arguments, the default beta scheme \eqref{beta} was used. Refer to Table \ref{fig:derivatives} below for more examples of the use of the \verb|Derivatives| option.

\subsection{Notebooks} \label{sec:notebooks}

The package is accompanied by a number of Mathematica notebooks which derive the results
presented in this paper.

\begin{itemize}
\item Notebooks \verb|BetaScheme.nb|, \verb|GeneralScheme.nb|, \verb|Checks.nb|, \verb|Renormalization.nb| are carried over from the previous version of the package. They contain calculations related to non-derivative amplitudes. Please refer to Section 7.2 of \cite{Bzowski:2022rlz} for more details.

\item \verb|BetaScheme_Derivatives.nb| contains calculations of derivative 3-, and 4-point functions regulated in the beta scheme \eqref{beta}. The notebook requires the \verb|HypExp| package \cite{Huber:2005yg}, which is included in the packet. The raw results of the calculations are then saved to \verb|Results_BetaScheme_Derivatives.nb|.

\item \verb|GeneralScheme_Derivatives.nb| contains calculations of derivative 3-, and 4-point functions regulated in a general scheme. The results of the calculations are saved to \verb|Results_GeneralScheme_Derivatives.nb|.

\item \verb|Checks_Derivatives.nb| provides checks on the results contained in the above notebooks. Furthermore, it compares these raw results with the expressions stored in the package file \verb|HandbooK.wl| and described in this paper.

\item  \verb|Renormalization_Derivatives.nb| contains the details of the renormalization procedure. We compare the expressions stored within the package file \verb|HandbooK.wl| with the expressions obtained by adding suitable counterterms in the asymmetric theory.

\end{itemize}

\section{Summary} \label{sec:summary}

In the paper I presented the comprehensive analysis of derivative 3- and 4-point amplitudes (Witten diagrams) in anti-de Sitter spacetime. After defining amplitudes in Section \ref{sec:amplitudes} I derived the number of relations between various amplitudes in Section \ref{sec:identities}. In Sections \ref{sec:asymmetric_theory} and \ref{sec:symmetric} I presented two Lagrangian theories, the asymmetric and the symmetry theory, leading to specific boundary correlation functions expressible in terms of the amplitudes. 

The defining property of the asymmetric theory is that it realizes a given single amplitude as a standalone correlator. Thus, by manipulating the action of the theory, one can obtain multiple identities between amplitudes, all at once. This creates a new method for deriving such identities that can be useful in the analysis of higher-point amplitudes in the future.

In Sections \ref{sec:3pt}, \ref{sec:4ptC} and \ref{sec:4ptX} I analyzed a number of derivative 3- and 4-point contact and exchange amplitudes in $d = 3$ boundary dimensions for operators of conformal dimensions $\Delta = 2,3$. Due to the sheer number of amplitudes calculated, the paper contains only a subset of 13 interesting contact and exchange 4-point amplitudes. The presented analysis includes the complete renormalization of the amplitudes.

This article is accompanied by the Mathematica package \verb|HandbooK|, which gathers all results from this paper as well as \cite{Bzowski:2022rlz} and the upcoming paper \cite{toappear} on cosmology. This includes all 2-, 3-, and 4-point amplitudes both in AdS as well as the de Sitter spacetime. Our hope is that the three papers together constitute the most comprehensive and useful set of tools available to a wide range of researchers. To this end, we provide full documentation including a set of accompanying Mathematica notebooks.

Apart from listing the results of our calculations, we discussed a number of new features and peculiarities discovered in the derivative amplitudes. In Section \ref{sec:finite_amps} we encountered finite, scheme-dependent 3-point amplitudes $\ireg_{[\contraction[0.5ex]{3}{3}{}{3} 333]}$ and $\ireg_{[\contraction[0.5ex]{2}{3}{}{2} 232]}$. We discovered more finite scheme-dependent amplitudes in Sections \ref{sec:4ptC} and \ref{sec:4ptX} and they are gathered in Tables \ref{fig:divs4pt2d} and \ref{fig:divs4pt4d}.

In Section \ref{sec:finite_amps} we saw how scheme-dependence emerges from derivatives acting on propagators and how it can be related to the existence of counterterms. We then argued that even finite amplitudes are not immune to renormalization effects. For example, there exist local terms which can make the renormalized amplitudes $\iren_{[\contraction[0.5ex]{3}{3}{}{3} 333]}$ and $\iren_{[\contraction[0.5ex]{2}{3}{}{2} 232]}$ vanish. 

The choice of the renormalization scheme for 3-point amplitudes influences the 4-point amplitudes in an interesting fashion. In particular, we found that the renormalized amplitude $\iren_{[22; \contraction[0.5ex]{}{3}{}{3} 33x3]}$ can be made either $s$-independent but scale-violating or scale-invariant but $s$-dependent by the specific choice of the renormalization scheme.

We conclude that one must be very careful when using finite amplitudes as they stand. Since even finite amplitudes undergo non-trivial renormalization, they are essentially non-unique. This important discovery goes against the naive expectation that finite amplitudes are fixed and do not require renormalization. In particular additional caution is required when deriving precise numerical predictions.

\section*{Acknowledgments}
AB would like to thank Paul McFadden and Kostas Skenderis for invaluable discussions and encouragement to finish this paper. AB is supported by the NCN POLS grant No.~2020/37/K/ST2/02768 financed from the Norwegian Financial Mechanism 2014-2021 \includegraphics[width=12pt]{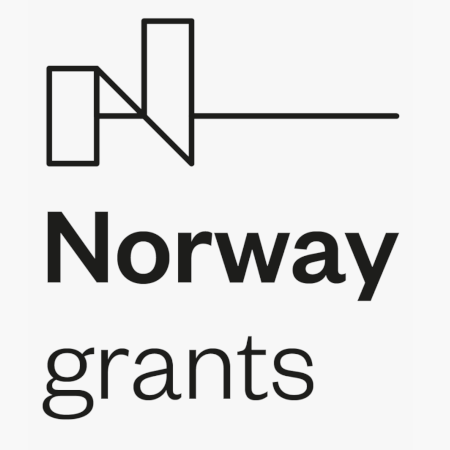} \includegraphics[width=12pt]{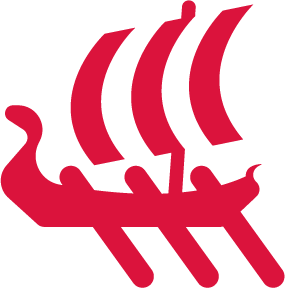}. 

\appendix

\section{Conventions and definitions}\label{sec:conventions}

\subsection{QFT conventions} \label{sec:QFT_conventions}

\begin{itemize}
\item We work almost exclusively in $d = 3$ Euclidean spacetime dimensions. 

\item Scaling dimensions are denoted throughout via square brackets, including where they appear as superscripts on operators and sources.  We assign the dimensions $[\bs{x}] = -1$ and  $[\partial_\mu] = 1$ for coordinates and their derivatives, $[\bs{k}] = 1$ for momenta,  $[\mu] = 1$ for the RG scale, $[\O_j] = \Delta$ for operators and $[\phi_j] = d - \Delta$ for their sources.

\item Our main focus will be scalar operators $\O^{[2]}$ and  $\O^{[3]}$ of dimensions $\Delta=2,3$,  their sources $\phi^{[1]}, \phi^{[0]}$ and the holographically dual bulk fields $\Phi^{[2]}$ and  $\Phi^{[3]}$. 

\item Regarding the relation of the generating functional of connected diagrams $W$ to the action $S$, we follow the conventions of \cite{Skenderis:2002wp} and \cite{Bzowski:2016kni} and define $W$ as
\begin{align}
Z = e^W = \< e^{-S} \>
\end{align}
from which it follows that
\begin{align}
\< \O_1(\bs{x}_1) \ldots \O_n(\bs{x}_n) \> = (-1)^n \frac{\delta^n W}{\delta \phi_1(\bs{x}_1) \ldots \delta \phi_n (\bs{x}_n)}.
\end{align}

\item When Fourier-transformed to momentum space correlation functions contain the momentum-conserving delta function. We introduce the double brackets to represent the stripped correlators,
\begin{align} \label{double_brackets}
\<O(\bs{k}_1)\ldots \O(\bs{k}_n)\> = \lla\O(\bs{k}_1)\ldots \O(\bs{k}_n)\rra \,(2\pi)^d\delta\big(\sum_{j=1}^n\bs{k}_j\big).
\end{align}

\item The most general regularization scheme we use is \eqref{general}.  
The parameterization is such that the natural parameters arising in holographic calculations \cite{Bzowski:2015pba},
\begin{align} \label{alphabeta}
& \alpha = \frac{d}{2} - 1, && \beta_j = \Delta_j - \frac{d}{2}
\end{align}
are regulated according to
\begin{align}
& \alpha \longmapsto \areg = \alpha + u \ep, && \beta_j \longmapsto \breg_j = \beta_j + v_j \ep.
\end{align}
Here, $\beta_j$ is the index associated with the Bessel function representing the bulk-boundary propagator for external leg $j$.

Any quantity $f = f(d, \Delta_j)$ depending on the dimensions $d$ and $\Delta_j$ is regulated according to the selected scheme in \eqref{general}. We denote the regulated version of $f$ as $\reg{f}$, \textit{i.e.}, $\reg{f} = f(\dreg, \Dreg_j)$.
\item Mostly, though, we work in the special \emph{beta scheme} \eqref{beta}.  This  corresponds to setting $u = 1$ and $v_j = 0$ for all  $j = 1,2,3,4,x$ in the general scheme \eqref{general}.
In this scheme the value of the $\beta_j$ parameters 
do not change:
\begin{align}
& \alpha \longmapsto \areg = \alpha + \ep, && \beta_j \longmapsto \breg_j = \beta_j.
\end{align}
This is a good renormalization scheme in $d = 3$ for operators of dimension $\Delta = 2, 3$ in the sense that it regulates all correlation functions of such operators.

\end{itemize}

\subsection{Conventions for momenta} \label{sec:momenta_conventions}

\begin{itemize}
\item External momenta are denoted $\bs{k}_j$, with lengths or magnitudes $k_j = |\bs{k}_j|$, where $j=1,2,\ldots$. The Mandelstam variables are
\begin{align} \label{def:stu}
& s = | \bs{k}_1 + \bs{k}_2 |, && t = | \bs{k}_1 + \bs{k}_3 |, && u = | \bs{k}_2 + \bs{k}_3 |
\end{align}
without squares. For convenience, we also adopt the convention $k_s = s$.

\item Mandelstam variables satisfy
\begin{align}
s^2 + t^2 + u^2 = k_1^2 + k_2^2 + k_3^2 + k_4^2.
\end{align}

\item The $3$- and $4$-point total magnitudes are denoted
\begin{align} \label{total}
& k_t = k_1 + k_2 + k_3, && k_T = k_1 + k_2 + k_3 + k_4.
\end{align}

\item We use $\s{m}{J}$ to denote  the corresponding  $m$-th symmetric polynomial on the set of indices $J$. 
 To be precise, let $J$ be an ordered set of indices and let $m$ be an integer such that $1 \leq m \leq |J|$. Then,
\begin{align} \label{def:sigma}
\s{m}{J} = \sum_{\substack{L \subseteq J\\|L|=m}} k_{L_1} \ldots k_{L_m},
\end{align}
where the sum is taken over all ordered subsets $L \subseteq J$ of cardinality $m$. In particular
\begin{align}
\s{1}{12} & = k_1 + k_2, & k_t = \s{1}{123} & = k_1 + k_2 + k_3, \\
\s{2}{12} & = k_1 k_2, & \s{2}{123} & = k_1 k_2 + k_1 k_3 + k_2 k_3, \\
&& \s{3}{123} & = k_1 k_2 k_3.
\end{align}

\item As we are interested mostly in 4-point functions, we use the shortened notation $\sigma_i = \s{i}{1234}$ for $J = {1,2,3,4}$,
\begin{align}
k_T = \sigma_1 = \s{1}{1234} & = k_1 + k_2 + k_3 + k_4, \\
\sigma_2 = \s{2}{1234} & = k_1 k_2 + k_1 k_3 + k_1 k_4 + k_2 k_3 + k_2 k_4 + k_3 k_4, \label{def:sigma2} \\
\sigma_3 = \s{3}{1234} & = k_1 k_2 k_3 + k_1 k_2 k_4 + k_1 k_3 k_4 + k_2 k_3 k_4, \\
\sigma_4 = \s{4}{1234} & = k_1 k_2 k_3 k_4.
\end{align}

\item We also allow for the indices to take the value $s$, so that, for example, $\sigma_{(1)12s} = k_1 + k_2 + s$ and so on.

\item Some derivative diagrams exhibit a smaller symmetry group, the dihedral subgroup $D_4 \leq S_4$. This group contains eight permutations generated by swapping the momenta $k_1 \leftrightarrow k_2$ or $k_3 \leftrightarrow k_4$ as well as exchanging the pairs, $(k_1, k_2) \leftrightarrow (k_3, k_4)$. The invariant polynomials for $D_4$ consist of the 4 symmetric polynomials $\sigma_i$ for $i=1,2,3,4$ as well as the additional invariant
\begin{align}
\tau = (k_1 + k_2) (k_3 + k_4).
\end{align}

\item Many exchange amplitudes depend on the following combination of dilogarithms
\begin{align}
\ifin_{[22,22x2]} & = - \frac{1}{2 s} \left[ \Li_2 \left( \frac{\m{34}}{k_T} \right) + \Li_2 \left( \frac{\m{12}}{k_T} \right) + \log \left( \frac{\p{12}}{k_T} \right) \log \left( \frac{\p{34}}{k_T} \right) - \frac{\pi^2}{6} \right].
\end{align}

\end{itemize}

\subsection{AdS conventions} \label{sec:AdS_conventions}

\begin{itemize}
\item $d+1$ dimensional Euclidean AdS spacetime in Poincar\'{e} coordinates is given by the metric
\begin{align}
g_{\mu\nu} \D x^\mu \D x^\nu = \frac{\D z^2 + \D \bs{x}^2}{z^2},
\end{align}
with $x = (z, \bs{x})$, where $z$ is the radial variable and $\bs{x}$ parameterizes the $d$-dimensional boundary.

\item A free bulk field $\Phi_{[\Delta]}$ is governed by the action
\begin{align}
S_{\text{free}} = \frac{1}{2} \int \D^{d+1} x \sqrt{g} \left[ \partial_{\mu} \Phi_{[\Delta]} \partial^\mu \Phi_{[\Delta]} + m_{\Delta}^2 \Phi_{[\Delta]}^2 \right],
\end{align}
where the mass of the field reads
\begin{align}
m_{\Delta}^2 = \Delta(\Delta - d).
\end{align}

\item The 1-point function with sources turned on of the boundary operator $\O_{[\Delta]}$ dual to the bulk field $\Phi_{[\Delta]}$ is equal
\begin{align}
\< \O_{[\Delta]} \>_{s, \text{reg}} = - (2 \Dreg - \dreg) \phi_{(\Dreg)},
\end{align}
where $\phi_{(\Dreg)}$ is the coefficient of $z^{\Dreg}$ in the power expansion of $\Phi_{[\Delta]}$ around $z = 0$.

\item The scalar bulk-to-boundary propagator is
\begin{align} \label{KPropagator}
	\K_{d, \Delta}(z, k) = \frac{k^{\Delta - \frac{d}{2}} z^{\frac{d}{2}} K_{\Delta - \frac{d}{2}}(k z)}{2^{\Delta - \frac{d}{2} - 1} \Gamma \left( \Delta - \frac{d}{2} \right)}
\end{align}
while the scalar bulk-to-bulk propagator is
\begin{align} \label{GPropagator}
	\G_{d, \Delta}(z, k; \z) = \left\{ \begin{array}{ll}
		(z \z)^{\frac{d}{2}} I_{\Delta - \frac{d}{2}}(k z) K_{\Delta - \frac{d}{2}}(k \z) & \text{ for } z < \z, \\
		(z \z)^{\frac{d}{2}} K_{\Delta - \frac{d}{2}}(k z) I_{\Delta - \frac{d}{2}}(k \z) & \text{ for } z > \z.
	\end{array} \right.	
\end{align}
where $I_{\beta}$ and $K_{\beta}$ are the modified Bessel functions.

\item To avoid clutter, we will use $\Kreg_{[\Delta]}$ and $\Greg_{[\Delta]}$ to denote the propagators for the regulated parameters $\dreg$ and $\Dreg$ defined in \eqref{general}, leaving the specific scheme implicit.

\item The near-boundary expansion of the bulk-to-bulk propagator reads
\begin{align} \label{holo1}
\Greg_{[\Delta]}(z, k; \zeta) = \frac{z^{\Dreg}}{2 \Dreg - \dreg} \Kreg_{[\Dreg]}(\zeta, k) + O(z^{\Dreg + 2}).
\end{align}

\item A Bessel function $K$ with a half-integral index is equal to
\begin{equation} \label{Khalfint}
K_{\beta}(x) = \frac{e^{-x}}{\sqrt{x}} \sum_{j=0}^{|\beta| - \frac{1}{2}} \frac{c_j(\beta)}{x^j}, \ \beta \in \Z + \frac{1}{2},
\end{equation}
where the coefficients are
\begin{equation} \label{e:cfa}
c_j(\beta) = \sqrt{\frac{\pi}{2}} \frac{ \left( |\beta| - \frac{1}{2} + j \right)!}{2^j j! \left(|\beta| - \frac{1}{2} - j \right)!}.
\end{equation}

\item The AdS Laplacian Fourier-transformed along the boundary direction reads
\begin{align}
\Box_{z, \bs{k}} = z^2 \partial_z^2 - (d-1) z \partial_z - z^2 k^2.
\end{align}

\item For derivative amplitudes we define the operator
\begin{align} \label{ap:Dm}
\mathcal{D}^{m}(z, \bs{k}) = z \left( \I \bs{k}^m + \hat{\bs{z}} \partial_z \right),
\end{align}
where $m = 1, \ldots, d$ represents the boundary coordinates, which are contracted with the Euclidean metric, $\delta_{mn}$. Furthermore, $\hat{\bs{z}}$ is a unit vector, $\hat{\bs{z}} \cdot \hat{\bs{z}} = 1$, orthogonal to the boundary directions, $\hat{\bs{z}} \cdot \bs{k} = 0$. We drop the arguments of the operators $\mathcal{D}^m$ if they are the same as the bulk-to-boundary propagator they act on, \textit{e.g.}, $[ \mathcal{D}^{m} \reg{\K}_{[\Delta_2]}](z, \bs{k}_2) = \mathcal{D}^{m}(z, \bs{k}_2) \reg{\K}_{[\Delta_2]}(z, k_2)$ and so on.

\end{itemize}

\section{Definitions of regulated amplitudes} \label{sec:amp_defs}

\begin{figure}[t]
\begin{tikzpicture}[scale=1.0]
\draw (0,0) circle [radius=3];
\draw [fill=black] (-2.121,-2.121) circle [radius=0.1];
\draw [fill=black] (-2.121, 2.121) circle [radius=0.1];
\draw [fill=black] ( 2.121,-2.121) circle [radius=0.1];
\draw [fill=black] ( 2.121, 2.121) circle [radius=0.1];
\draw [fill=black] ( 0, 0) circle [radius=0.1];
\draw (-2.121,-2.121) -- ( 2.121, 2.121);
\draw ( 2.121,-2.121) -- (-2.121, 2.121);
\draw [red] (0.4,-0.3) -- (0.1,0) -- (0.4,0.3);
\node [left] at (-2.121, 2.2) {$\O_1(\bs{k}_1)$}; 
\node [left] at (-2.121,-2.2) {$\O_2(\bs{k}_2)$}; 	
\node [right] at ( 2.121,-2.2) {$\O_3(\bs{k}_3)$}; 
\node [right] at ( 2.121, 2.2) {$\O_4(\bs{k}_4)$}; 	
\node [above] at (-0.9, 1.06) {$\K_{[\Delta_1]}$};
\node [above] at (-1.3,-1.06) {$\K_{[\Delta_2]}$};
\node [above] at ( 1.2,-1.06) {$\K_{[\Delta_3]}$};
\node [above] at ( 0.8, 1.06) {$\K_{[\Delta_4]}$};
\end{tikzpicture}
\qquad
\begin{tikzpicture}[scale=1.0]
\draw (0,0) circle [radius=3];
\draw [fill=black] (-2.121,-2.121) circle [radius=0.1];
\draw [fill=black] (-2.121, 2.121) circle [radius=0.1];
\draw [fill=black] ( 2.121,-2.121) circle [radius=0.1];
\draw [fill=black] ( 2.121, 2.121) circle [radius=0.1];
\draw [fill=black] ( 0, 0) circle [radius=0.1];
\draw (-2.121,-2.121) -- ( 2.121, 2.121);
\draw ( 2.121,-2.121) -- (-2.121, 2.121);
\draw [red] (0.4,-0.3) -- (0.1,0) -- (0.4,0.3);
\draw [blue] (-0.4,-0.3) -- (-0.1,0) -- (-0.4,0.3);
\node [left] at (-2.121, 2.2) {$\O_1(\bs{k}_1)$}; 
\node [left] at (-2.121,-2.2) {$\O_2(\bs{k}_2)$}; 	
\node [right] at ( 2.121,-2.2) {$\O_3(\bs{k}_3)$}; 
\node [right] at ( 2.121, 2.2) {$\O_4(\bs{k}_4)$}; 	
\node [above] at (-0.9, 1.06) {$\K_{[\Delta_1]}$};
\node [above] at (-1.3,-1.06) {$\K_{[\Delta_2]}$};
\node [above] at ( 1.2,-1.06) {$\K_{[\Delta_3]}$};
\node [above] at ( 0.8, 1.06) {$\K_{[\Delta_4]}$};
\end{tikzpicture}
\centering
\caption{Witten diagram representing the 4-point amplitudes $\ino_{[\Delta_1 \Delta_2 \contraction[0.5ex]{}{\Delta}{{}_3}{\Delta} \Delta_3 \Delta_4]}$ and $\ino_{[\contraction[0.5ex]{}{\Delta}{{}_1}{\Delta} \Delta_1 \Delta_2 \contraction[0.5ex]{}{\Delta}{{}_3}{\Delta} \Delta_3 \Delta_4]}$. The vertex in the left panel is $\Phi_1 \Phi_2 \nabla_\mu \Phi_3 \nabla^\mu \Phi_4$, while the vertex on the right is $\nabla_\nu \Phi_1 \nabla^\nu \Phi_2 \nabla_\mu \Phi_3 \nabla^\mu \Phi_4$. The red and blue lines indicate the contractions of bulk derivatives.\label{fig:intro4ptC}}
\end{figure}

\begin{figure}[t]
\begin{tikzpicture}[scale=1.0]
\draw (0,0) circle [radius=3];
\draw [fill=black] (-2.121,-2.121) circle [radius=0.1];
\draw [fill=black] (-2.121, 2.121) circle [radius=0.1];
\draw [fill=black] ( 2.121,-2.121) circle [radius=0.1];
\draw [fill=black] ( 2.121, 2.121) circle [radius=0.1];
\draw [fill=black] (-1, 0) circle [radius=0.1];
\draw [fill=black] ( 1, 0) circle [radius=0.1];
\draw (-2.121,-2.121) -- (-1,0) -- (-2.121, 2.121);
\draw ( 2.121, 2.121) -- ( 1,0) -- ( 2.121,-2.121);
\draw (-1,0) -- (1,0);
\draw [red] (1.4,-0.52) -- (1.1,0) -- (1.4,0.52);
\node [left] at (-2.121, 2.2) {$\O_1(\bs{k}_1)$}; 
\node [left] at (-2.121,-2.2) {$\O_2(\bs{k}_2)$}; 	
\node [right] at ( 2.121,-2.2) {$\O_3(\bs{k}_3)$}; 
\node [right] at ( 2.121, 2.2) {$\O_4(\bs{k}_4)$}; 	
\node [right] at (-1.5, 1.2) {$\K_{[\Delta_1]}$};
\node [right] at (-1.5, -1.2) {$\K_{[\Delta_2]}$};
\node [left] at ( 1.5, -1.2) {$\K_{[\Delta_3]}$};
\node [left] at ( 1.5, 1.2) {$\K_{[\Delta_4]}$};
\node [above] at (0,0) {$\G_{[\Delta_x]}$};
\end{tikzpicture}
\qquad
\begin{tikzpicture}[scale=1.0]
\draw (0,0) circle [radius=3];
\draw [fill=black] (-2.121,-2.121) circle [radius=0.1];
\draw [fill=black] (-2.121, 2.121) circle [radius=0.1];
\draw [fill=black] ( 2.121,-2.121) circle [radius=0.1];
\draw [fill=black] ( 2.121, 2.121) circle [radius=0.1];
\draw [fill=black] (-1, 0) circle [radius=0.1];
\draw [fill=black] ( 1, 0) circle [radius=0.1];
\draw (-2.121,-2.121) -- (-1,0) -- (-2.121, 2.121);
\draw ( 2.121, 2.121) -- ( 1,0) -- ( 2.121,-2.121);
\draw (-1,0) -- (1,0);
\draw [red] (-1.4,-0.52) -- (-1.1,0) -- (-1.4,0.52);
\node [left] at (-2.121, 2.2) {$\O_1(\bs{k}_1)$}; 
\node [left] at (-2.121,-2.2) {$\O_2(\bs{k}_2)$}; 	
\node [right] at ( 2.121,-2.2) {$\O_3(\bs{k}_3)$}; 
\node [right] at ( 2.121, 2.2) {$\O_4(\bs{k}_4)$}; 	
\node [right] at (-1.5, 1.2) {$\K_{[\Delta_1]}$};
\node [right] at (-1.5, -1.2) {$\K_{[\Delta_2]}$};
\node [left] at ( 1.5, -1.2) {$\K_{[\Delta_3]}$};
\node [left] at ( 1.5, 1.2) {$\K_{[\Delta_4]}$};
\node [above] at (0,0) {$\G_{[\Delta_x]}$};
\end{tikzpicture}
\qquad
\begin{tikzpicture}[scale=1.0]
\draw (0,0) circle [radius=3];
\draw [fill=black] (-2.121,-2.121) circle [radius=0.1];
\draw [fill=black] (-2.121, 2.121) circle [radius=0.1];
\draw [fill=black] ( 2.121,-2.121) circle [radius=0.1];
\draw [fill=black] ( 2.121, 2.121) circle [radius=0.1];
\draw [fill=black] (-1, 0) circle [radius=0.1];
\draw [fill=black] ( 1, 0) circle [radius=0.1];
\draw (-2.121,-2.121) -- (-1,0) -- (-2.121, 2.121);
\draw ( 2.121, 2.121) -- ( 1,0) -- ( 2.121,-2.121);
\draw (-1,0) -- (1,0);
\draw [red] (1.4,-0.52) -- (1.1,0) -- (1.4,0.52);
\draw [blue] (-1.4,-0.52) -- (-1.1,0) -- (-1.4,0.52);
\node [left] at (-2.121, 2.2) {$\O_1(\bs{k}_1)$}; 
\node [left] at (-2.121,-2.2) {$\O_2(\bs{k}_2)$}; 	
\node [right] at ( 2.121,-2.2) {$\O_3(\bs{k}_3)$}; 
\node [right] at ( 2.121, 2.2) {$\O_4(\bs{k}_4)$}; 	
\node [right] at (-1.5, 1.2) {$\K_{[\Delta_1]}$};
\node [right] at (-1.5, -1.2) {$\K_{[\Delta_2]}$};
\node [left] at ( 1.5, -1.2) {$\K_{[\Delta_3]}$};
\node [left] at ( 1.5, 1.2) {$\K_{[\Delta_4]}$};
\node [above] at (0,0) {$\G_{[\Delta_x]}$};
\end{tikzpicture}
\qquad
\begin{tikzpicture}[scale=1.0]
\draw (0,0) circle [radius=3];
\draw [fill=black] (-2.121,-2.121) circle [radius=0.1];
\draw [fill=black] (-2.121, 2.121) circle [radius=0.1];
\draw [fill=black] ( 2.121,-2.121) circle [radius=0.1];
\draw [fill=black] ( 2.121, 2.121) circle [radius=0.1];
\draw [fill=black] (-1, 0) circle [radius=0.1];
\draw [fill=black] ( 1, 0) circle [radius=0.1];
\draw (-2.121,-2.121) -- (-1,0) -- (-2.121, 2.121);
\draw ( 2.121, 2.121) -- ( 1,0) -- ( 2.121,-2.121);
\draw (-1,0) -- (1,0);
\draw [red] (1.2,0.52) -- (0.95,0.1) -- (0.5,0.1);
\draw [blue] (-1.2,0.52) -- (-0.95,0.1) -- (-0.5,0.1);
\node [left] at (-2.121, 2.2) {$\O_1(\bs{k}_1)$}; 
\node [left] at (-2.121,-2.2) {$\O_2(\bs{k}_2)$}; 	
\node [right] at ( 2.121,-2.2) {$\O_3(\bs{k}_3)$}; 
\node [right] at ( 2.121, 2.2) {$\O_4(\bs{k}_4)$}; 	
\node [right] at (-1.5, 1.2) {$\K_{[\Delta_1]}$};
\node [right] at (-1.5, -1.2) {$\K_{[\Delta_2]}$};
\node [left] at ( 1.5, -1.2) {$\K_{[\Delta_3]}$};
\node [left] at ( 1.5, 1.2) {$\K_{[\Delta_4]}$};
\node [above] at (0,0) {$\G_{[\Delta_x]}$};
\end{tikzpicture}
\centering
\caption{Witten diagrams representing all derivative exchange 4-point amplitudes analyzed in this paper: $\ino_{[\Delta_1 \Delta_2; \contraction[0.5ex]{}{\Delta}{{}_3}{\Delta} \Delta_3 \Delta_4 x \Delta_x]}$, $\ino_{[\contraction[0.5ex]{}{\Delta}{{}_1}{\Delta} \Delta_1 \Delta_2; \Delta_3 \Delta_4 x \Delta_x]}$, $\ino_{[\contraction[0.5ex]{}{\Delta}{{}_1}{\Delta} \Delta_1 \Delta_2; \contraction[0.5ex]{}{\Delta}{{}_3}{\Delta} \Delta_3 \Delta_4 x \Delta_x]}$ and $\ino_{[\contraction[1.0ex]{}{\Delta}{{}_1 \Delta_2; \Delta_3 \Delta_4 x \,}{\Delta} \contraction[0.5ex]{\Delta_1 \Delta_2;}{\Delta}{{}_3 \Delta_4 x \!\!}{\Delta} \Delta_1 \Delta_2; \Delta_3 \Delta_4 x \Delta_x]}$.\label{fig:intro4ptX}}
\end{figure}

\subsection{2-point amplitude}

\begin{itemize}
\item For convenience,  regulated 2-point amplitudes are normalized so as  to match the holographic 2-point functions:
\begin{align} \label{amp2}
\ireg_{[\Delta \Delta]}(k) = (2 \Dreg - \dreg) \times\text{coefficient of } z^{\Dreg} \text{ in } \Kreg_{[\Delta]}(z, k).
\end{align}
All non-diagonal 2-point amplitudes $\ireg_{[\Delta \Delta']}$ with $\Delta \neq \Delta'$ vanish.
\end{itemize}

\subsection{3-point amplitudes}

\begin{itemize}
\item We define the regulated non-derivative 3-point amplitudes as
\begin{align} \label{ap:amp3}
\ireg_{[\Delta_1 \Delta_2 \Delta_3]}(k_1, k_2, k_3) = \int_0^{\infty} \D z \, z^{-\dreg-1} \, \Kreg_{[\Delta_1]}(z, k_1) \Kreg_{[\Delta_2]}(z, k_2) \Kreg_{[\Delta_3]}(z, k_3).
\end{align}
The corresponding Witten diagram is presented in Figure [x].

\item We define the regulated 2-derivative 3-point amplitudes as
\begin{align} \label{ap:amp3_d2}
\ireg_{[\Delta_1 \contraction[0.5ex]{}{\Delta}{{}_2}{\Delta} \Delta_2 \Delta_3]} & = \int_0^{\infty} \D z \, z^{-\dreg-1} \reg{\K}_{[\Delta_1]}(z, k_1) [ \mathcal{D}^{m} \reg{\K}_{[\Delta_2]}](z, \bs{k}_2) [ \mathcal{D}_{m} \reg{\K}_{[\Delta_3]}](z, \bs{k}_3).
\end{align}
The corresponding Witten diagram is presented in Figure [x].

\end{itemize}

\subsection{4-point contact amplitudes}

\begin{itemize}
\item We use $\ino_{[\Delta_1 \Delta_2 \Delta_3 \Delta_4]}$ to  denote the non-derivative amplitudes with four external scalars of dimensions $\Delta_1, \Delta_2, \Delta_3, \Delta_4$, as presented in Figure [x]. The regulated expression is
\begin{align} \label{4ptC}
& \ireg_{[\Delta_1 \Delta_2 \Delta_3 \Delta_4]}(k_1, k_2, k_3, k_4)  \nn\\
& \qquad = \int_0^\infty \D z \, z^{-\dreg-1} \Kreg_{[\Delta_1]}(z, k_1) \Kreg_{[\Delta_2]}(z, k_2) \Kreg_{[\Delta_3]}(z, k_3) \Kreg_{[\Delta_4]}(z, k_4).
\end{align}

\item The 2-derivative 4-point contact amplitudes are defined as,
\begin{align} \label{ap:amp4c_2d}
& \ireg_{[\Delta_1 \Delta_2 \contraction[0.5ex]{}{\Delta}{{}_3}{\Delta} \Delta_3 \Delta_4]}(k_1, k_2, k_3, k_4) \nn\\
& \qquad = \int_0^\infty \D z \, z^{-\dreg-1} \Kreg_{[\Delta_1]}(z, k_1) \Kreg_{[\Delta_2]}(z, k_2) [\mathcal{D}^{m} \Kreg_{[\Delta_3]}](z, \bs{k}_3) [\mathcal{D}_{m} \Kreg_{[\Delta_4]}](z, \bs{k}_4)
\end{align}

\item The 4-derivative 4-point contact amplitudes considered in this paper are defined as,
\begin{align} \label{4ptC4d}
& \ireg_{[\contraction[0.5ex]{}{\Delta}{{}_1}{\Delta} \Delta_1 \Delta_2 \contraction[0.5ex]{}{\Delta}{{}_3}{\Delta} \Delta_3 \Delta_4]}(k_1, k_2, k_3, k_4) \nn\\
& \qquad = \int_0^\infty \D z \, z^{-\dreg-1} [ \mathcal{D}^{m} \Kreg_{[\Delta_1]} ](z, k_1) [ \mathcal{D}_{m} \Kreg_{[\Delta_2]} ](z, k_2) [\mathcal{D}^{n} \Kreg_{[\Delta_3]}](z, k_3) [\mathcal{D}_{n} \Kreg_{[\Delta_4]}](z, k_4).
\end{align}
\end{itemize}

\subsection{4-point exchange amplitudes}

There is a number of 4-point exchange amplitudes, depending on which lines the derivatives act. We have the following amplitudes considered in this paper:

\begin{itemize}
\item No-derivative amplitudes,
\begin{align} \label{4ptX}
& \ireg_{[\Delta_1 \Delta_2; \Delta_3 \Delta_4 x \Delta_x]}(k_1, k_2, k_3, k_4, s)  \nn\\
& \qquad = \int_0^\infty \D z \, z^{-\dreg-1} \Kreg_{[\Delta_1]}(z, \bs{k}_1) \Kreg_{[\Delta_2]}(z, \bs{k}_2) \times\nn\\
& \qquad\qquad \times \int_0^\infty \D \z \, \z^{-\dreg-1} \Greg_{[\Delta_x]}(z, s; \z) \Kreg_{[\Delta_3]}(\z, k_3) \Kreg_{[\Delta_4]}(\z, k_4).
\end{align}
\item 2-derivative amplitudes,
\begin{align}
& \ireg_{[\contraction[0.5ex]{}{\Delta}{{}_1}{\Delta} \Delta_1 \Delta_2; \Delta_3 \Delta_4 x \Delta_x]}(k_1, k_2, k_3, k_4, s)  \nn\\
& \qquad = \int_0^\infty \D z \, z^{-\dreg-1} [\mathcal{D}^m \Kreg_{[\Delta_1]}](z, \bs{k}_1) [\mathcal{D}_m \Kreg_{[\Delta_2]}](z, \bs{k}_2) \times\nn\\
& \qquad\qquad \times \int_0^\infty \D \z \, \z^{-\dreg-1} \Greg_{[\Delta_x]}(z, s; \z) \Kreg_{[\Delta_3]}(\z, k_3) \Kreg_{[\Delta_4]}(\z, k_4),
\end{align}
and
\begin{align} \label{4ptX2da}
& \ireg_{[\Delta_1 \Delta_2; \contraction[0.5ex]{}{\Delta}{{}_3 \Delta_4 x}{\Delta} \Delta_3 \Delta_4 x \Delta_x]}(k_1, k_2, k_3, k_4, s)  \nn\\
& \qquad = \int_0^\infty \D z \, z^{-\dreg-1} \Kreg_{[\Delta_1]}(z, \bs{k}_1) \Kreg_{[\Delta_2]}(z, \bs{k}_2) \times\nn\\
& \qquad\qquad \times \int_0^\infty \D \z \, \z^{-\dreg-1} \mathcal{D}^m(\z, \bs{s}) \Greg_{[\Delta_x]}(z, s; \z) [\mathcal{D}_m \Kreg_{[\Delta_3]}(\z, k_3)] \Kreg_{[\Delta_4]}(\z, k_4).
\end{align}
\item 4-derivative amplitudes,
\begin{align}
& \ireg_{[\contraction[0.5ex]{}{\Delta}{{}_1}{\Delta} \Delta_1 \Delta_2; \contraction[0.5ex]{}{\Delta}{{}_3}{\Delta} \Delta_3 \Delta_4 x \Delta_x]}(k_1, k_2, k_3, k_4, s)  \nn\\
& \qquad = \int_0^\infty \D z \, z^{-\dreg-1} [\mathcal{D}^m \Kreg_{[\Delta_1]}](z, \bs{k}_1) [\mathcal{D}_m \Kreg_{[\Delta_2]}](z, \bs{k}_2) \times\nn\\
& \qquad\qquad \times \int_0^\infty \D \z \, \z^{-\dreg-1} \Greg_{[\Delta_x]}(z, s; \z) [\mathcal{D}^n \Kreg_{[\Delta_3]}](\z, \bs{k}_3) [\mathcal{D}_n \Kreg_{[\Delta_4]}](\z, \bs{k}_4)
\end{align}
and
\begin{align} \label{4ptX4da}
& \ireg_{[\contraction[0.5ex]{}{\Delta}{{}_1}{\Delta} \Delta_1 \Delta_2; \contraction[0.5ex]{}{\Delta}{{}_3 \Delta_4 x}{\Delta} \Delta_3 \Delta_4 x \Delta_x]}(k_1, k_2, k_3, k_4, s)  \nn\\
& \qquad = \int_0^\infty \D z \, z^{-\dreg-1} [\mathcal{D}^m \Kreg_{[\Delta_1]}](z, \bs{k}_1) [\mathcal{D}_m \Kreg_{[\Delta_2]}](z, k_2) \times\nn\\
& \qquad\qquad \times \int_0^\infty \D \z \, \z^{-\dreg-1} \mathcal{D}^n(\z, \bs{s}) \Greg_{[\Delta_x]}(z, s; \z) [\mathcal{D}^n \Kreg_{[\Delta_3]}](\z, \bs{k}_3) [ \mathcal{D}_n \Kreg_{[\Delta_4]}](\z, k_4)
\end{align}
and
\begin{align} \label{4ptX4db}
& \ireg_{[\contraction[1.0ex]{}{\Delta}{{}_1 \Delta_2; \Delta_3 \Delta_4 x \,}{\Delta} \contraction[0.5ex]{\Delta_1 \Delta_2;}{\Delta}{{}_3 \Delta_4 x \!\!}{\Delta} \Delta_1 \Delta_2; \Delta_3 \Delta_4 x \Delta_x]}(k_1, k_2, k_3, k_4, s)  \nn\\
& \qquad = \int_0^\infty \D z \, z^{-\dreg-1} [\mathcal{D}^m \Kreg_{[\Delta_1]}](z, \bs{k}_1) \Kreg_{[\Delta_2]}(z, k_2) \times\nn\\
& \qquad\qquad \times \int_0^\infty \D \z \, \z^{-\dreg-1} \mathcal{D}_m(z, \bs{s}) \mathcal{D}_n(\z, \bs{s}) \Greg_{[\Delta_x]}(z, s; \z) [\mathcal{D}^n \Kreg_{[\Delta_3]}](\z, \bs{k}_3) \Kreg_{[\Delta_4]}(\z, k_4).
\end{align}
\end{itemize}

\begin{figure}[t]
\begin{tikzpicture}[scale=1.0]
\draw (0,0) circle [radius=3];
\draw [fill=black] (-2.121,-2.121) circle [radius=0.1];
\draw [fill=black] (-2.121, 2.121) circle [radius=0.1];
\draw [fill=black] ( 2.121,-2.121) circle [radius=0.1];
\draw [fill=black] ( 2.121, 2.121) circle [radius=0.1];
\draw [fill=black] ( 0, 0) circle [radius=0.1];
\draw (-2.121,-2.121) -- ( 2.121, 2.121);
\draw ( 2.121,-2.121) -- (-2.121, 2.121);	
\node [left] at (-2.121, 2.2) {$\O_1(\bs{k}_1)$}; 
\node [left] at (-2.121,-2.2) {$\O_2(\bs{k}_2)$}; 	
\node [right] at ( 2.121,-2.2) {$\O_3(\bs{k}_3)$}; 
\node [right] at ( 2.121, 2.2) {$\O_4(\bs{k}_4)$}; 	
\node [above] at (-0.9, 1.06) {$\K_{[\Delta_1]}$};
\node [above] at (-1.3,-1.06) {$\K_{[\Delta_2]}$};
\node [above] at ( 1.2,-1.06) {$\K_{[\Delta_3]}$};
\node [above] at ( 0.8, 1.06) {$\K_{[\Delta_4]}$};
\end{tikzpicture}
\qquad
\begin{tikzpicture}[scale=1.0]
\draw (0,0) circle [radius=3];
\draw [fill=black] (-2.121,-2.121) circle [radius=0.1];
\draw [fill=black] (-2.121, 2.121) circle [radius=0.1];
\draw [fill=black] ( 2.121,-2.121) circle [radius=0.1];
\draw [fill=black] ( 2.121, 2.121) circle [radius=0.1];
\draw [fill=black] (-1, 0) circle [radius=0.1];
\draw [fill=black] ( 1, 0) circle [radius=0.1];
\draw (-2.121,-2.121) -- (-1,0) -- (-2.121, 2.121);
\draw ( 2.121, 2.121) -- ( 1,0) -- ( 2.121,-2.121);
\draw (-1,0) -- (1,0);
\node [left] at (-2.121, 2.2) {$\O_1(\bs{k}_1)$}; 
\node [left] at (-2.121,-2.2) {$\O_2(\bs{k}_2)$}; 	
\node [right] at ( 2.121,-2.2) {$\O_3(\bs{k}_3)$}; 
\node [right] at ( 2.121, 2.2) {$\O_4(\bs{k}_4)$}; 	
\node [right] at (-1.5, 1.2) {$\K_{[\Delta_1]}$};
\node [right] at (-1.5, -1.2) {$\K_{[\Delta_2]}$};
\node [left] at ( 1.5, -1.2) {$\K_{[\Delta_3]}$};
\node [left] at ( 1.5, 1.2) {$\K_{[\Delta_4]}$};
\node [above] at (0,0) {$\G_{[\Delta_x]}$};
\end{tikzpicture}
\centering
\caption{Witten diagrams representing the contact and exchange 4-point amplitudes $\ino_{[\Delta_1 \Delta_2 \Delta_3 \Delta_4]}$ and $\ino_{[\Delta_1 \Delta_2, \Delta_3 \Delta_4 x \Delta_x]}$ given in \eqref{4ptC} and \eqref{4ptX}.\label{fig:intro4pt}}
\end{figure}
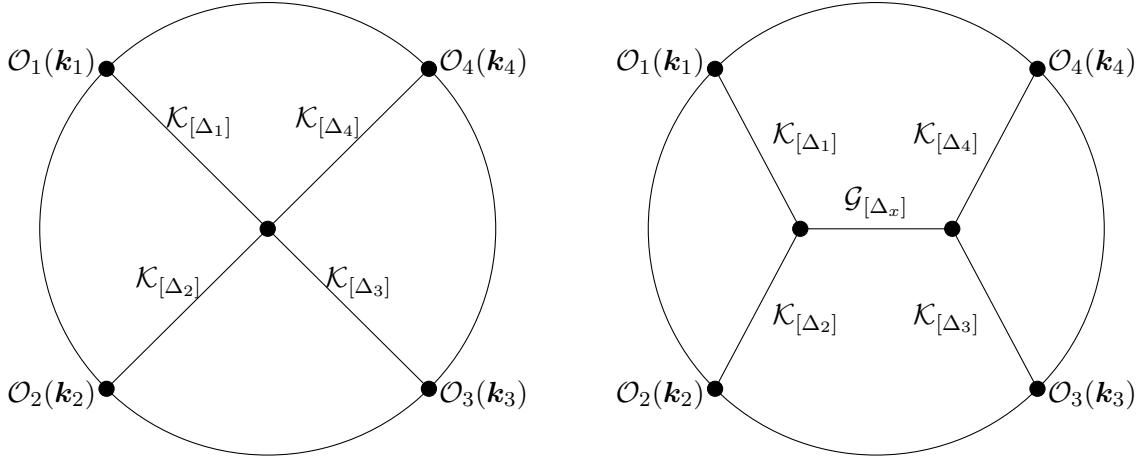

\providecommand{\href}[2]{#2}\begingroup\raggedright\endgroup


\begin{thebibliography}{10}

\bibitem{Bzowski:2022rlz}
A.~Bzowski, P.~McFadden and K.~Skenderis, \emph{{A handbook of holographic
  4-point functions}}, {\emph{JHEP} {\bfseries 12} (2022) 039}
  [\href{https://arxiv.org/abs/2207.02872}{{\ttfamily 2207.02872}}].

\bibitem{toappear}
A.~Bzowski, P.~McFadden and K.~Skenderis, \emph{{Holography for de Sitter
  4-point functions}},  to appear.

\bibitem{McFadden:2009fg}
P.~McFadden and K.~Skenderis, \emph{{Holography for Cosmology}},
  \href{https://doi.org/10.1103/PhysRevD.81.021301}{\emph{Phys. Rev. D}
  {\bfseries 81} (2010) 021301}
  [\href{https://arxiv.org/abs/0907.5542}{{\ttfamily 0907.5542}}].

\bibitem{McFadden:2010na}
P.~McFadden and K.~Skenderis, \emph{{The Holographic Universe}},
  \href{https://doi.org/10.1088/1742-6596/222/1/012007}{\emph{J. Phys. Conf.
  Ser.} {\bfseries 222} (2010) 012007}
  [\href{https://arxiv.org/abs/1001.2007}{{\ttfamily 1001.2007}}].

\bibitem{Maldacena:2011nz}
J.M.~Maldacena and G.L.~Pimentel, \emph{{On graviton non-Gaussianities during
  inflation}}, \href{https://doi.org/10.1007/JHEP09(2011)045}{\emph{JHEP}
  {\bfseries 09} (2011) 045} [\href{https://arxiv.org/abs/1104.2846}{{\ttfamily
  1104.2846}}].

\bibitem{Bzowski:2011ab}
A.~Bzowski, P.~McFadden and K.~Skenderis, \emph{{Holographic predictions for
  cosmological 3-point functions}},
  \href{https://doi.org/10.1007/JHEP03(2012)091}{\emph{JHEP} {\bfseries 03}
  (2012) 091} [\href{https://arxiv.org/abs/1112.1967}{{\ttfamily 1112.1967}}].

\bibitem{Bzowski:2012ih}
A.~Bzowski, P.~McFadden and K.~Skenderis, \emph{{Holography for inflation using
  conformal perturbation theory}},
  \href{https://doi.org/10.1007/JHEP04(2013)047}{\emph{JHEP} {\bfseries 04}
  (2013) 047} [\href{https://arxiv.org/abs/1211.4550}{{\ttfamily 1211.4550}}].

\bibitem{Mata:2012bx}
I.~Mata, S.~Raju and S.~Trivedi, \emph{{CMB from CFT}},
  \href{https://doi.org/10.1007/JHEP07(2013)015}{\emph{JHEP} {\bfseries 07}
  (2013) 015} [\href{https://arxiv.org/abs/1211.5482}{{\ttfamily 1211.5482}}].

\bibitem{McFadden:2013ria}
P.~McFadden, \emph{{On the power spectrum of inflationary cosmologies dual to a
  deformed CFT}}, \href{https://doi.org/10.1007/JHEP10(2013)071}{\emph{JHEP}
  {\bfseries 10} (2013) 071} [\href{https://arxiv.org/abs/1308.0331}{{\ttfamily
  1308.0331}}].

\bibitem{Anninos:2014lwa}
D.~Anninos, T.~Anous, D.Z.~Freedman and G.~Konstantinidis, \emph{{Late-time
  Structure of the Bunch-Davies De Sitter Wavefunction}},
  \href{https://doi.org/10.1088/1475-7516/2015/11/048}{\emph{JCAP} {\bfseries
  1511} (2015) 048} [\href{https://arxiv.org/abs/1406.5490}{{\ttfamily
  1406.5490}}].

\bibitem{Arkani-Hamed:2015bza}
N.~Arkani-Hamed and J.~Maldacena, \emph{{Cosmological Collider Physics}},
  \href{https://arxiv.org/abs/1503.08043}{{\ttfamily 1503.08043}}.

\bibitem{Arkani-Hamed:2018kmz}
N.~Arkani-Hamed, D.~Baumann, H.~Lee and G.L.~Pimentel, \emph{{The Cosmological
  Bootstrap: Inflationary Correlators from Symmetries and Singularities}},
  \href{https://doi.org/10.1007/JHEP04(2020)105}{\emph{JHEP} {\bfseries 04}
  (2020) 105} [\href{https://arxiv.org/abs/1811.00024}{{\ttfamily
  1811.00024}}].

\bibitem{Baumann:2019oyu}
D.~Baumann, C.~Duaso~Pueyo, A.~Joyce, H.~Lee and G.L.~Pimentel, \emph{{The
  cosmological bootstrap: weight-shifting operators and scalar seeds}},
  \href{https://doi.org/10.1007/JHEP12(2020)204}{\emph{JHEP} {\bfseries 12}
  (2020) 204} [\href{https://arxiv.org/abs/1910.14051}{{\ttfamily
  1910.14051}}].

\bibitem{Baumann:2020dch}
D.~Baumann, C.~Duaso~Pueyo, A.~Joyce, H.~Lee and G.L.~Pimentel, \emph{{The
  Cosmological Bootstrap: Spinning Correlators from Symmetries and
  Factorization}},
  \href{https://doi.org/10.21468/SciPostPhys.11.3.071}{\emph{SciPost Phys.}
  {\bfseries 11} (2021) 071}
  [\href{https://arxiv.org/abs/2005.04234}{{\ttfamily 2005.04234}}].

\bibitem{Sleight:2019hfp}
C.~Sleight and M.~Taronna, \emph{{Bootstrapping Inflationary Correlators in
  Mellin Space}},  \href{https://arxiv.org/abs/1907.01143}{{\ttfamily
  1907.01143}}.

\bibitem{Wang:2022eop}
D.-G.~Wang, G.L.~Pimentel and A.~Ach\'ucarro, \emph{{Bootstrapping multi-field
  inflation: non-Gaussianities from light scalars revisited}},
  \href{https://doi.org/10.1088/1475-7516/2023/05/043}{\emph{JCAP} {\bfseries
  05} (2023) 043} [\href{https://arxiv.org/abs/2212.14035}{{\ttfamily
  2212.14035}}].

\bibitem{Arkani-Hamed:2023kig}
N.~Arkani-Hamed, D.~Baumann, A.~Hillman, A.~Joyce, H.~Lee and G.L.~Pimentel,
  \emph{{Differential Equations for Cosmological Correlators}},
  \href{https://arxiv.org/abs/2312.05303}{{\ttfamily 2312.05303}}.

\bibitem{Baumann:2021fxj}
D.~Baumann, W.-M.~Chen, C.~Duaso~Pueyo, A.~Joyce, H.~Lee and G.L.~Pimentel,
  \emph{{Linking the Singularities of Cosmological Correlators}},
  \href{https://arxiv.org/abs/2106.05294}{{\ttfamily 2106.05294}}.

\bibitem{Sleight:2019mgd}
C.~Sleight, \emph{{A Mellin Space Approach to Cosmological Correlators}},
  \href{https://arxiv.org/abs/1906.12302}{{\ttfamily 1906.12302}}.

\bibitem{Sleight:2020obc}
C.~Sleight and M.~Taronna, \emph{{From AdS to dS Exchanges: Spectral
  Representation, Mellin Amplitudes and Crossing}},
  \href{https://arxiv.org/abs/2007.09993}{{\ttfamily 2007.09993}}.

\bibitem{Sleight:2021plv}
C.~Sleight and M.~Taronna, \emph{{From dS to AdS and back}},
  \href{https://doi.org/10.1007/JHEP12(2021)074}{\emph{JHEP} {\bfseries 12}
  (2021) 074} [\href{https://arxiv.org/abs/2109.02725}{{\ttfamily
  2109.02725}}].

\bibitem{Stefanyszyn:2023qov}
D.~Stefanyszyn, X.~Tong and Y.~Zhu, \emph{{Cosmological Correlators Through the
  Looking Glass: Reality, Parity, and Factorisation}},
  \href{https://arxiv.org/abs/2309.07769}{{\ttfamily 2309.07769}}.

\bibitem{DiPietro:2021sjt}
L.~Di~Pietro, V.~Gorbenko and S.~Komatsu, \emph{{Analyticity and Unitarity for
  Cosmological Correlators}},
  \href{https://arxiv.org/abs/2108.01695}{{\ttfamily 2108.01695}}.

\bibitem{Meltzer:2021zin}
D.~Meltzer, \emph{{The inflationary wavefunction from analyticity and
  factorization}},
  \href{https://doi.org/10.1088/1475-7516/2021/12/018}{\emph{JCAP} {\bfseries
  12} (2021) 018} [\href{https://arxiv.org/abs/2107.10266}{{\ttfamily
  2107.10266}}].

\bibitem{Iacobacci:2022yjo}
L.~Iacobacci, C.~Sleight and M.~Taronna, \emph{{From celestial correlators to
  AdS, and back}}, \href{https://doi.org/10.1007/JHEP06(2023)053}{\emph{JHEP}
  {\bfseries 06} (2023) 053}
  [\href{https://arxiv.org/abs/2208.01629}{{\ttfamily 2208.01629}}].

\bibitem{Goodhew:2022ayb}
H.~Goodhew, \emph{{Rational wavefunctions in de Sitter spacetime}},
  \href{https://doi.org/10.1088/1475-7516/2023/03/036}{\emph{JCAP} {\bfseries
  03} (2023) 036} [\href{https://arxiv.org/abs/2210.09977}{{\ttfamily
  2210.09977}}].

\bibitem{Salcedo:2022aal}
S.A.~Salcedo, M.H.G.~Lee, S.~Melville and E.~Pajer, \emph{{The Analytic
  Wavefunction}}, \href{https://doi.org/10.1007/JHEP06(2023)020}{\emph{JHEP}
  {\bfseries 06} (2023) 020}
  [\href{https://arxiv.org/abs/2212.08009}{{\ttfamily 2212.08009}}].

\bibitem{Agui-Salcedo:2023wlq}
S.~Agui-Salcedo and S.~Melville, \emph{{The Cosmological Tree Theorem}},
  \href{https://arxiv.org/abs/2308.00680}{{\ttfamily 2308.00680}}.

\bibitem{Melville:2023kgd}
S.~Melville and G.L.~Pimentel, \emph{{A de Sitter $S$-matrix for the masses}},
  \href{https://arxiv.org/abs/2309.07092}{{\ttfamily 2309.07092}}.

\bibitem{Cespedes:2023aal}
S.~C\'espedes, A.-C.~Davis and D.-G.~Wang, \emph{{On the IR Divergences in de
  Sitter Space: loops, resummation and the semi-classical wavefunction}},
  \href{https://arxiv.org/abs/2311.17990}{{\ttfamily 2311.17990}}.

\bibitem{Armstrong:2022vgl}
C.~Armstrong, A.~Lipstein and J.~Mei, \emph{{Enhanced soft limits in de Sitter
  space}}, \href{https://doi.org/10.1007/JHEP12(2022)064}{\emph{JHEP}
  {\bfseries 12} (2022) 064}
  [\href{https://arxiv.org/abs/2210.02285}{{\ttfamily 2210.02285}}].

\bibitem{Raju:2010by}
S.~Raju, \emph{{BCFW for Witten Diagrams}},
  \href{https://doi.org/10.1103/PhysRevLett.106.091601}{\emph{Phys. Rev. Lett.}
  {\bfseries 106} (2011) 091601}
  [\href{https://arxiv.org/abs/1011.0780}{{\ttfamily 1011.0780}}].

\bibitem{Raju:2012zr}
S.~Raju, \emph{{New Recursion Relations and a Flat Space Limit for AdS/CFT
  Correlators}}, \href{https://doi.org/10.1103/PhysRevD.85.126009}{\emph{Phys.
  Rev. D} {\bfseries 85} (2012) 126009}
  [\href{https://arxiv.org/abs/1201.6449}{{\ttfamily 1201.6449}}].

\bibitem{Raju:2012zs}
S.~Raju, \emph{{Four point functions of the stress tensor and conserved
  currents in AdS$_4$/CFT$_3$}},
  \href{https://doi.org/10.1103/PhysRevD.85.126008}{\emph{Phys.Rev.} {\bfseries
  D85} (2012) 126008} [\href{https://arxiv.org/abs/1201.6452}{{\ttfamily
  1201.6452}}].

\bibitem{Arkani-Hamed:2017fdk}
N.~Arkani-Hamed, P.~Benincasa and A.~Postnikov, \emph{{Cosmological Polytopes
  and the Wavefunction of the Universe}},
  \href{https://arxiv.org/abs/1709.02813}{{\ttfamily 1709.02813}}.

\bibitem{Albayrak:2019asr}
S.~Albayrak, C.~Chowdhury and S.~Kharel, \emph{{New relation for AdS
  amplitudes}},  \href{https://arxiv.org/abs/1904.10043}{{\ttfamily
  1904.10043}}.

\bibitem{Armstrong:2022mfr}
C.~Armstrong, H.~Gomez, R.~Lipinski~Jusinskas, A.~Lipstein and J.~Mei,
  \emph{{New recursion relations for tree-level correlators in
  anti\textendash{}de Sitter spacetime}},
  \href{https://doi.org/10.1103/PhysRevD.106.L121701}{\emph{Phys. Rev. D}
  {\bfseries 106} (2022) L121701}
  [\href{https://arxiv.org/abs/2209.02709}{{\ttfamily 2209.02709}}].

\bibitem{Chen:2023xlt}
Q.~Chen and Y.-X.~Tao, \emph{{Notes on weight-shifting operators and unifying
  relations for cosmological correlators}},
  \href{https://doi.org/10.1103/PhysRevD.108.105005}{\emph{Phys. Rev. D}
  {\bfseries 108} (2023) 105005}
  [\href{https://arxiv.org/abs/2307.00870}{{\ttfamily 2307.00870}}].

\bibitem{Lee:2022fgr}
H.~Lee and X.~Wang, \emph{{Cosmological double-copy relations}},
  \href{https://doi.org/10.1103/PhysRevD.108.L061702}{\emph{Phys. Rev. D}
  {\bfseries 108} (2023) L061702}
  [\href{https://arxiv.org/abs/2212.11282}{{\ttfamily 2212.11282}}].

\bibitem{Groote:2018rpb}
S.~Groote and J.G.~K\"orner, \emph{{Coordinate space calculation of two- and
  three-loop sunrise-type diagrams, elliptic functions and truncated Bessel
  integral identities}},
  \href{https://doi.org/10.1016/j.nuclphysb.2018.11.023}{\emph{Nucl. Phys. B}
  {\bfseries 938} (2019) 416}
  [\href{https://arxiv.org/abs/1804.10570}{{\ttfamily 1804.10570}}].

\bibitem{Isono:2018rrb}
H.~Isono, T.~Noumi and G.~Shiu, \emph{{Momentum space approach to crossing
  symmetric CFT correlators}},
  \href{https://doi.org/10.1007/JHEP07(2018)136}{\emph{JHEP} {\bfseries 07}
  (2018) 136} [\href{https://arxiv.org/abs/1805.11107}{{\ttfamily
  1805.11107}}].

\bibitem{Albayrak:2018tam}
S.~Albayrak and S.~Kharel, \emph{{Towards the higher point holographic momentum
  space amplitudes}},
  \href{https://doi.org/10.1007/JHEP02(2019)040}{\emph{JHEP} {\bfseries 02}
  (2019) 040} [\href{https://arxiv.org/abs/1810.12459}{{\ttfamily
  1810.12459}}].

\bibitem{Maglio:2019grh}
C.~Corianò and M.M.~Maglio, \emph{{On Some Hypergeometric Solutions of the
  Conformal Ward Identities of Scalar 4-point Functions in Momentum Space}},
  \href{https://arxiv.org/abs/1903.05047}{{\ttfamily 1903.05047}}.

\bibitem{Isono:2019wex}
H.~Isono, T.~Noumi and G.~Shiu, \emph{{Momentum space approach to crossing
  symmetric CFT correlators. Part II. General spacetime dimension}},
  \href{https://doi.org/10.1007/JHEP10(2019)183}{\emph{JHEP} {\bfseries 10}
  (2019) 183} [\href{https://arxiv.org/abs/1908.04572}{{\ttfamily
  1908.04572}}].

\bibitem{Coriano:2019nkw}
C.~Corianò, M.M.~Maglio and D.~Theofilopoulos, \emph{{Four-Point Functions in
  Momentum Space: Conformal Ward Identities in the Scalar/Tensor case}},
  \href{https://doi.org/10.1140/epjc/s10052-020-8089-1}{\emph{Eur. Phys. J. C}
  {\bfseries 80} (2020) 540}
  [\href{https://arxiv.org/abs/1912.01907}{{\ttfamily 1912.01907}}].

\bibitem{Albayrak:2020isk}
S.~Albayrak, C.~Chowdhury and S.~Kharel, \emph{{Study of momentum space scalar
  amplitudes in AdS spacetime}},
  \href{https://doi.org/10.1103/PhysRevD.101.124043}{\emph{Phys. Rev. D}
  {\bfseries 101} (2020) 124043}
  [\href{https://arxiv.org/abs/2001.06777}{{\ttfamily 2001.06777}}].

\bibitem{Bonifacio:2022vwa}
J.~Bonifacio, H.~Goodhew, A.~Joyce, E.~Pajer and D.~Stefanyszyn, \emph{{The
  graviton four-point function in de Sitter space}},
  \href{https://doi.org/10.1007/JHEP06(2023)212}{\emph{JHEP} {\bfseries 06}
  (2023) 212} [\href{https://arxiv.org/abs/2212.07370}{{\ttfamily
  2212.07370}}].

\bibitem{Albayrak:2023kfk}
S.~Albayrak, S.~Kharel and X.~Wang, \emph{{Momentum-space formulae for AdS
  correlators for diverse theories in diverse dimensions}},
  \href{https://arxiv.org/abs/2312.02154}{{\ttfamily 2312.02154}}.

\bibitem{Albayrak:2023jzl}
S.~Albayrak and S.~Kharel, \emph{{All plus four point (A)dS graviton function
  using generalized on-shell recursion relation}},
  \href{https://doi.org/10.1007/JHEP05(2023)151}{\emph{JHEP} {\bfseries 05}
  (2023) 151} [\href{https://arxiv.org/abs/2302.09089}{{\ttfamily
  2302.09089}}].

\bibitem{Chowdhury:2023khl}
C.~Chowdhury and K.~Singh, \emph{{Analytic Results for Loop-Level Momentum
  Space Witten Diagrams}},  \href{https://arxiv.org/abs/2305.18529}{{\ttfamily
  2305.18529}}.

\bibitem{Bzowski:2015pba}
A.~Bzowski, P.~McFadden and K.~Skenderis, \emph{{Scalar 3-point functions in
  CFT: renormalisation, beta functions and anomalies}},
  \href{https://doi.org/10.1007/JHEP03(2016)066}{\emph{JHEP} {\bfseries 03}
  (2016) 066} [\href{https://arxiv.org/abs/1510.08442}{{\ttfamily
  1510.08442}}].

\bibitem{Bzowski:2015yxv}
A.~Bzowski, P.~McFadden and K.~Skenderis, \emph{{Evaluation of conformal
  integrals}}, \href{https://doi.org/10.1007/JHEP02(2016)068}{\emph{JHEP}
  {\bfseries 02} (2016) 068}
  [\href{https://arxiv.org/abs/1511.02357}{{\ttfamily 1511.02357}}].

\bibitem{Bzowski:2017poo}
A.~Bzowski, P.~McFadden and K.~Skenderis, \emph{{Renormalised 3-point functions
  of stress tensors and conserved currents in CFT}},
  \href{https://doi.org/10.1007/JHEP11(2018)153}{\emph{JHEP} {\bfseries 11}
  (2018) 153} [\href{https://arxiv.org/abs/1711.09105}{{\ttfamily
  1711.09105}}].

\bibitem{Bzowski:2018fql}
A.~Bzowski, P.~McFadden and K.~Skenderis, \emph{{Renormalised CFT 3-point
  functions of scalars, currents and stress tensors}},
  \href{https://doi.org/10.1007/JHEP11(2018)159}{\emph{JHEP} {\bfseries 11}
  (2018) 159} [\href{https://arxiv.org/abs/1805.12100}{{\ttfamily
  1805.12100}}].

\bibitem{Bzowski:2016kni}
A.~Bzowski, \emph{{Dimensional renormalization in AdS/CFT}},
  \href{https://arxiv.org/abs/1612.03915}{{\ttfamily 1612.03915}}.

\bibitem{Mack:2009mi}
G.~Mack, \emph{{D-independent representation of conformal field theories in D
  dimensions via transformation to auxiliary dual resonance models. Scalar
  amplitudes}},  \href{https://arxiv.org/abs/0907.2407}{{\ttfamily 0907.2407}}.

\bibitem{Penedones:2010ue}
J.~Penedones, \emph{{Writing CFT correlation functions as AdS scattering
  amplitudes}}, \href{https://doi.org/10.1007/JHEP03(2011)025}{\emph{JHEP}
  {\bfseries 03} (2011) 025} [\href{https://arxiv.org/abs/1011.1485}{{\ttfamily
  1011.1485}}].

\bibitem{Fitzpatrick:2011ia}
A.L.~Fitzpatrick, J.~Kaplan, J.~Penedones, S.~Raju and B.C.~van Rees, \emph{{A
  Natural Language for AdS/CFT Correlators}},
  \href{https://doi.org/10.1007/JHEP11(2011)095}{\emph{JHEP} {\bfseries 11}
  (2011) 095} [\href{https://arxiv.org/abs/1107.1499}{{\ttfamily 1107.1499}}].

\bibitem{Coriano:2013jba}
C.~Corian\`{o}, L.~Delle~Rose, E.~Mottola and M.~Serino, \emph{{Solving the
  Conformal Constraints for Scalar Operators in Momentum Space and the
  Evaluation of Feynman's Master Integrals}},
  \href{https://doi.org/10.1007/JHEP07(2013)011}{\emph{JHEP} {\bfseries 07}
  (2013) 011} [\href{https://arxiv.org/abs/1304.6944}{{\ttfamily 1304.6944}}].

\bibitem{Bzowski:2013sza}
A.~Bzowski, P.~McFadden and K.~Skenderis, \emph{{Implications of conformal
  invariance in momentum space}},
  \href{https://doi.org/10.1007/JHEP03(2014)111}{\emph{JHEP} {\bfseries 03}
  (2014) 111} [\href{https://arxiv.org/abs/1304.7760}{{\ttfamily 1304.7760}}].

\bibitem{Coriano:2019sth}
C.~Corian\`o and M.M.~Maglio, \emph{{On Some Hypergeometric Solutions of the
  Conformal Ward Identities of Scalar 4-point Functions in Momentum Space}},
  \href{https://doi.org/10.1007/JHEP09(2019)107}{\emph{JHEP} {\bfseries 09}
  (2019) 107} [\href{https://arxiv.org/abs/1903.05047}{{\ttfamily
  1903.05047}}].

\bibitem{paule2008algorithms}
P.~Paule and B.~Sturmfels, \emph{Algorithms in Invariant Theory}, Texts \&
  Monographs in Symbolic Computation, Springer Vienna (2008).

\bibitem{Boos:1987bg}
E.E.~Boos and A.I.~Davydychev, \emph{{A Method of the Evaluation of the Vertex
  Type Feynman Integrals}}, {\emph{Moscow Univ. Phys. Bull.} {\bfseries 42N3}
  (1987) 6}.

\bibitem{Davydychev:1992xr}
A.I.~Davydychev, \emph{{Recursive algorithm of evaluating vertex type Feynman
  integrals}}, {\emph{J. Phys. A} {\bfseries 25} (1992) 5587}.

\bibitem{tHooft:1978jhc}
G.~'t~Hooft and M.J.G.~Veltman, \emph{{Scalar One Loop Integrals}},
  \href{https://doi.org/10.1016/0550-3213(79)90605-9}{\emph{Nucl. Phys. B}
  {\bfseries 153} (1979) 365}.

\bibitem{Cheung:2007st}
C.~Cheung, P.~Creminelli, A.L.~Fitzpatrick, J.~Kaplan and L.~Senatore,
  \emph{{The Effective Field Theory of Inflation}},
  \href{https://doi.org/10.1088/1126-6708/2008/03/014}{\emph{JHEP} {\bfseries
  03} (2008) 014} [\href{https://arxiv.org/abs/0709.0293}{{\ttfamily
  0709.0293}}].

\bibitem{Weinberg:2008hq}
S.~Weinberg, \emph{{Effective Field Theory for Inflation}},
  \href{https://doi.org/10.1103/PhysRevD.77.123541}{\emph{Phys. Rev. D}
  {\bfseries 77} (2008) 123541}
  [\href{https://arxiv.org/abs/0804.4291}{{\ttfamily 0804.4291}}].

\bibitem{Huber:2005yg}
T.~Huber and D.~Maitre, \emph{{HypExp: A Mathematica package for expanding
  hypergeometric functions around integer-valued parameters}},
  \href{https://doi.org/10.1016/j.cpc.2006.01.007}{\emph{Comput. Phys. Commun.}
  {\bfseries 175} (2006) 122}
  [\href{https://arxiv.org/abs/hep-ph/0507094}{{\ttfamily hep-ph/0507094}}].

\bibitem{Skenderis:2002wp}
K.~Skenderis, \emph{{Lecture notes on holographic renormalization}},
  \href{https://doi.org/10.1088/0264-9381/19/22/306}{\emph{Class. Quant. Grav.}
  {\bfseries 19} (2002) 5849}
  [\href{https://arxiv.org/abs/hep-th/0209067}{{\ttfamily hep-th/0209067}}].

\end{thebibliography}
\end{document}